\documentstyle{mn2e}
\input psfig

  \makeatletter
  \ifx\CUP@mtlplain@loaded\undefined  % this is a font switch for CUP
  \makeatother  
    
  \else
    
  \fi

\loadboldmathitalic % loads bold math italic
\loadboldgreek      % loads bold greek
  
\makeatletter
\ifx\CUP@mtlplain@loaded\undefined  % this is a font switch for CUP
  \newfont\bit{cmbxti10 at 9pt}
\else
  \newfont\bit{mtbxti10 at 9pt}
\fi
\makeatother

\def\LaTeX{L\kern-.36em\raise.3ex\hbox{a}\kern-.15em
    T\kern-.1667em\lower.7ex\hbox{E}\kern-.125emX}

\newcommand{\gsim}{\mathrel{\hbox{\rlap{\lower.55ex \hbox {$\sim$}}
                   \kern-.3em \raise.4ex \hbox{$>$}}}}
\newcommand{\lsim}{\mathrel{\hbox{\rlap{\lower.55ex \hbox {$\sim$}}
                   \kern-.3em \raise.4ex \hbox{$<$}}}}
\newcommand{\msun}{\mbox{M$_\odot$}}
\newcommand{\me}{\mbox{M$_{\rm \oplus}$}}
\newcommand{\mj}{\mbox{M$_{\rm J}$}}
\newcommand{\mpl}{M_{\rm p}}

\title[Three-dimensional calculations of planets in discs]{Three-dimensional calculations of high and low-mass planets embedded in protoplanetary discs}

\author[M. R. Bate et~al.]
  {M. R. Bate,$^{1}$\thanks{E-mail: mbate@astro.ex.ac.uk}
  S. H. Lubow,$^2$
  G. I. Ogilvie,$^3$
  and K. A. Miller.$^{4}$\\
  $^1$School of Physics, University of Exeter, Stocker Road,
    Exeter EX4 4QL \\
  $^2$Space Telescope Science Institute, 3700 San Martin Drive, Balitmore, MD 21218, USA \\
  $^3$Institute of Astronomy, University of Cambridge, Madingley Road,
    Cambridge CB3 0HA \\
  $^4$Astronomy Department, University of Maryland, College Park, MD 20742, U.S.A.
}

\date{Accepted by MNRAS}

\begin{document}

\maketitle

\begin{abstract}

We analyse the non-linear, three-dimensional response  of a gaseous,   
viscous protoplanetary disc to the presence of a planet of mass ranging
from one Earth mass (1 \me) to one Jupiter mass (1 \mj) by using the   
ZEUS hydrodynamics code.  We determine the gas flow pattern, and the accretion
and migration rates of the planet.  The planet is assumed to be in a 
fixed circular orbit about the central star. It is also 
assumed to be able to accrete gas without expansion on the scale
of its Roche radius.   Only planets with masses $\mpl\gsim 0.1$ \mj\  
produce significant perturbations in the disc's surface density.  The
flow within the Roche lobe of the planet is fully three-dimensional.   
Gas streams generally enter the Roche lobe close to the disc midplane, but
produce much weaker shocks than the streams in two-dimensional models.
The streams supply material to a circumplanetary disc that rotates in
the same sense as the planet's orbit.  Much of the mass supply to the
circumplanetary disc comes from non-coplanar flow.  The accretion rate peaks
with a planet mass of approximately 0.1 \mj\ and is highly efficient, occurring
at the local viscous rate. The migration timescales for planets of mass
less than 0.1 \mj, based on torques from disc material outside the
planets' Roche lobes, are in excellent agreement with the linear theory
of Type I (non-gap) migration for three-dimensional discs. The
transition from Type I to Type II (gap) migration is smooth, with
changes in migration times of about a factor of 2.  Starting with
a core which can undergo runaway growth, a planet can gain up to a
few \mj\ with little migration.  Planets with final masses of order 
10 \mj\ would undergo large migration, which makes formation and 
survival difficult.

\end{abstract}


\section{Introduction}

Young planets interact with the surrounding discs from which they form
by accreting mass and exerting torques.  In the core-accretion model
for planet formation, the initial growth proceeds as solids accumulate
to form a planetary core (Lissauer 1995, Wuchterl, Guillot \& 
Lissauer 2000).  The
core may become a terrestrial planet or may further develop to become a
gas giant through accretion of gas. The torques
resulting from planet-disc interactions cause a planet to migrate
typically inwards (Lin et al.\ 2000; Ward and Hahn 2000).

Numerical hydrodynamical calculations have been carried out in 
two dimensions in order to
understand better the dynamics of a circular orbit planet embedded in
a gaseous disc (e.g.\ Bryden et al.\ 1999; Kley 1999; 
Lubow, Seibert \& Artymowicz 1999; Nelson et al.\ 2000).
These studies concentrated on the interaction between a 1 \mj\ planet
and a gas disc that orbit a 1 \msun\ star. The tidal forces
caused by the planet create a gap in the disc. In spite of the
presence of the gap, disc mass flow on to the planet continues through
the gap with high efficiency.  Nearly all the flow through the gap is
accreted by the planet at a rate comparable to
the rate at which mass accretion would occur in the disc in the
absence of a planet.

The flow within the Roche lobe of the planet is highly non-axisymmetric
and involves shocks produced by colliding gas streams (Lubow et al.\ 
1999; D'Angelo, Henning \& Kley 2002). 
Torques on the planet are exerted by circumstellar disc
material that lies outside the gap. 
In addition, torques are exerted locally by the
material that flows close to the planet. The net torque results in
inward migration. Gas accretion can continue to planet masses of order
10 \mj, at which point tidal forces are sufficiently strong to 
prevent flow into the gap.

These earlier studies were limited by their neglect of effects in the
vertical direction (perpendicular to the orbit plane). The Roche lobe
radius of a 1 \mj\ planet orbiting a 1 \msun\ star is comparable
to the local disc thickness. The effects in the vertical direction
are even more important for lower-mass planets. In addition,
sufficiently low-mass planets do not open a gap in the disc.  Analytic
models of planet migration in the non-gap case, sometimes called Type I
migration, were carried out in two dimensions (Ward 1997). The
results implied that planetary migration timescales are much shorter
than the disc lifetimes.  Planets would then be accreted by the
central star.  Alternatively, the planets could reside at the
circumstellar disc inner edge, if there is a central hole in the disc.
This situation poses problems for planet survival and for gas
giant planet formation.  Recent two and three dimensional 
analytic calculations by Tanaka,
Takeuchi, and Ward (2002) of the migration rates of low-mass planets 
obtained smaller values (by about an order of magnitude), making
survival more plausible.  

The two-dimensional numerical results also indicated that accretion 
within the Roche lobe of a 1 \mj\ planet is driven by shocks
(Lubow et al.\ 1999; D'Angelo et al.\ 2002). It is not clear
whether the same flow pattern would persist in three dimensions. Recently,
Kley, D'Angelo, and Henning (2001) computed
the three-dimensional flow for 0.5 and 1 \mj\ planets and 
found minor differences in the accretion and migration rates 
compared to the two-dimensional case. However,
the flow within the Roche lobe was unresolved.

In this paper, we analyze the non-linear interactions
between a planet and a gaseous disc by using global 
three-dimensional numerical simulations
that resolve the flow within the Roche lobe of a 1 \mj\ planet.
We are interested in determining the flow patterns, the accretion rates,
and migration rates for planets whose masses range between 1 \me\ (Earth
mass) and 1 \mj\ (Jupiter mass).  Each planet is assumed to be 
in a circular orbit about the central star. 
The mass and orbital radius of each planet are fixed during the simulation;
consequently, the results do not include the effects on the flow
of planetary migration.

The outline of the paper is as follows. Section 2 describes the computational
procedure. Section 3 provides the results. In Section 4, we discuss the
implications of the results for giant planet formation.
Section 5 contains our conclusions.

\section{Computational method}

\subsection{Basic equations}

We use a computational method that is similar to that of
Lubow et al.\ \shortcite{LubSeiArt1999}, except that they
performed two-dimensional vertically averaged calculations 
whereas we solve the problem in three dimensions.

We assume a viscous model for the disc turbulence, with
the usual $\alpha$ prescription.
The origin of the coordinate system is taken to be the centre 
of the star.  (We ignore the slight centre-of-mass
shift caused by the planet.)  The disc self-gravity is ignored.
The flow is modelled in the orbital frame
that rotates with the angular speed of the planet 
$\Omega_{\rm p}= \sqrt{GM_*/r_{\rm p}^3}$, where $M_*$ is the mass
of the star, $r_{\rm p}$ is the orbital radius of the planet,
$G$ is the gravitational constant, and we have neglected the
mass of the planet.
In this frame, the flow achieves a near steady state.  We adopt
spherical coordinates $(r,\theta,\phi)$ with associated flow
velocities in the rotating frame 
${\bf u} = (u_{\rm r}, u_{\theta}, u_{\rm \phi})$.  
The equations of motion for the disc are
\begin{equation}
\label{cons-mass}
\frac{\partial \rho}{\partial t} + \nabla\cdot(\rho {\bf u})=0,
\end{equation}
\begin{eqnarray}
\label{rad-force}
\frac{\partial s_{\rm r}}{\partial t} + 
\nabla \cdot(s_{\rm r} {\bf u}) & = & 
\rho r \sin^2\theta\left(\frac{u_{\phi}}{r \sin\theta} + \Omega_p\right)^2  +{{\rho u_\theta^2}\over{r}} \nonumber \\ & & - \frac{\partial p}{\partial r}
- \rho \frac{\partial \Phi}{\partial r} + f_{r},
\end{eqnarray}
\begin{eqnarray}
\label{theta-force}
\frac{\partial s_{\rm \theta}}{\partial t} + 
\nabla \cdot(s_{\rm \theta} {\bf u}) & = &
\rho r^2 \sin\theta\cos\theta\left(\frac{u_{\phi}}{r \sin\theta} + \Omega_p\right)^2 \nonumber \\ & &  - \frac{\partial p}{\partial \theta}
- {\rho} \frac{\partial \Phi}{\partial \theta} + r f_{\theta},
\end{eqnarray}
and
\begin{equation}
\label{ang-mom}
\frac{\partial s_\phi}{\partial t} +
\nabla \cdot(s_\phi {\bf u}) =
- \frac{\partial p}{ \partial \phi}
- \rho \frac{\partial \Phi}{ \partial \phi} + r \sin \theta f_{\phi},
\end{equation}
where $\rho$ is the gas density, $p$ is the gas pressure, 
${s}_{\rm r} = \rho {u}_{\rm r}$ is the radial momentum per unit volume,   
${s}_{\rm \theta} = \rho r {u}_{\rm \theta}$ is the meridional momentum 
per unit volume,   
$s_{\rm \phi} = \rho r \sin \theta (u_{\phi} + \Omega_p r\sin \theta ) $ is the 
azimuthal angular
momentum per unit volume,
$\Phi$
is the gravitational potential due 
to the central star and the planet,
and  ${\bf f}=(f_{\rm r}, f_{\theta}, f_{\phi})$ is the viscous 
force per unit volume that describes
the effects of disc turbulence.
We use an unsoftened gravitational potential
\begin{equation}
\Phi({\bf r})=-\frac{GM_{\rm *}}{r} - \frac{GM_{\rm p}}{|{\bf r} - {\bf r}_{\rm p}|},
\end{equation}
where $M_{\rm p}$ is the planet mass.  This is possible because the location
of the planet is such that all gravitational force evaluations are made at
a finite distance from the planet.  We also performed some test 
calculations with gravity softened on the length scale of twice the grid
resolution near the planet, but found no significant difference between the
softened and unsoftened results.

Equations (\ref{cons-mass}), (\ref{rad-force}), (\ref{theta-force}),
and (\ref{ang-mom}) express conservation of mass, radial momentum,
meridional momentum, and azimuthal angular momentum, 
respectively.
Equation (\ref{ang-mom}) is written in terms of the
total azimuthal angular momentum
$s_{\phi}$, rather than that in the rotating 
frame.  The reason is that the $s_{\phi}$ equation provides better 
numerical stability \cite{Kley1998}.

The equation of state is taken to be locally isothermal, 
$p\propto \rho T$, with the temperature expressed as a specified 
function of radius, $T(r)$.  This equation of state is 
appropriate for a gas that radiates internal energy gained by 
shocks with high efficiency.  The viscosity force 
${\bf f}$ is assumed to be the standard Navier-Stokes force
(see eq. [15.3] of Landau \& Lifshitz 1975; Klahr, Henning \& Kley
1999).  The coefficient 
of shear viscoity $\mu$ represents the effects of disc turbulence,
while the bulk viscosity coefficient $\zeta$ is set to zero.  
The value for the kinematic turbulent viscosity $\nu=\mu/\rho$
is assumed to be constant in space and time.
It can be expressed in terms of the usual $\alpha$ prescription of 
Shakura \& Sunyaev \shortcite{ShaSun1973}.  Namely,
for a disc with local isothermal sound speed $c_{\rm s}$ 
and vertical scaleheight $H$,
dimensionless parameter $\alpha$ is defined through
\begin{equation}
\label{nu}
\alpha(r) = \frac{\nu}{ c_{\rm s} H}.
\end{equation}

The above equations are non-dimensionalized so that the unit
of time is the inverse of the planetary orbital frequency 
$\Omega_{\rm p}$, the unit of distance is the orbital radius
of the planet $r_{\rm p}$, and $G=1$.

\subsection{Numerical method}

The equations are solved using a three-dimensional spherical 
coordinate version of the ZEUS-2D code \cite{StoNor1992}.
The code was written by K.~A.\ Miller and J.~M.\ Stone.  
It was modified to include a standard three-dimensional 
Navier-Stokes viscous force term.  The code allows
for variably-spaced gridding, which permits us to obtain higher
resolution in the vicinity of the planet.  The timesteps satisfy
the usual Courant condition, for which we have adopted Courant
number 0.3.  The code provides an artificial viscosity term,
but since we introduce a Navier-Stokes viscous force, we 
suppress this artificial viscosity.  Of course, there is 
some intrinsic numerical viscosity because of the finite 
gridding.  For a uniformly spaced mesh, the code is formally 
second-order accurate in space and first-order accurate in time.
For variably spaced meshes, it is
formally first-order accurate in space.  However, a high level of
accuracy and resolution can be attained by limiting the fractional
change in mesh spacing between adjacent cells to be small, of 
order 1\%.  The code uses van Leer interpolation.

\subsection{Numerical grid and initial conditions}

The planet was fixed at location $(r,\theta,\phi)=(1,\pi/2,\pi)$.
We modelled the disc in the region $r\in[0.3,4.0]$, 
$\theta\in[\pi/2 - 4H/r, \pi/2]$, and $\phi\in[0,2\pi]$.
We imposed reflective boundary conditions at the radial and 
$\theta$ grid boundaries and periodic boundary conditions
at the azimuthal boundary.
The radial boundaries were sufficiently far from the planet
that the reflected waves were not noticeable.
The grid was uniform in $\theta$, but non-uniform in $r$ and 
$\phi$.
Both the $r$ and $\phi$ grids were uniform in the vicinity of the
planet, but the grid spacing increased logarithmically away from
this uniform region (i.e.\ each zone was a certain percentage 
larger or smaller than the zone preceding it).  
In addition, the last two zones inside 
$\phi=0$ and $\phi=2\pi$ were uniform to allow periodic boundaries
to be implemented easily.  Each region of the radial grid had
$[0.3,0.8965]=28$, $[0.8965,1.1035]=72$, $[1.1035,4.0]=80$ zones.
The $\phi$ grid had $[0,0.19]=[2\pi-0.19,2\pi]=2$, 
$[0.19,\pi-0.1035]=[\pi+0.1035,2\pi-0.19]=106$,
$[\pi-0.1035,\pi+0.1035]=72$ zones in each region.  The $\theta$
grid modelled 4 scale heights above the disc midplane using 
36 zones.  Thus, the 
zones in the vicinity of the planet had dimensions of 
0.002875 in $r$ and $\phi$, and twice this value in $\theta$.  
This grid was decided upon after extensive testing using two and 
three-dimensional calculations.  In particular, we found that if the
change in size between two neighbouring zones is too large or if
the $r$ and $\phi$ dimensions of zones are too different, numerical
instabilities occur.  The instabilities are worst in two-dimensional
calculations, but also appear in three-dimensional calculations 
that have the same gridding in $r$ and $\phi$.  We also performed
two-dimensional calculations with increasing resolution in 
$r$ and $\phi$ to test for convergence.  Lowering the resolution in either 
$r$ or $\phi$ tends to increase the accretion rates.  
Convergence testing of three-dimensional calculations was not practical
due to the factor of 16 increase in computational time that would have
been required to double the above resolution in all three directions.  
The above resolution was chosen as the minimum required both 
to avoid numerical instabilities and to give quasi-steady-state 
accretion rates that had converged to within a few percent in the
two-dimensional calculations.

The Roche lobe of the planet had a radius
\begin{equation}
\label{rocherad}
r_{\rm R} = \left(\frac{M_{\rm p}}{3 M_{\rm *}}\right)^{1/3}r_{\rm p}.
\end{equation}
We modelled planets from 1 Earth mass ($3\times 10^{-6}$ \msun) 
to 1 Jupiter mass ($1\times 10^{-3}$ \msun).  With the mass of the
star equal to 1 \msun, the planets' Roche 
radii ranged from $r_{\rm R}/r_{\rm p}=0.010$ to 0.069.  
Thus, the Roche radius of the 
Jupiter-mass planet was resolved by 24 zones (i.e.\ the Roche 
lobe contains about $3\times 10^4$ zones), while that of 
the Earth-mass planet
was marginally resolved by 3.5 zones (i.e.\ the Roche lobe 
contains about $80$ zones).  

The planet was assumed to be able 
to accrete material without substantial expansion of its radius 
on the scale of its Roche lobe.  
Some models of planet evolution suggest that this assumption is valid
only for planets of mass greater than about 50 \me\ (Pollack et
al 1996).  However, these models are subject to number of major
simplifying assumptions, as discussed in Wuchterl et al.\ (2000). We
return to this point in Section 4. In any case, the accretion
assumption provides a simple prescription for handling the accretion
flow that allows us to compute torques on the planet due to its
interaction with the disc.

To simulate the accretion on to the planet, we nearly fully removed
material in the four grid 
zones that surrounded the location of the planet in each timestep. 
A residual density was retained in these
zones to avoid numerical divergences.  The removed material was 
assumed to be accreted on to the planet, although the mass of the
planet was not increased.  There was a pressure force directed 
toward the planet at the edge of the evacuated region.  This 
force was small compared with other dynamical forces for massive planets,
but was noticable for the lowest mass planets that we modelled 
(see Section \ref{gasflow}).

A temperature profile $T(r)\propto r^{-1}$ was used at all times.
It was normalised such that 
$H/r_{\rm p}=c_{\rm s}/(\Omega_{\rm p}r_{\rm p})=0.05$.  
This profile
gives a constant value of $H/r$ throughout the unperturbed disc.
It is numerically convenient when using a spherical grid, since 
the vertical resolution of the disc is constant.  We set the 
kinematic disc viscosity to $\nu=10^{-5}$ in our dimensionless units, 
which corresponds to $\alpha=4\times 10^{-3}$ at $r=r_{\rm p}=1$ 
(equation \ref{nu}).

The underlying initial disc density profile was chosen to be 
axisymmetric and follow $\rho(r, \theta, \phi) \propto r^{-3/2} 
\exp{[-(\theta-\pi/2)^2 r^2/(2 H^2)]}$.  Thus, the disc surface density
$\Sigma(r)\propto r^{-1/2}$.  
For planets with masses $M_p \geq 0.1$\mj,  we imposed an initial
gap near the planet.  The initial gap size and
structure were estimated by an approximate torque balance 
condition between viscous and tidal torques near the planet.  
For the Jupiter-mass planet, the density at the middle of the gap
was 1\% of the unperturbed density.  For the 0.3 \mj\ and 0.1 \mj\ planets
the initial gap densities were 4.2\% and 35\% of the unperturbed
density, respectively.

In order to confirm that the presence of an initial
gap did not affect the final results, we performed two 
calculations of the 0.1 \mj\ planet, one with an initial
gap and one without.  The accretion and migration rates 
of the calculation with the initial gap reached steady values 
after about 30 orbits.  The calculation without an initial
gap initially had much higher accretion and migration rates, 
but these converged to the values given by the other
calculation after approximately 100 orbits.  Thus, the presence
of an initial gap for the high-mass planets does not affect 
the final results, but it significantly reduced the 
computational time.

\begin{figure}
\centerline{\psfig{figure=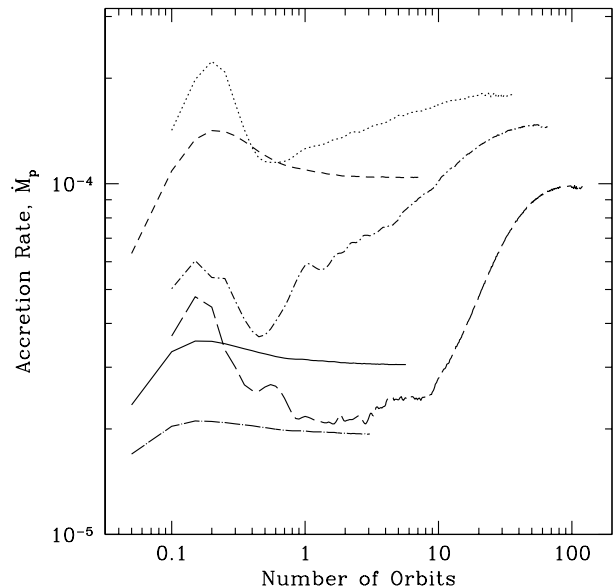,width=8.5truecm}}
\caption{\label{accvstime} The accretion rate versus time for each of the 6 planet calculations.  The accretion rate is averaged over every 1/20 of an orbit and plotted in units of the disc mass (7.5 \mj) per orbit.  The calculations were run until the accretion rates reached quasi-steady values.  The lines give the accretion rates for planets with masses of 1 (long-dashed), 0.3 (dot-dashed), 0.1 (dotted), 0.03 (short-dashed), 0.01 (solid), and 0.003 (dot-long-dashed) \mj.  Low mass planets reach steady accretion rates after only a few orbits, while the highest mass planets require approximately 100 orbits.}
\end{figure}

\section{Results}

\subsection{Calculations}

We performed calculations of 6 planets with masses of 
1, 0.3, 0.1, 0.03, 0.01, and 0.003 Jupiter masses, \mj\
(i.e.\ 330, 100, 33, 10, 3.3, and 1 Earth masses, \me).  
The results are scaled so that the planet is at a distance of 
5.2 AU from a 1 \msun\ star.  The disc mass, $M_{\rm d}$, between 
1.56 and 20.8 AU (the boundaries of the grid) is taken to be
7.5 Jupiter masses (i.e.\ 0.0075 \msun).  This gives an
unperturbed disc surface density of 75 g~cm$^{-2}$ at the radius
of the planet.

The calculations were run until the accretion rate on to the 
planet reached a quasi-steady state (Figure \ref{accvstime}).  
The number of orbits required depended on the mass of the planet.  
Lower-mass planets reached a steady state after only about 5 orbits, 
while the 1 \mj\ planet required approximately 100 orbits.  
The calculations never quite reach a true steady state because the 
protoplanetary
disc is evolving due to the Navier-Stokes viscosity.  Thus, the accretion
rates of some of the high-mass planets can be seen to decrease slightly
with time after the quasi-steady-state accretion is reached.

The calculations were performed on the United Kingdom 
Astrophysical Fluids Facility (UKAFF), a 128-processor
SGI Origin 3800 computer, and on GRAND, a 24-processor
SGI Origin 2000 computer.  The computational time required 
for each calculation was relatively independent of the planet's
mass and required approximately 160 CPU hours per orbit on UKAFF.  
The total CPU time required for all the calculations was 
approximately 50000 CPU hours.

\subsection{Density structure and gas flow}
\label{gasflow}

The interaction of the planet with the disc alters the density
and flow of the gas in the vicinity of the planet.  This interaction
was analysed by means of two-dimensional simulations for high-mass planets
by Lubow et al.\ \shortcite{LubSeiArt1999}.  
Recently, D'Angelo et al.\ \shortcite{DAnHenKle2002}
studied the gas flow near low-mass planets by using two-dimensional 
simulations.  Here, we study the
problem in three dimensions for both high and low-mass planets.

\begin{figure}
\centerline{\psfig{figure=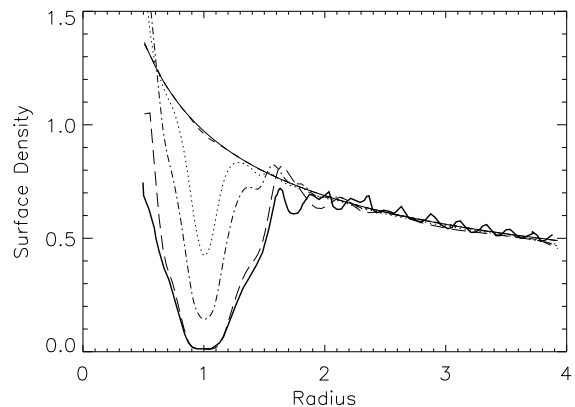,width=8.5truecm}}
\caption{\label{surfdens} The final azimuthally averaged disc surface density for 1 (long-dashed), 0.3 (dot-dashed), 0.1 (dotted), 0.03 (short-dashed), and 0.01 (thin solid) \mj\ planets.  Only planets with masses $\mpl \gsim 0.1$ \mj\ ($\mpl\gsim 30$ \me) produce significant perturbations.  The thick solid line gives the result for a 1 \mj\ planet from the two-dimensional calculations of Lubow et al.\ (1999). }
\end{figure}

\subsubsection{Global features}

Figure \ref{surfdens} plots the azimuthally averaged surface
density as a function of radius for the planets with masses
ranging from 1 to 0.01 \mj.  Also plotted (with
the thick solid line) is the two-dimensional result for a 1 \mj\ planet
after 150 orbits from Figure 1 of Lubow et al.\ (1999).  
The surface density in the vicinity a 1 \mj\ planet is well 
modelled by two-dimensional calculations (see also Kley et al.\ 2001).  
Only planets with masses
$\mpl\gsim 0.1$ \mj\ produce significant perturbations in the
disc's surface density; $0.03$ \mj\ (10 \me) is insufficient.
This result is consistent with the expectation that a gap will start
to open when the Roche radius of the planet is comparable to
the disc scaleheight $H$.  The Roche radii of planets with
masses 0.3, 0.1, and 0.03 \mj\ are $r_{\rm R}/r_{\rm p}=0.046,
0.032,$ and 0.022, respectively.  Comparing these to $H/r=0.05$,
we find that $r_{\rm R}>H/2$ is required for a significant
surface density perturbation.  Even for $H=r_{\rm R}$
(the 0.3 \mj\ case), the gap only drops to 15\% of the
unperturbed density.

\begin{figure*}
\centerline{\psfig{figure=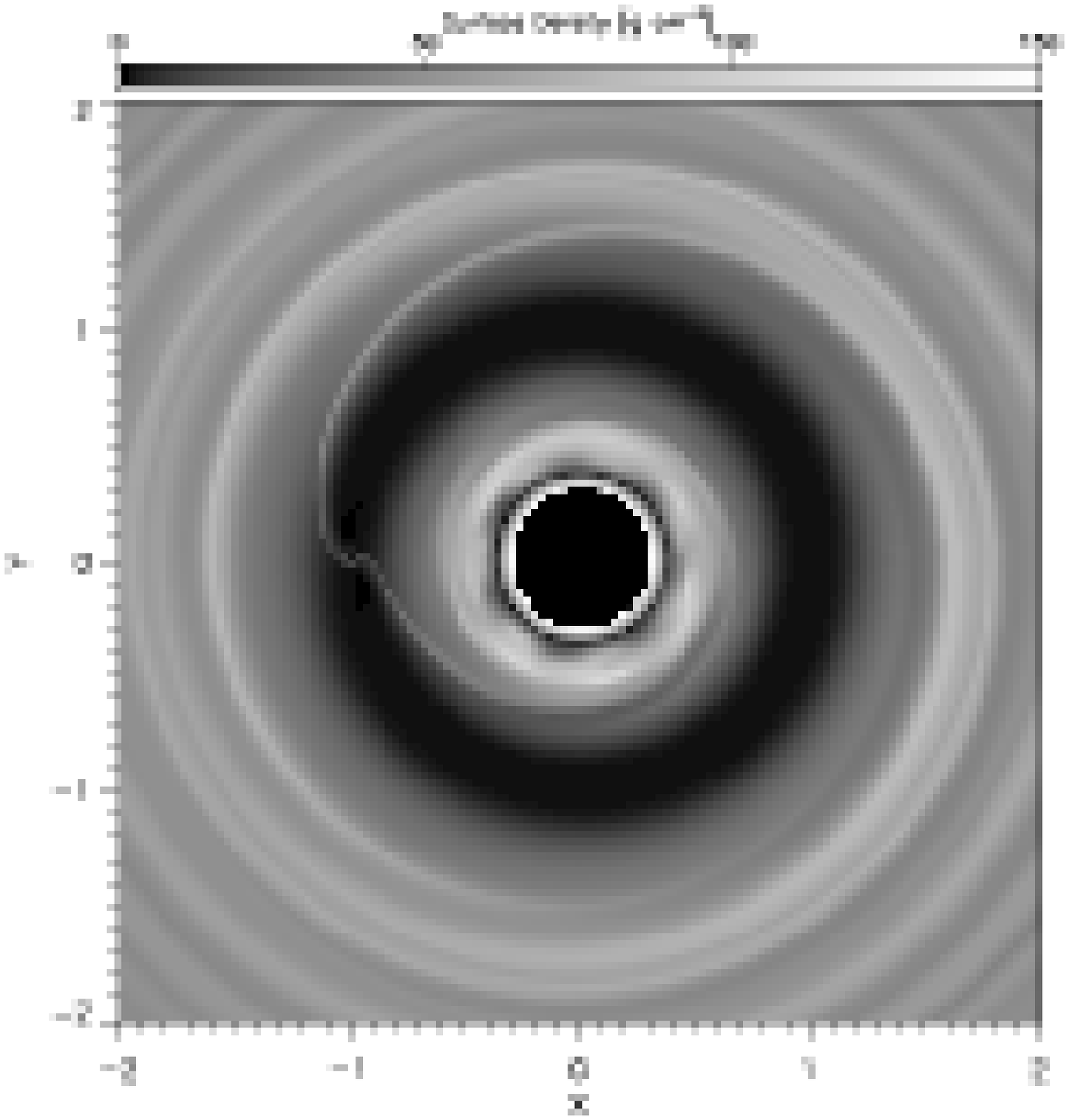,width=7.5truecm}\psfig{figure=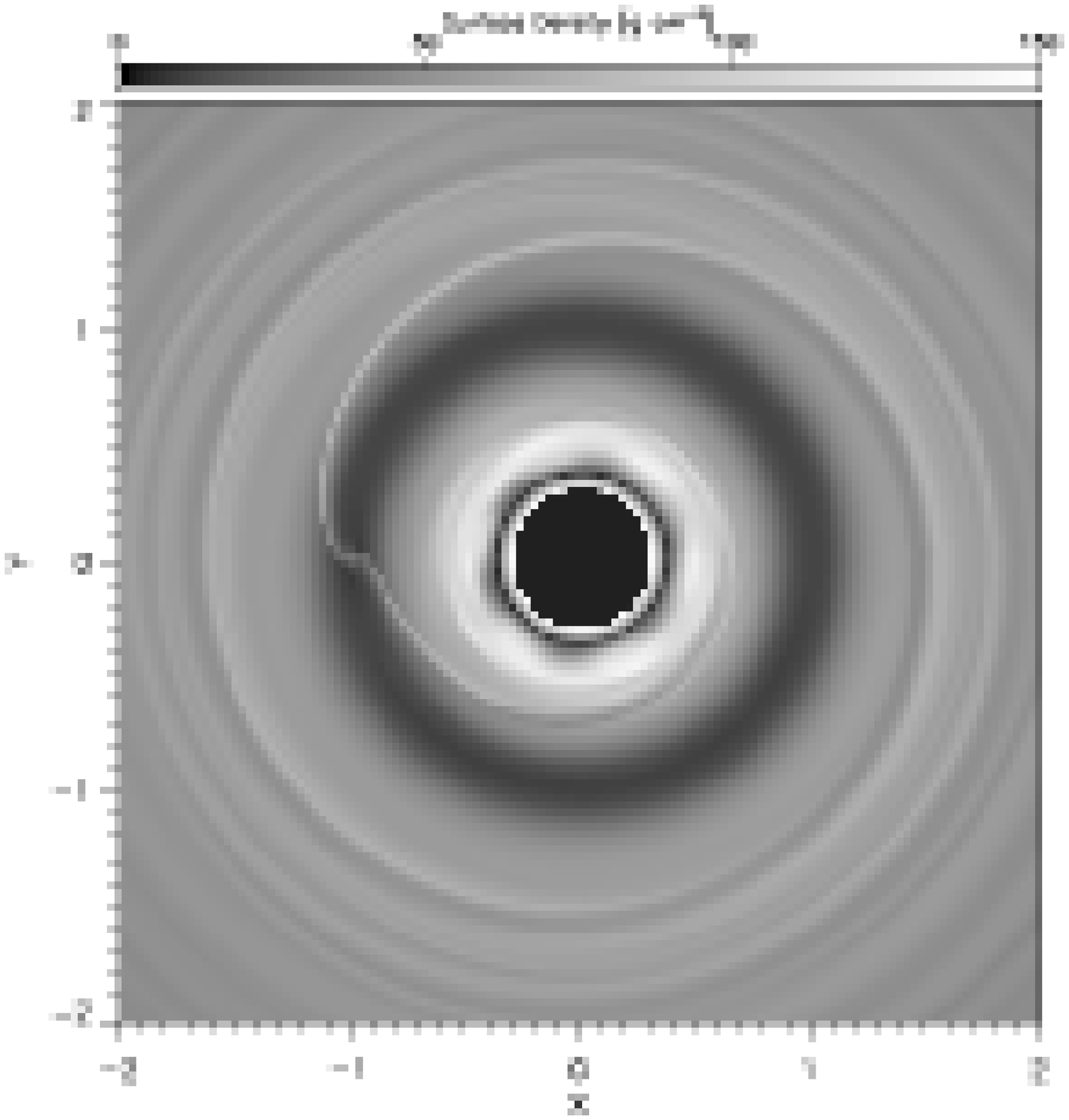,width=7.5truecm}}
\centerline{\psfig{figure=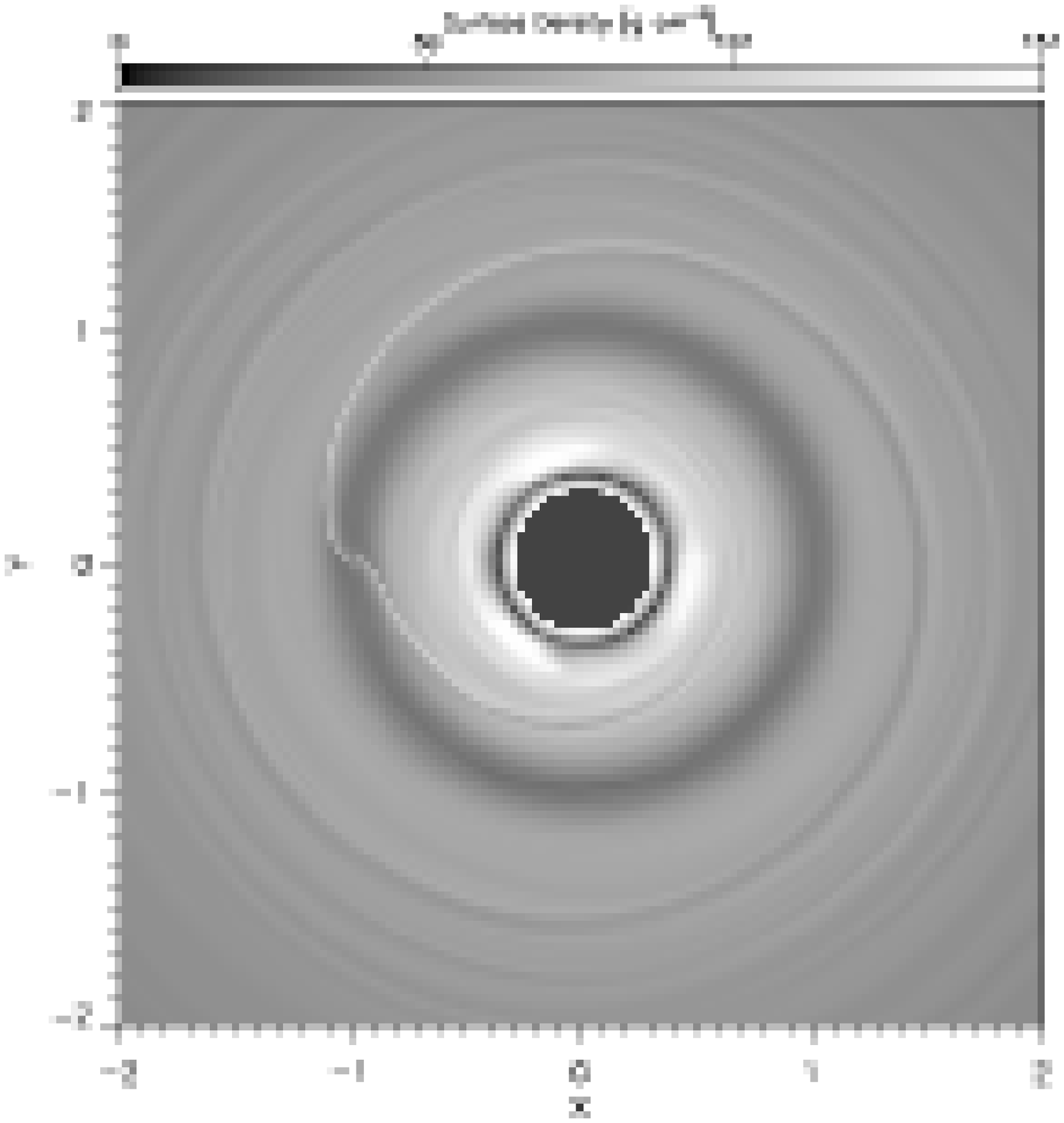,width=7.5truecm}\psfig{figure=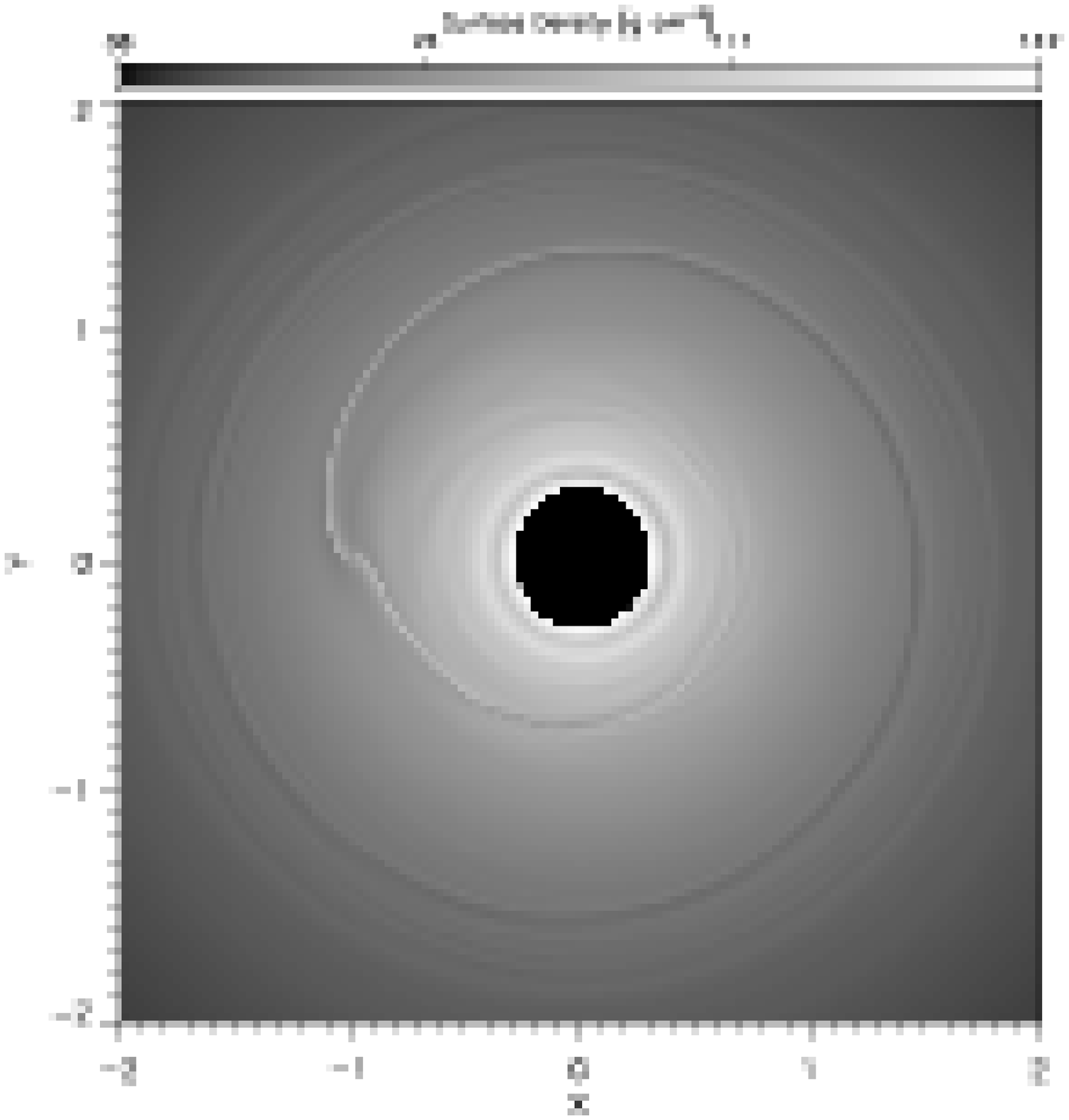,width=7.5truecm}}
\centerline{\psfig{figure=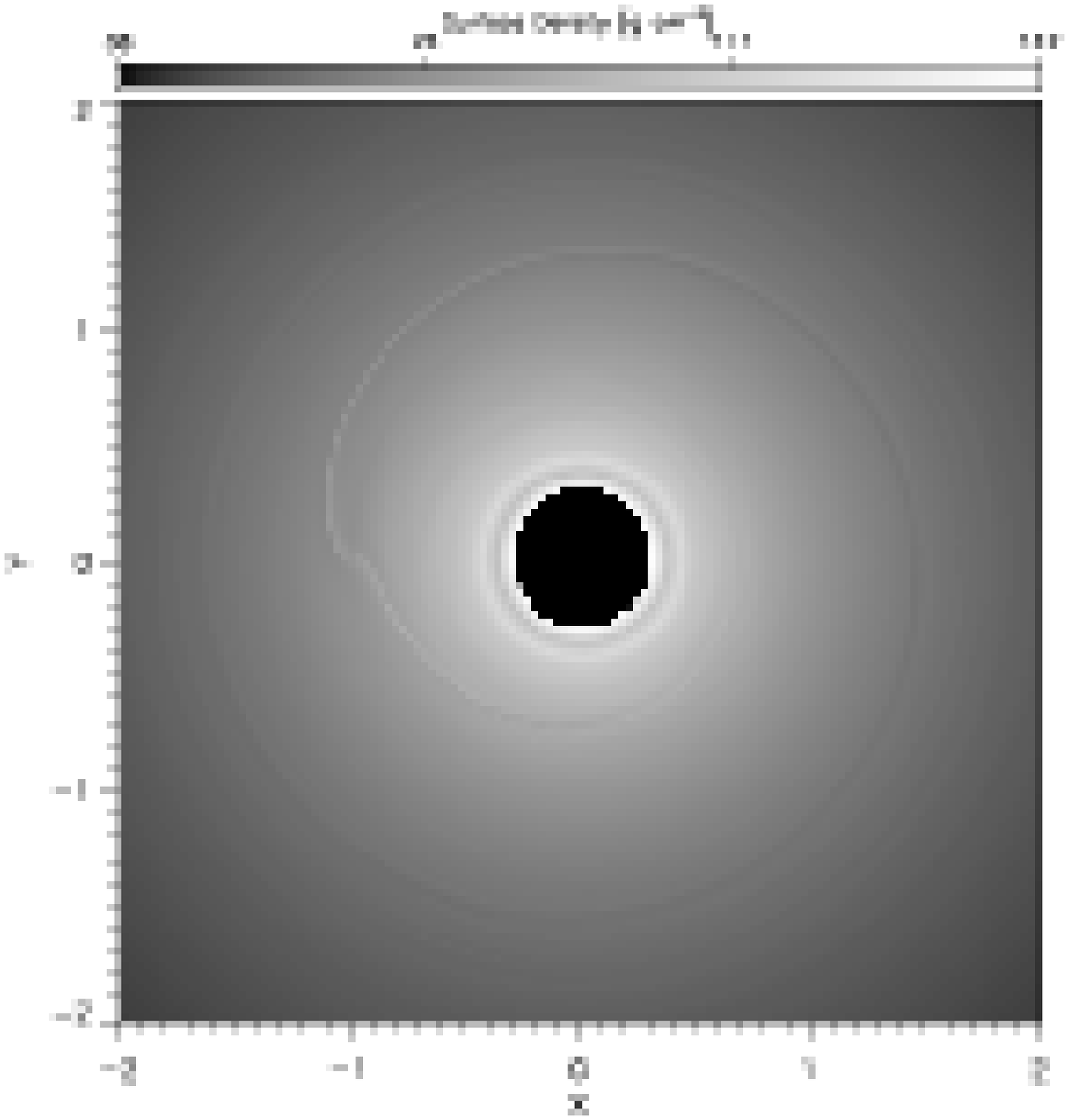,width=7.5truecm}\psfig{figure=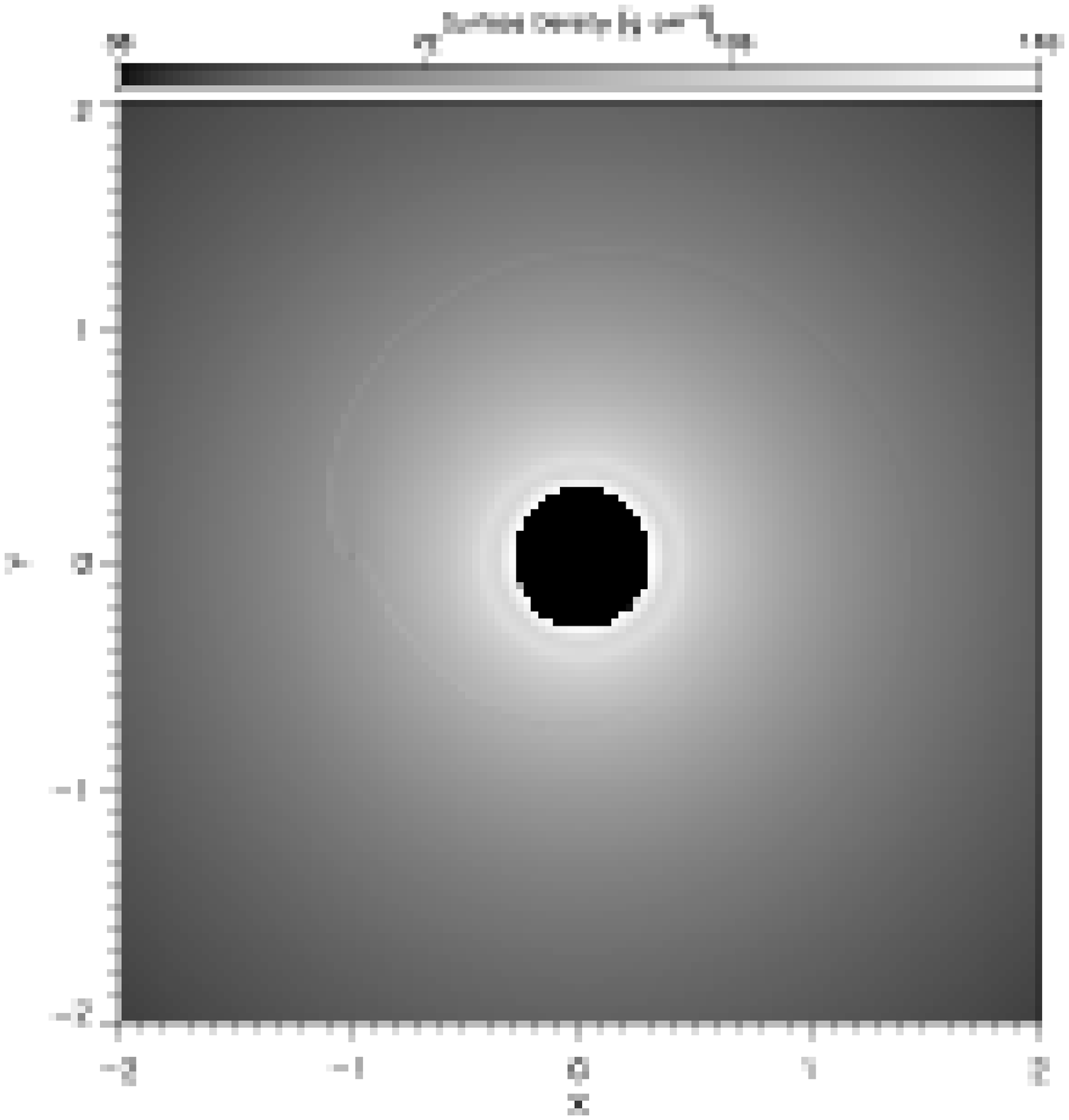,width=7.5truecm}}
\caption{\label{global} Disc surface densities, $\Sigma$, for 1, 0.3, 0.1, 0.03, 0.01, and 0.003 \mj\ planets (top-left to bottom-right).}
\end{figure*}

In Figure \ref{global}, we plot the global surface density
of the disc at the end of the simulations for each of the 6 planets.  
The (partial) clearing of gaps by the planets is clearly 
visible in the global surface density plots for planets 
with masses $\mpl \geq 0.1$ \mj.  Spiral density
waves launched by the planet propagate inwards and
outwards from the planet's radius into the disc.  For masses
$\mpl \geq 0.1$ \mj, these density waves are strong enough to appear 
even in the azimuthally averaged surface density profiles
(Figure \ref{surfdens}).  The strength
of the surface density perturbations decreases with lower mass
planets, but their form is essentially independent of 
the planet's mass.  For the lowest-mass planet, the spiral
density perturbations are too small to be visible over the
underlying density gradient in the disc.  These waves
are essentially two-dimensional in nature and have been 
studied in the past using two and three-dimensional nonlinear calculations 
(e.g.\ Artymowicz 1992; Kley 1999; Bryden et al.\ 1999;
Lubow et al.\ 1999; Nelson et al.\ 2000; Kley et al.\ 2001)
as well as analytically and through linearized numerical calculations
(Ogilvie \& Lubow 2002).

\begin{figure*}
\centerline{\psfig{figure=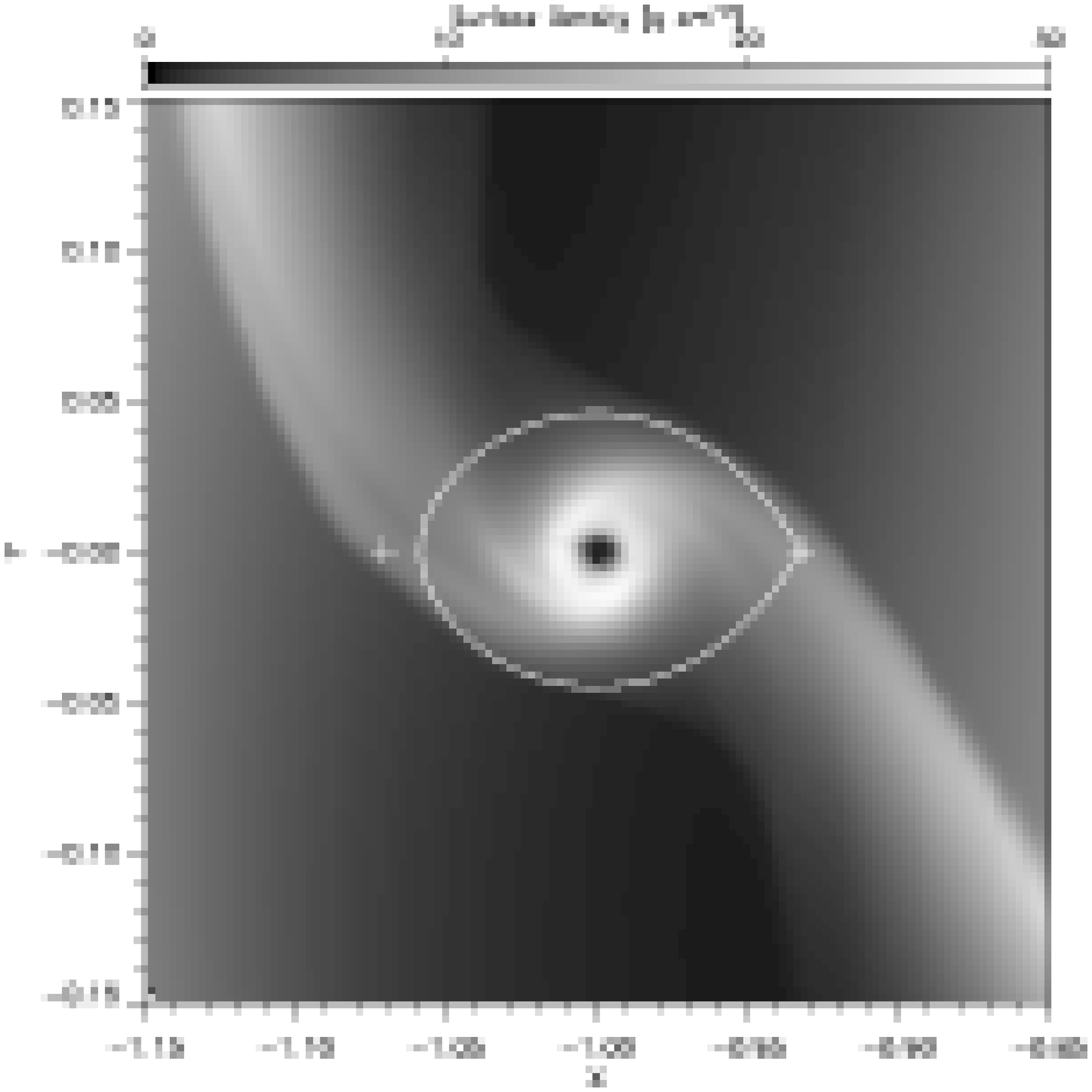,width=7.5truecm}\psfig{figure=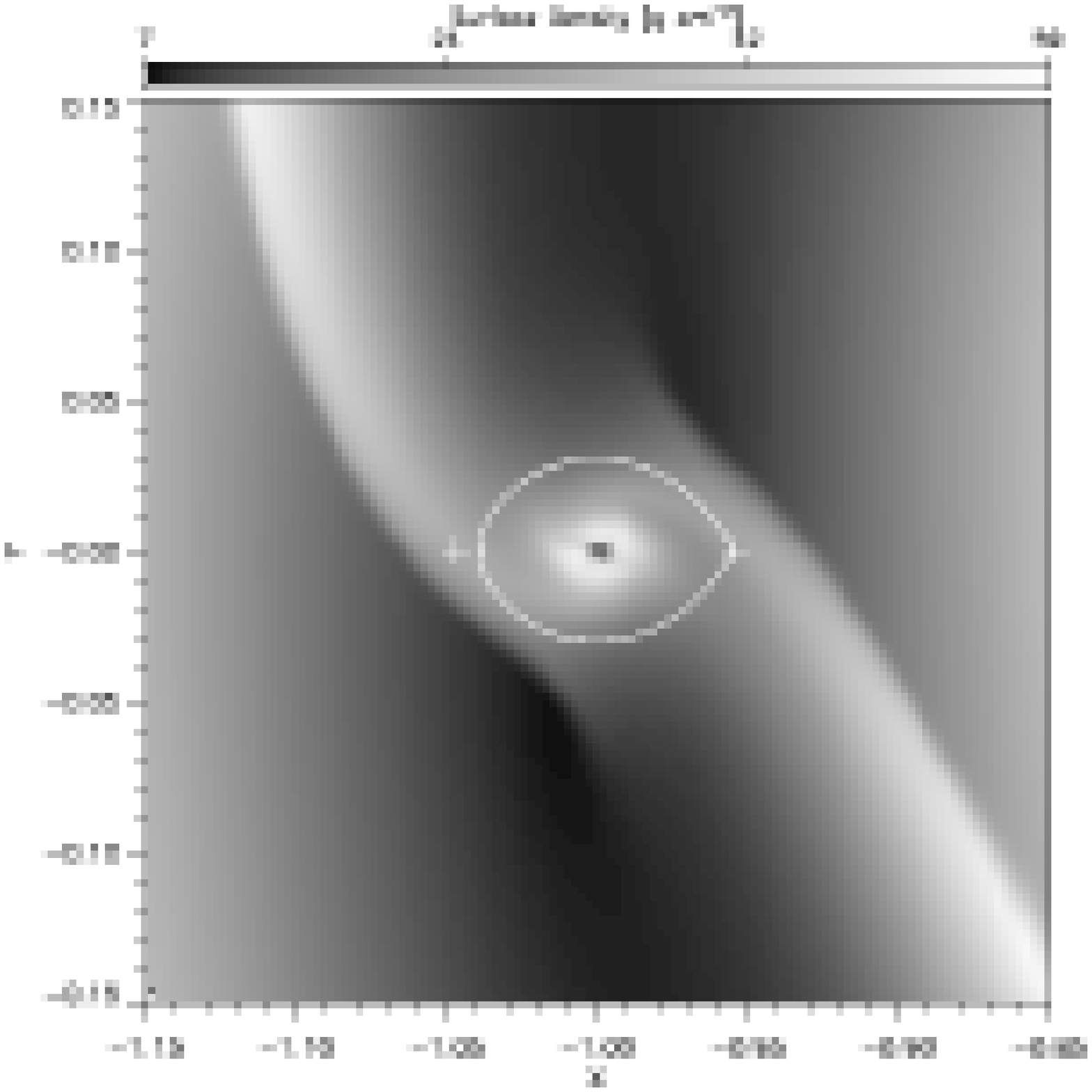,width=7.5truecm}}
\centerline{\psfig{figure=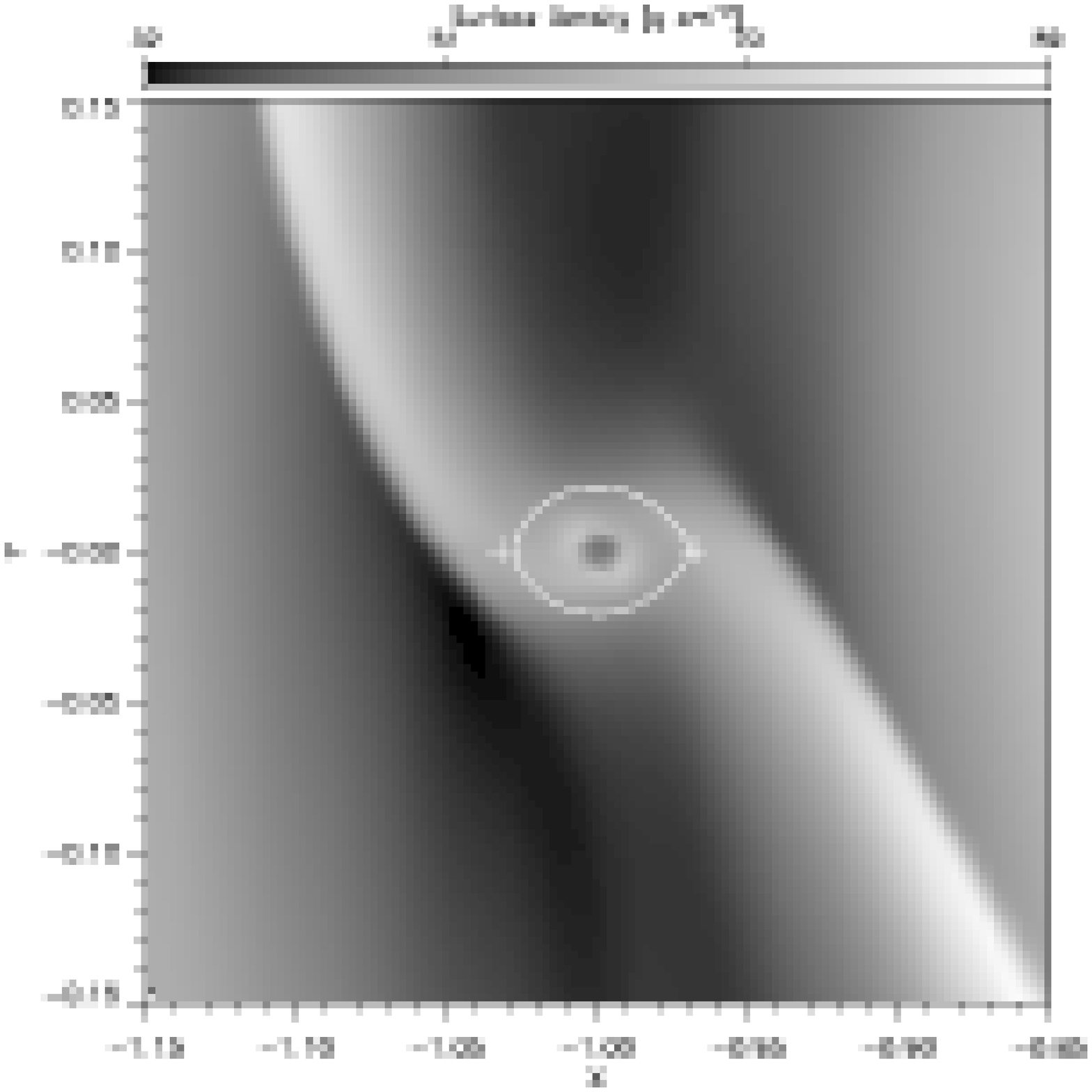,width=7.5truecm}\psfig{figure=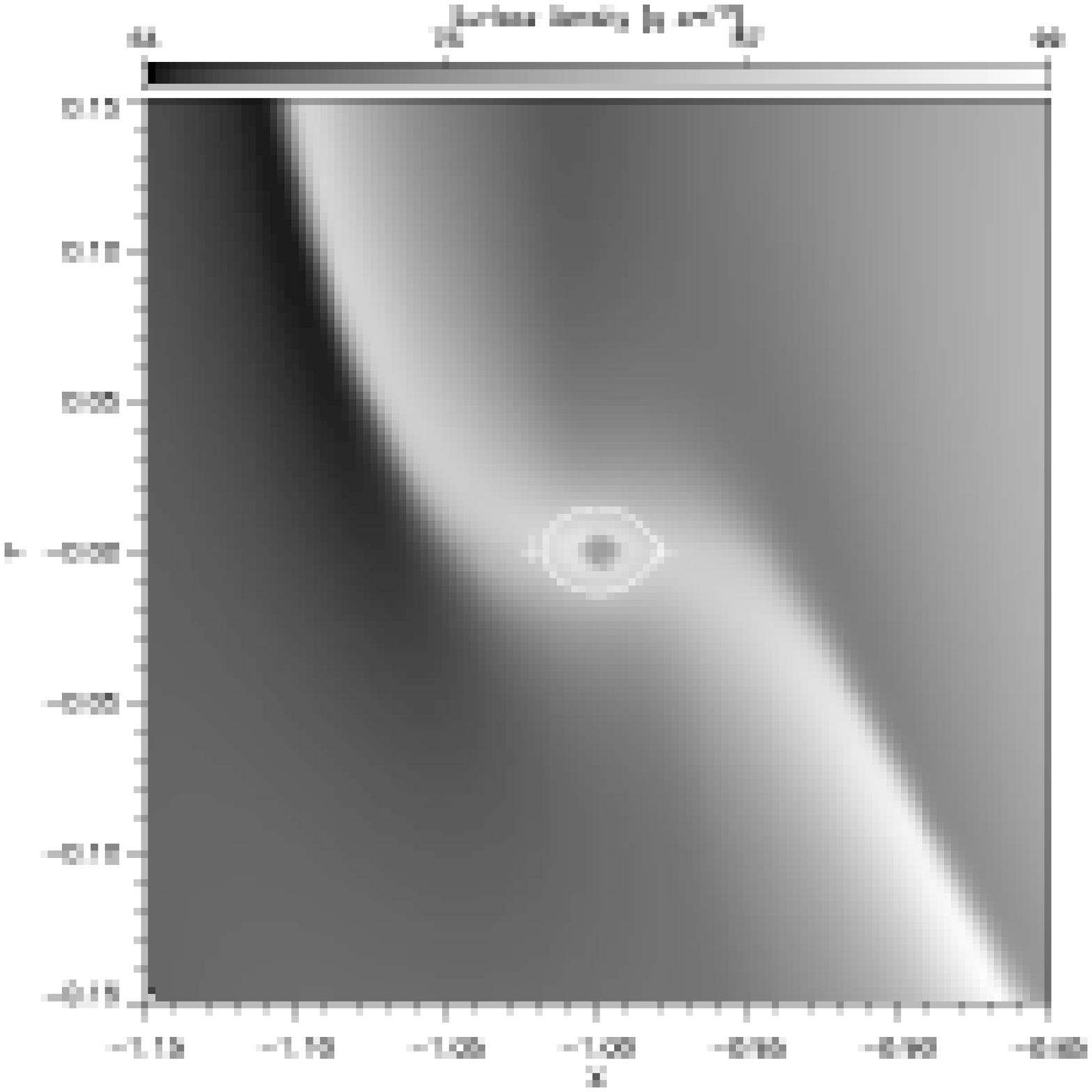,width=7.5truecm}}
\centerline{\psfig{figure=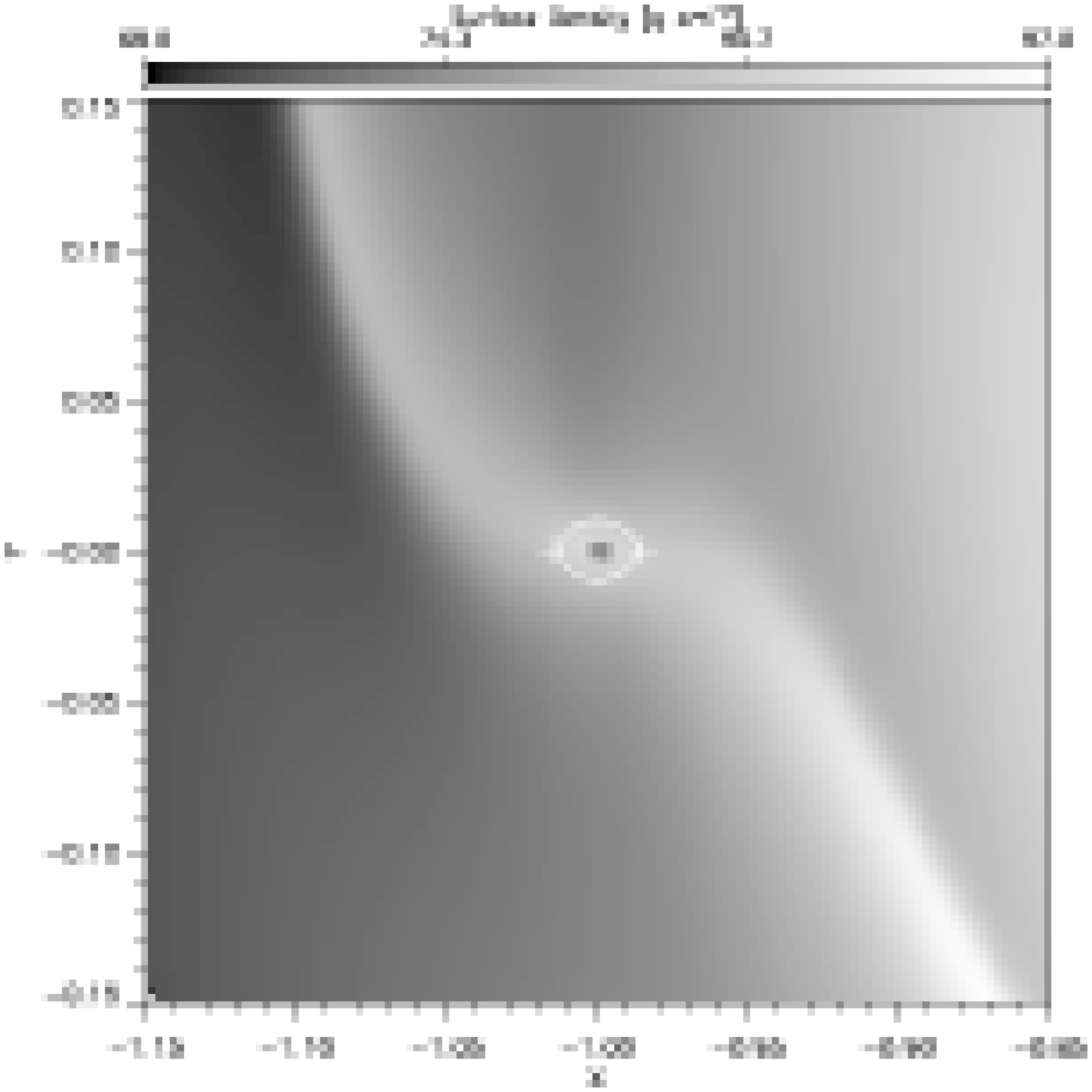,width=7.5truecm}\psfig{figure=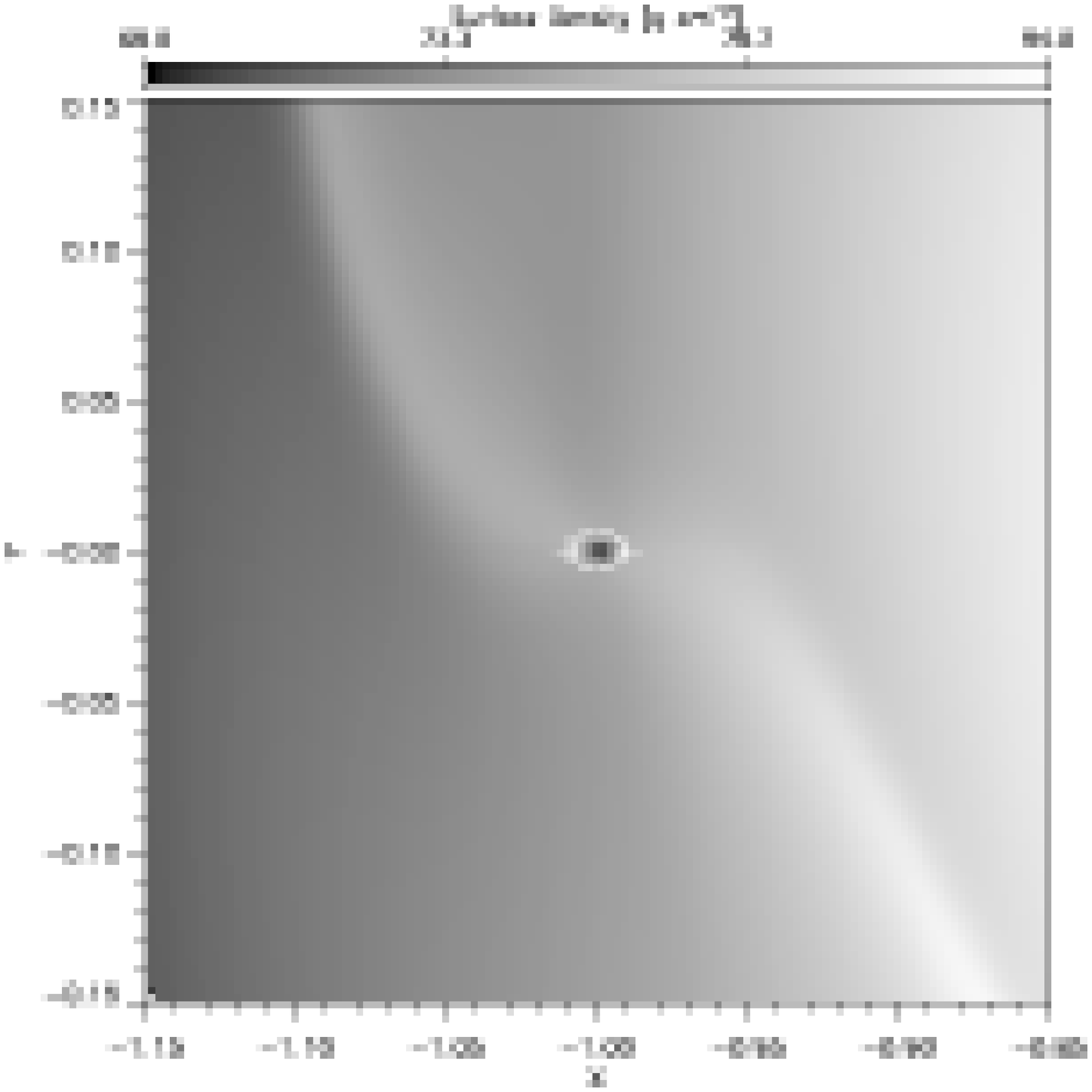,width=7.5truecm}}
\caption{\label{zoomsurf} Surface density, $\Sigma$, in the vicinity of the planet for 1, 0.3, 0.1, 0.03, 0.01, and 0.003 \mj\ planets (top-left to bottom-right).  Also plotted are the Roche lobes (white curves) and the inner (L1) and outer (L2) Lagrangian points (crosses).}
\end{figure*}

\begin{figure*}
\centerline{\psfig{figure=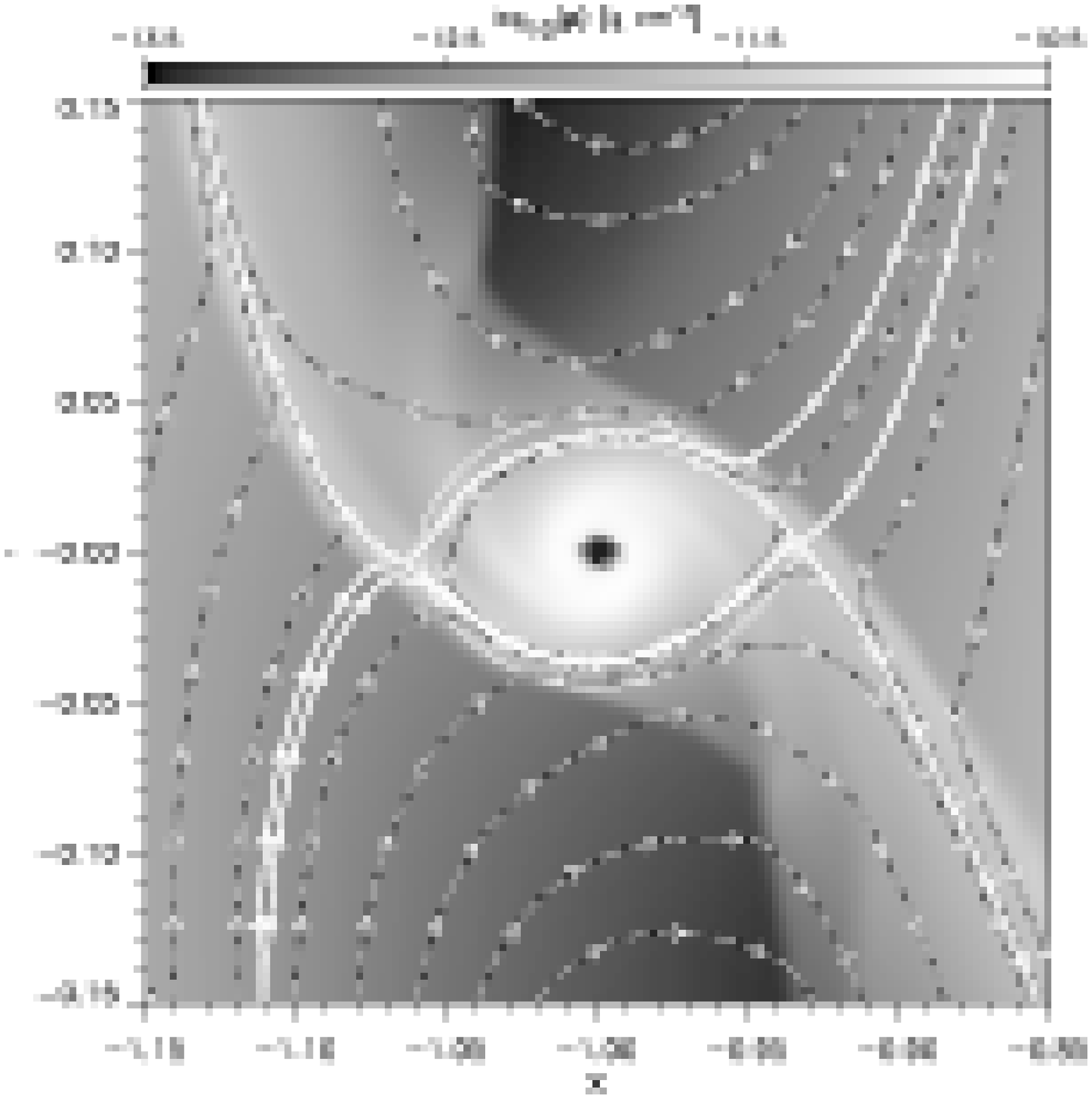,width=7.5truecm}\psfig{figure=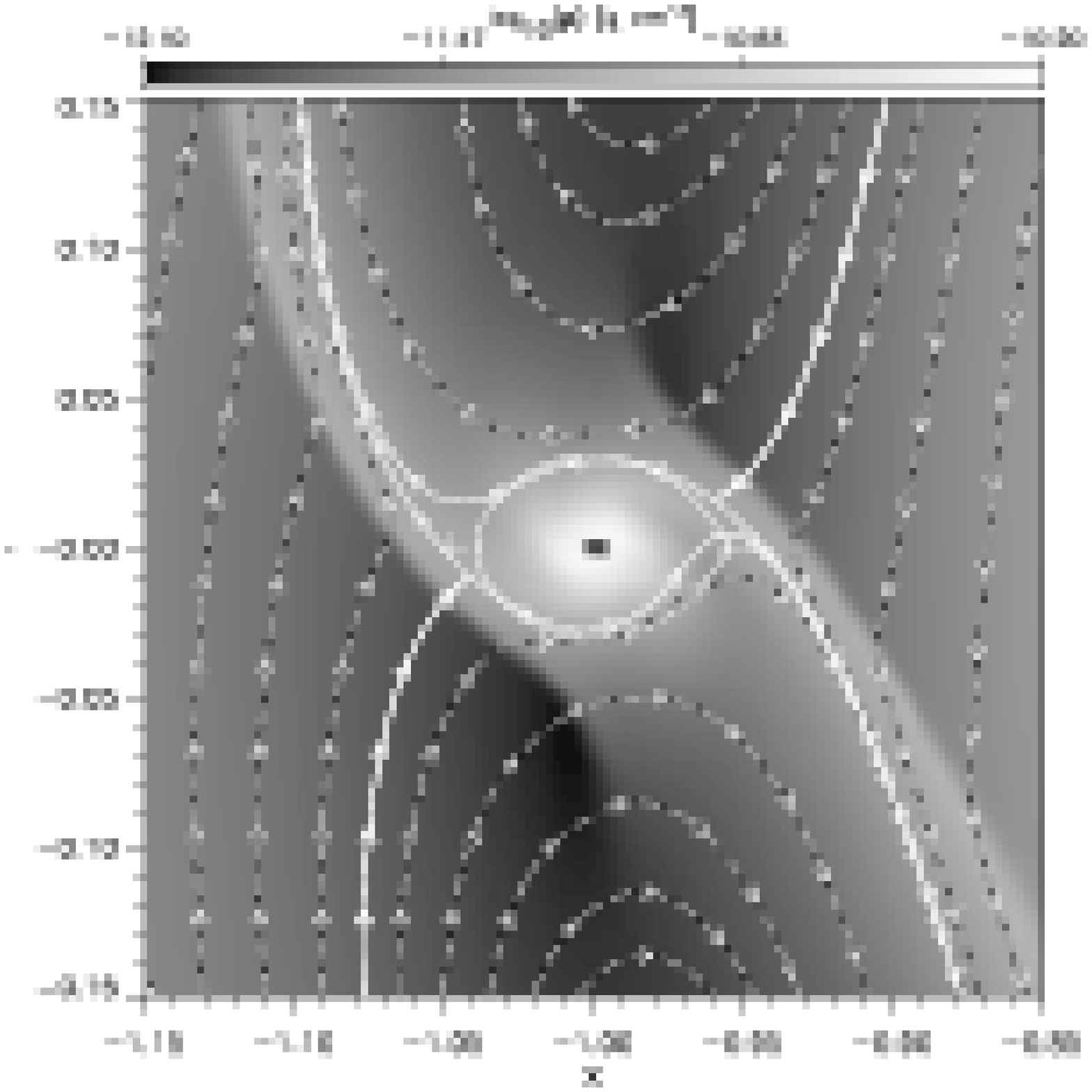,width=7.5truecm}}
\centerline{\psfig{figure=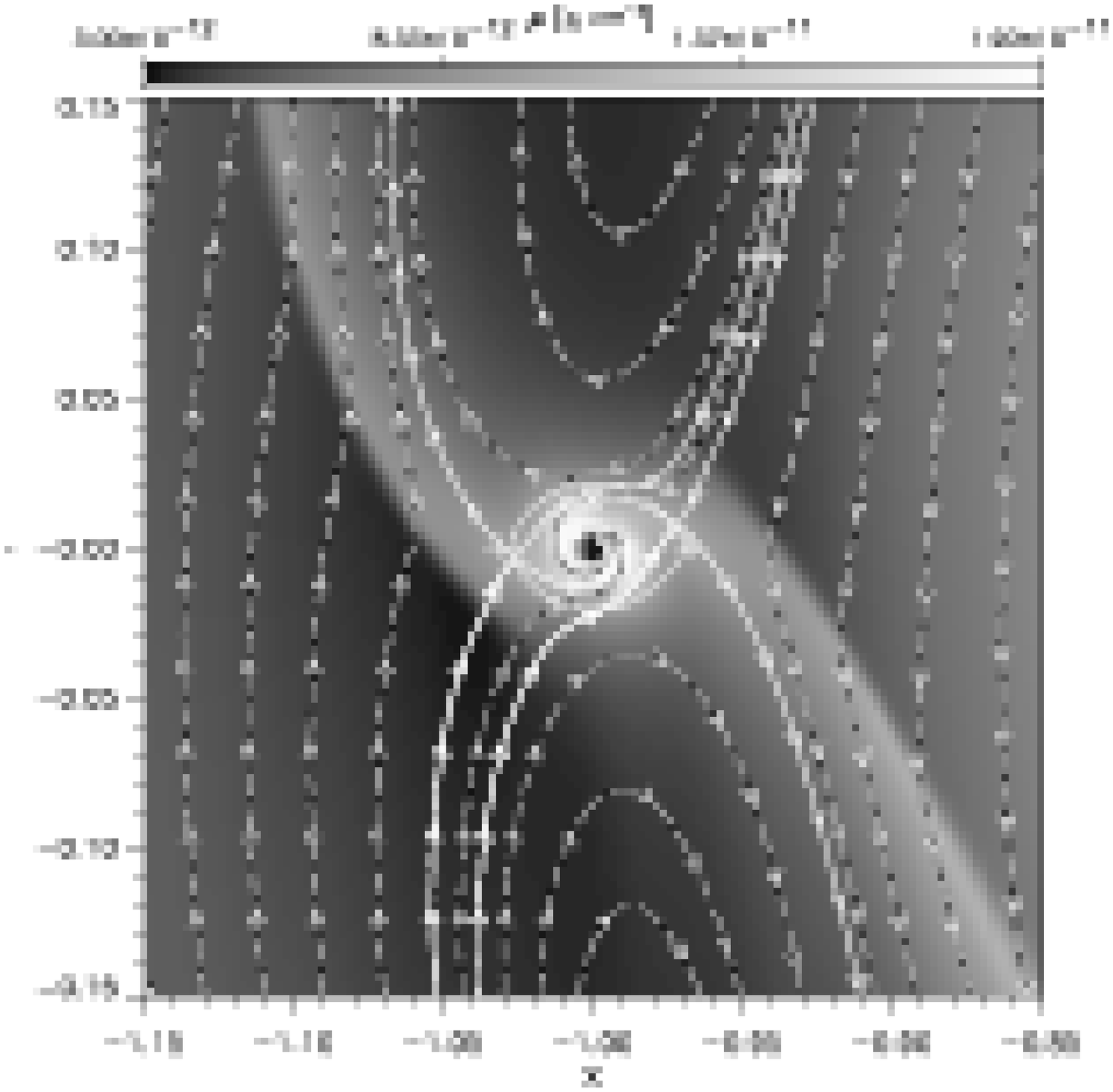,width=7.5truecm}\psfig{figure=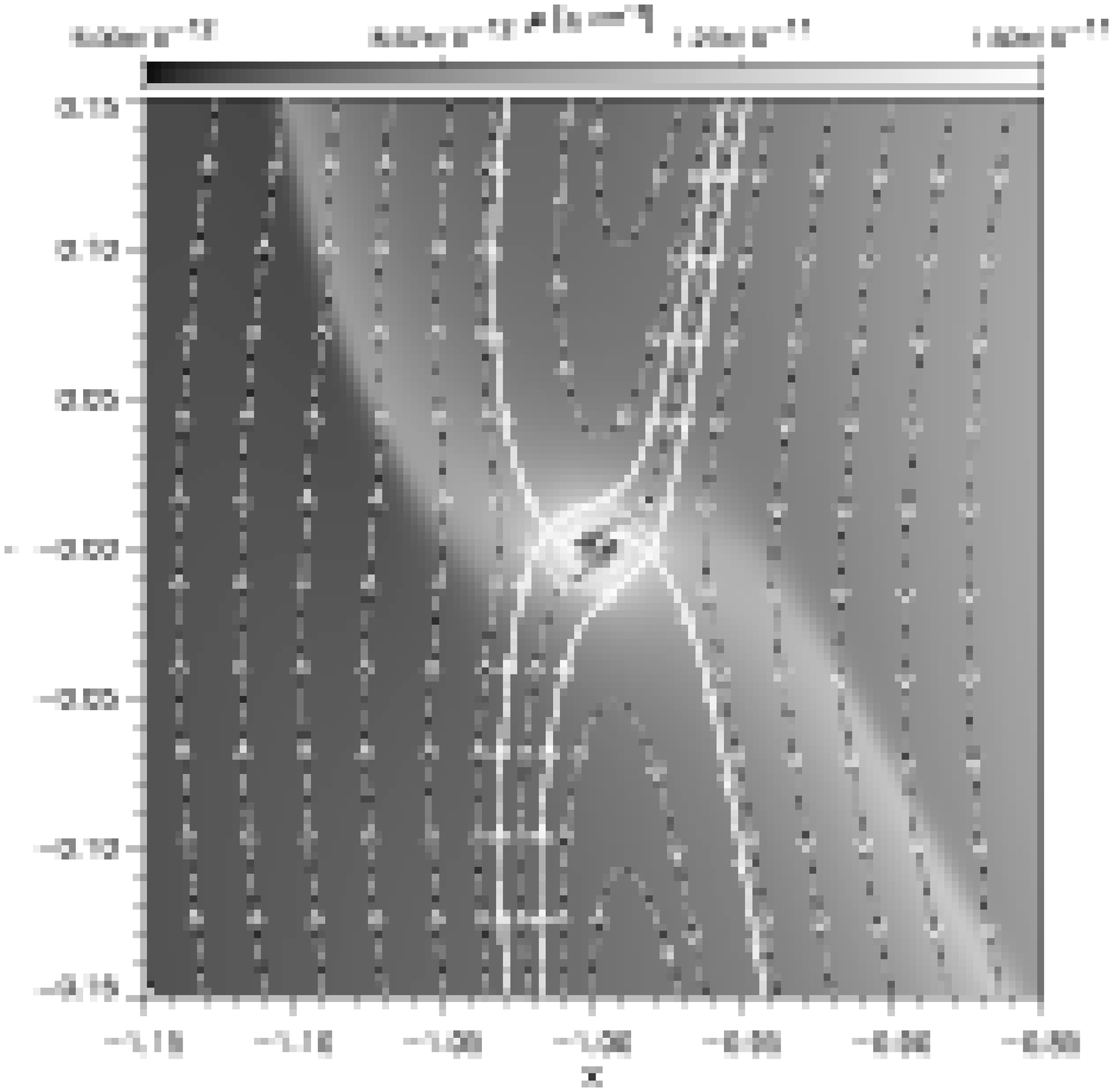,width=7.5truecm}}
\centerline{\psfig{figure=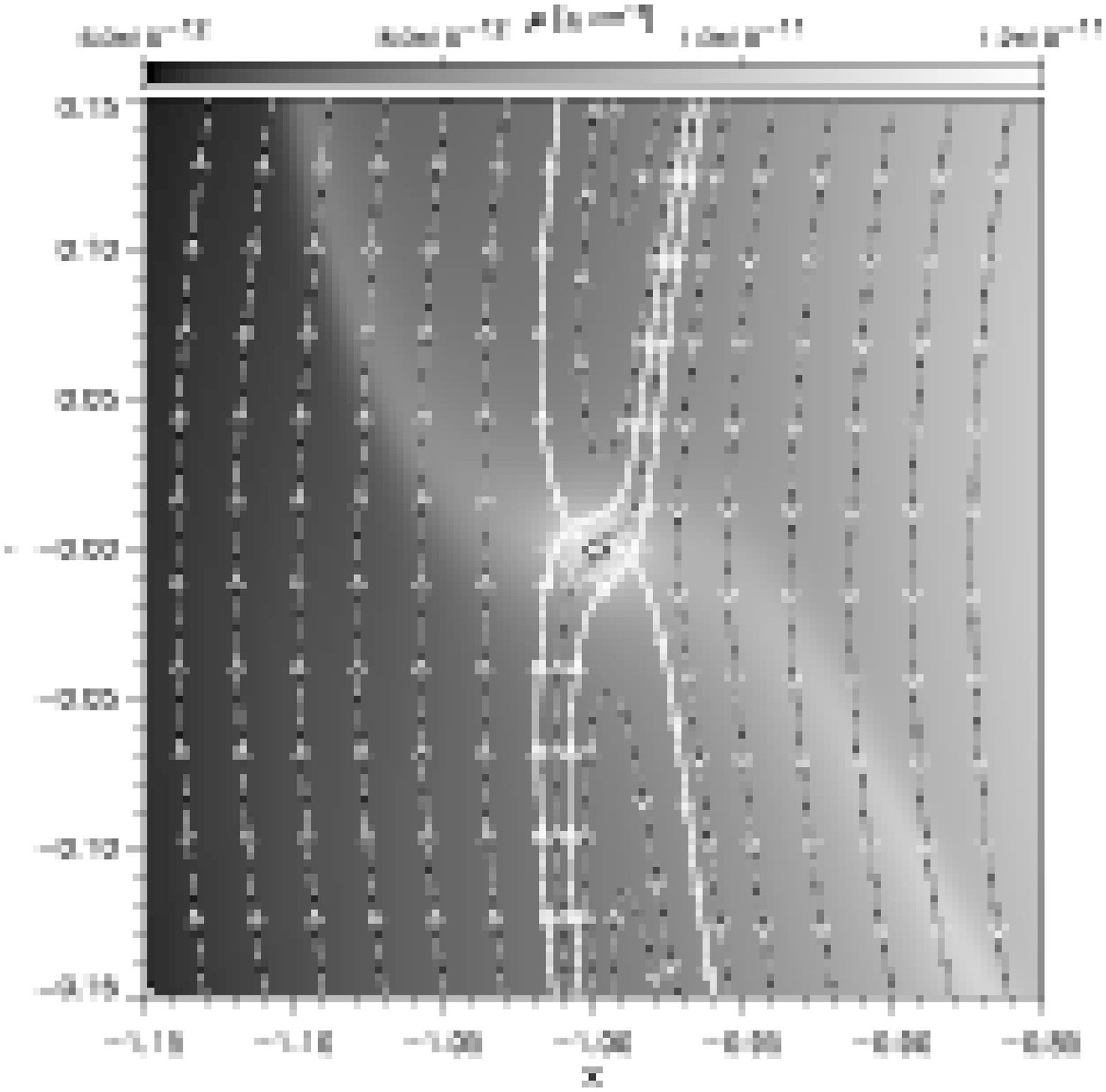,width=7.5truecm}\psfig{figure=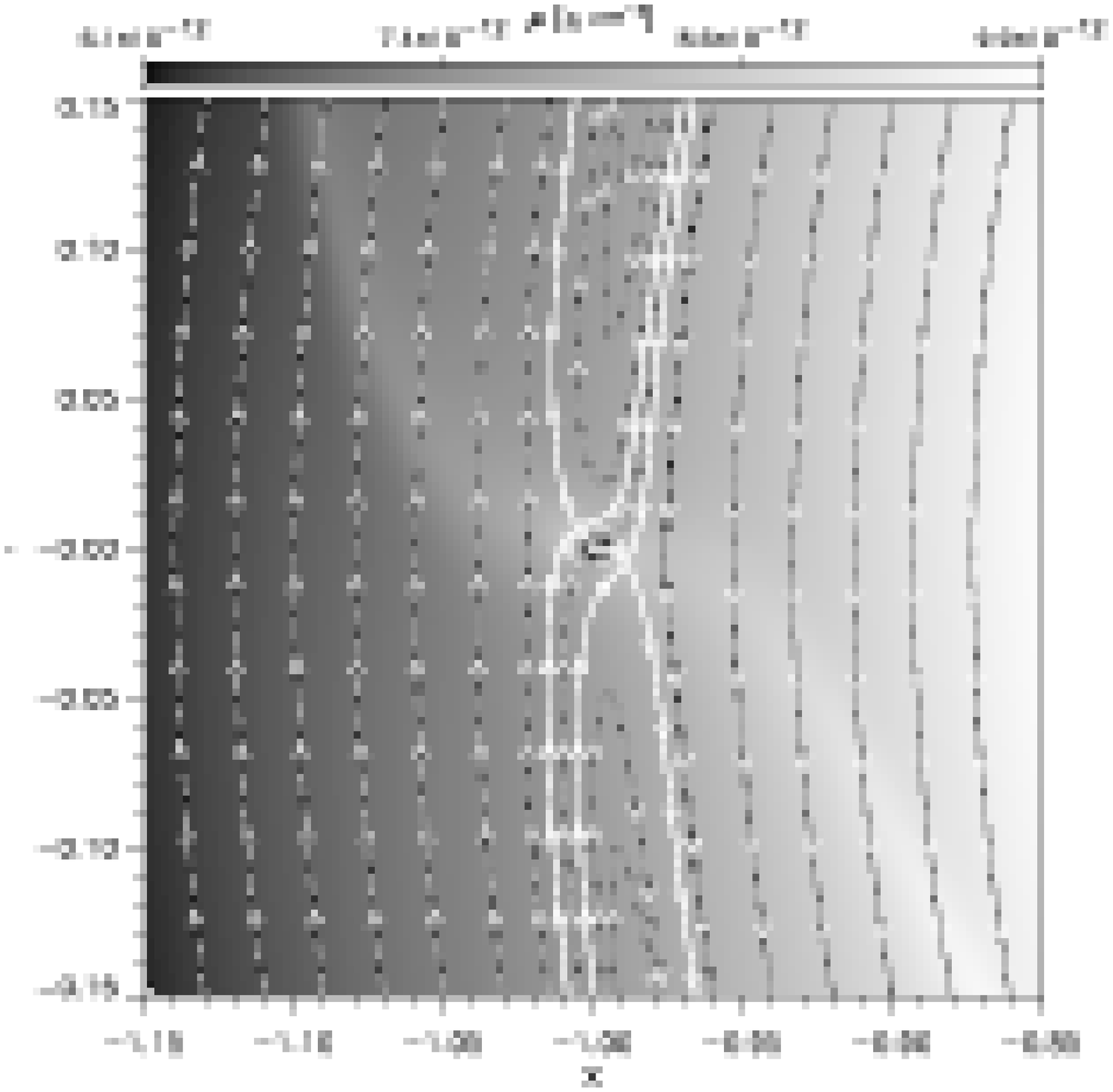,width=7.5truecm}}
\caption{\label{hmid} Disc density, $\rho$ (greyscale), and streamlines (dashed and solid lines) on the disc midplane for 1, 0.3, 0.1, 0.03, 0.01, and 0.003 \mj\ planets (top-left to bottom-right).  The white solid streamlines are the critical streamlines that mark the boundaries between the outer/inner disc, the accretion streams into the Roche lobe, and the horseshoe orbits.  Notice that the top 2 figures give the logarithm of the density.}
\end{figure*}

\begin{figure*}
\centerline{\psfig{figure=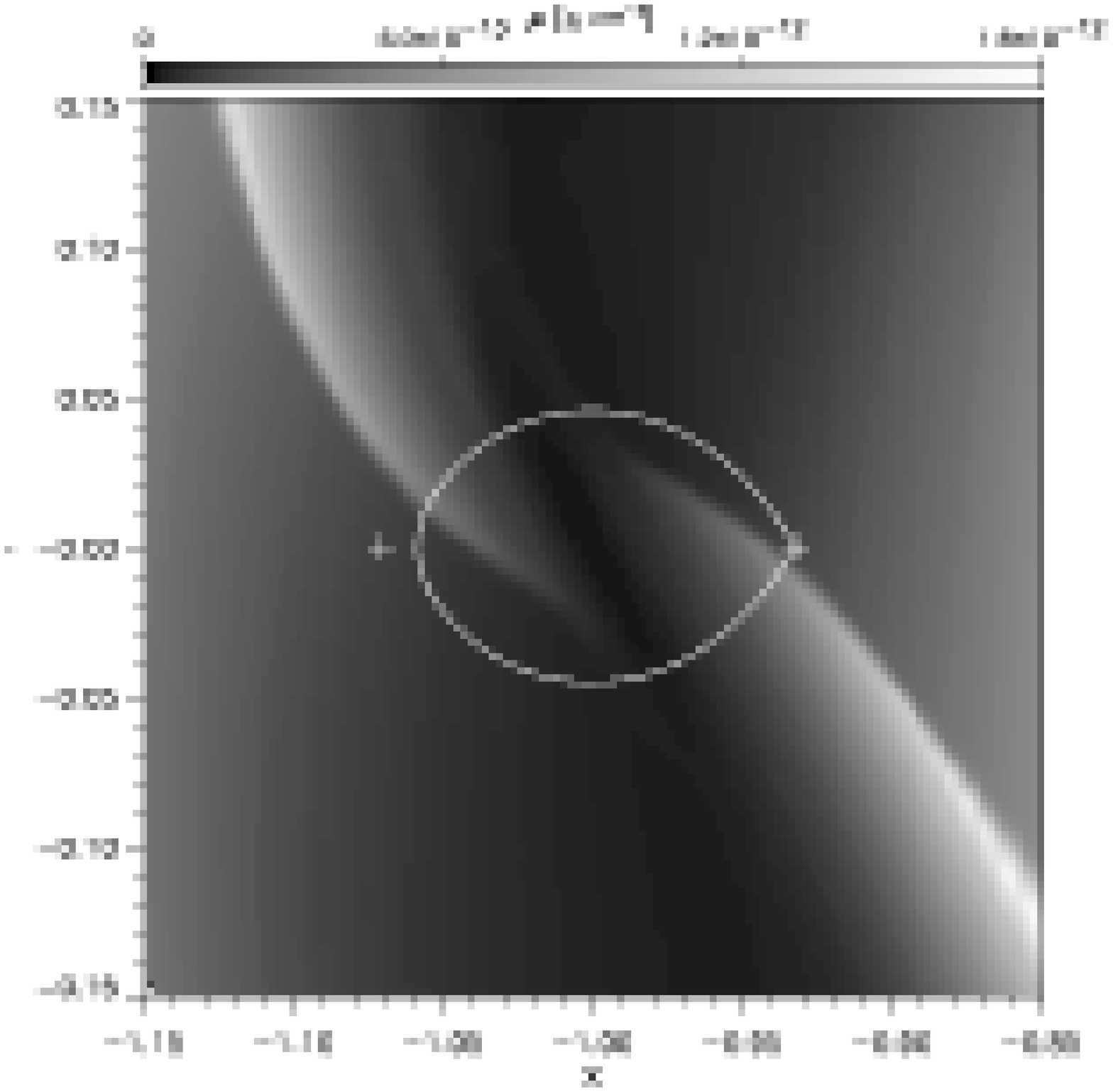,width=7.5truecm}\psfig{figure=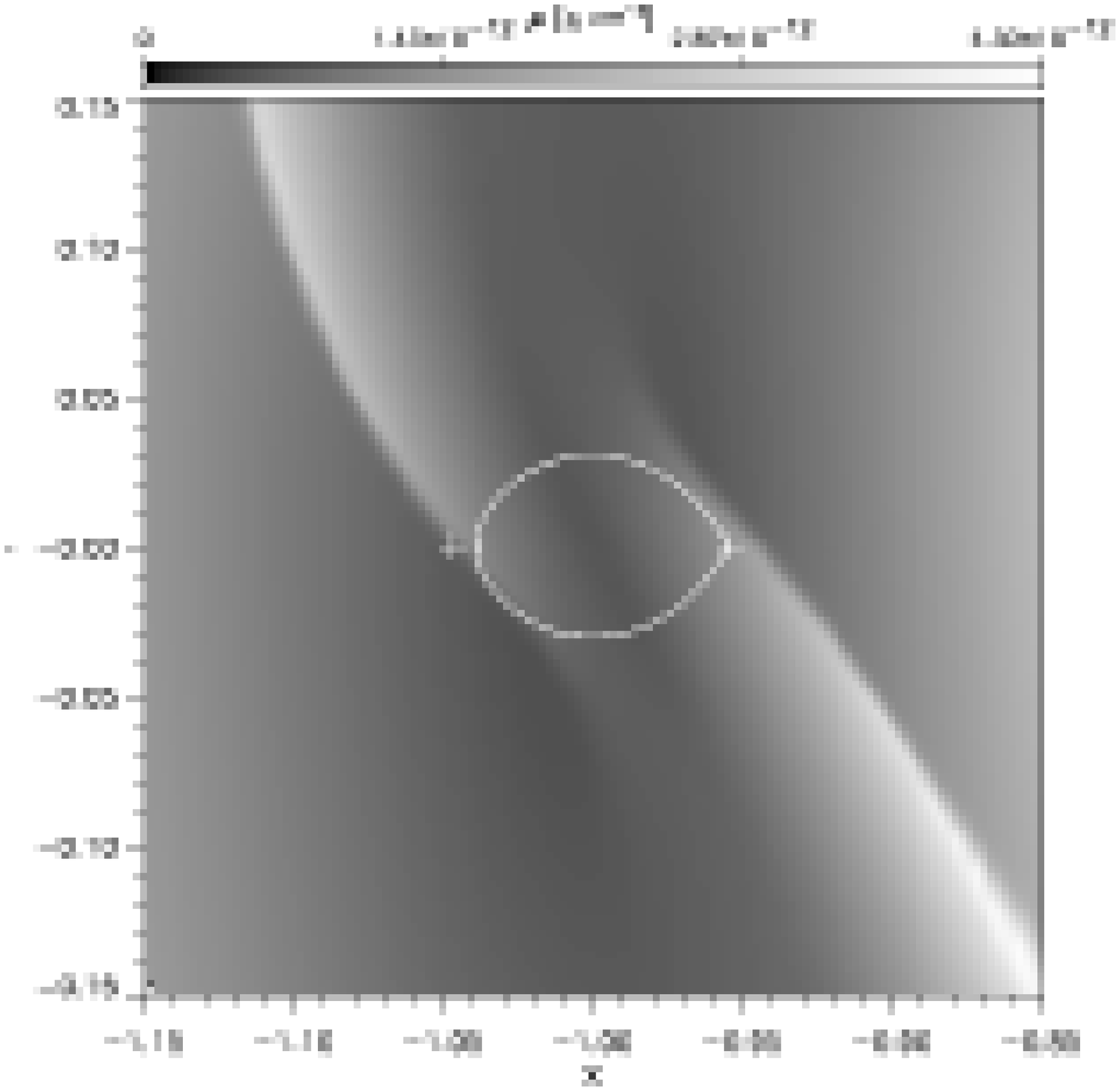,width=7.5truecm}}
\centerline{\psfig{figure=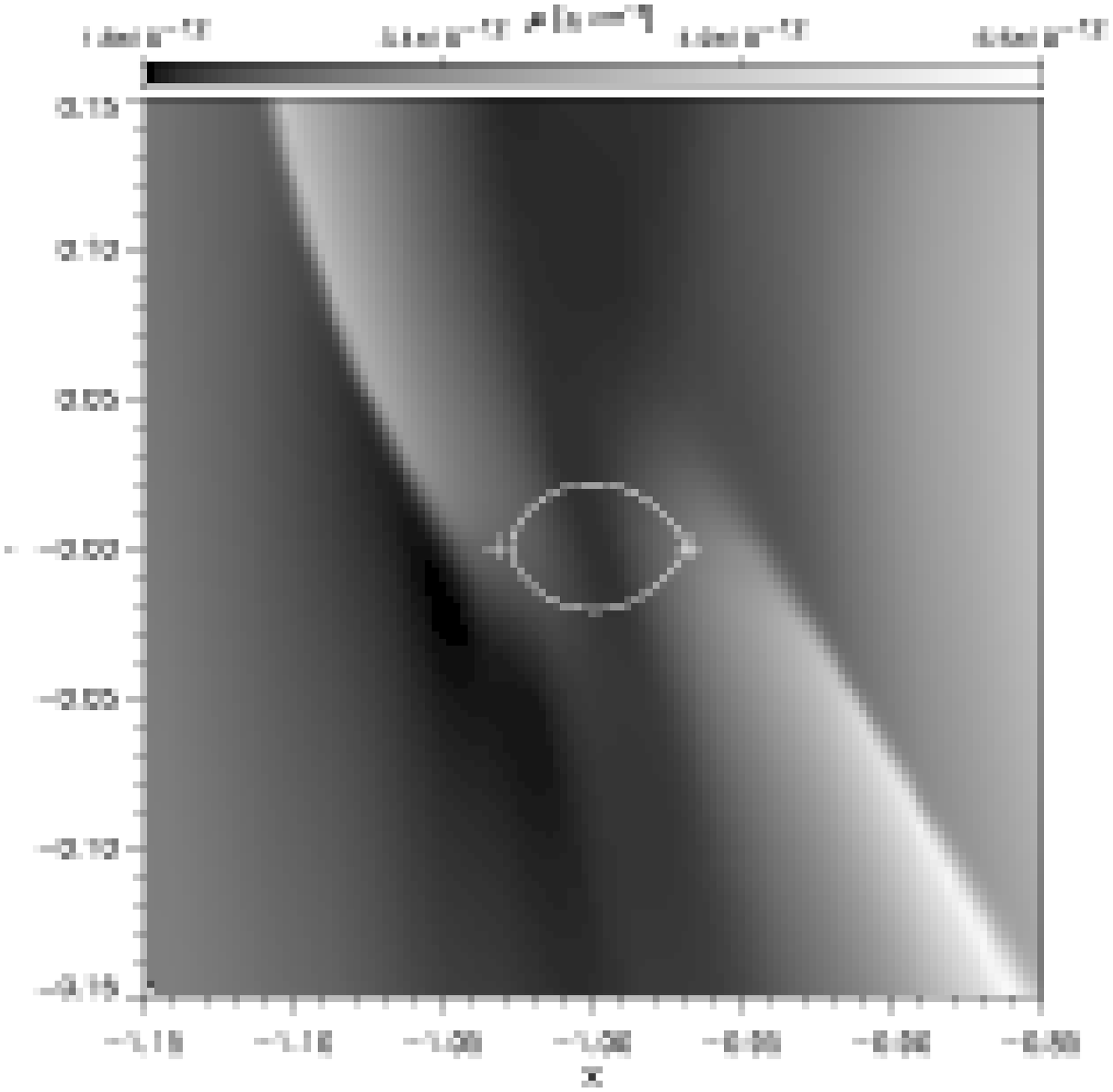,width=7.5truecm}\psfig{figure=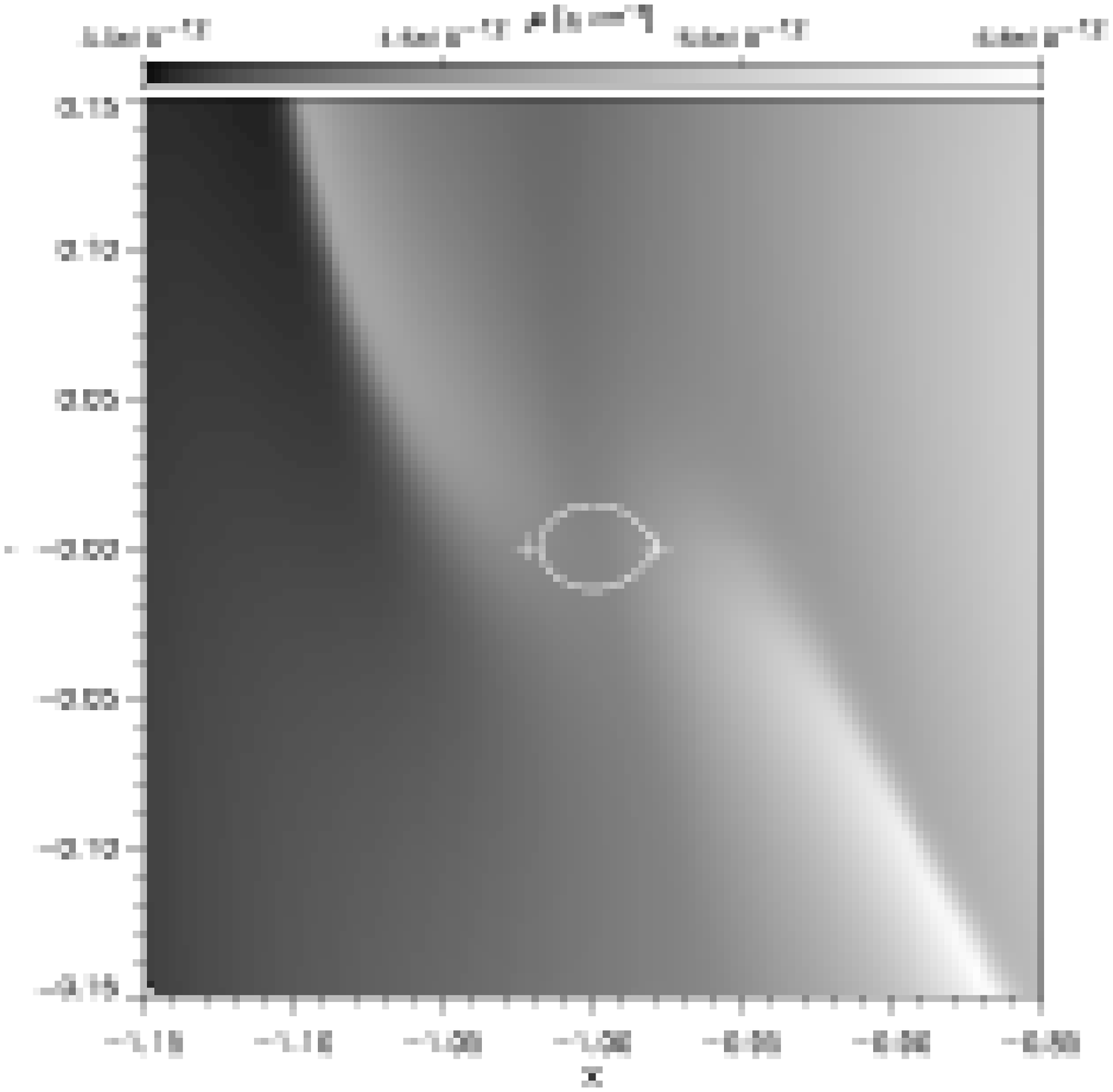,width=7.5truecm}}
\centerline{\psfig{figure=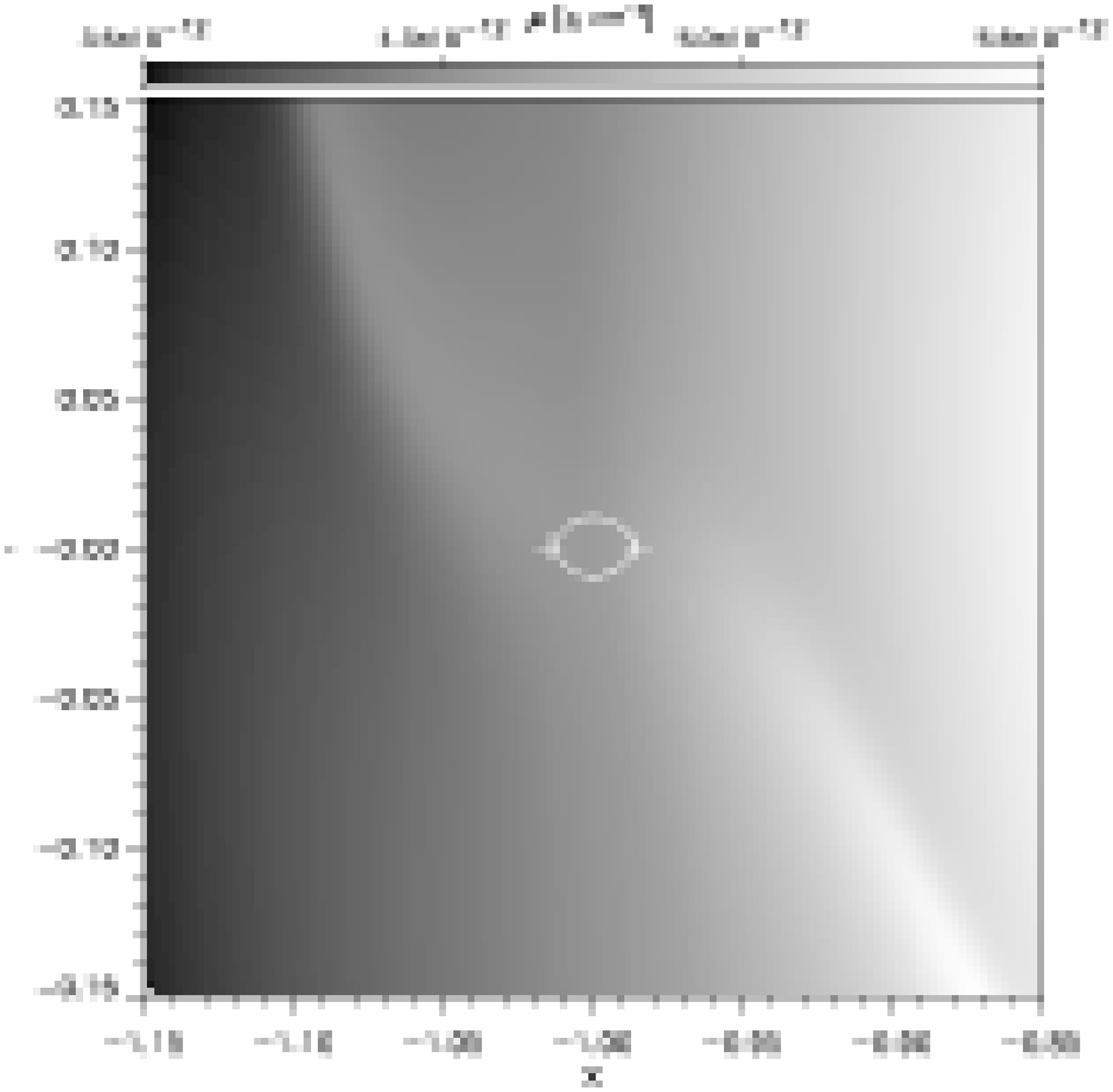,width=7.5truecm}\psfig{figure=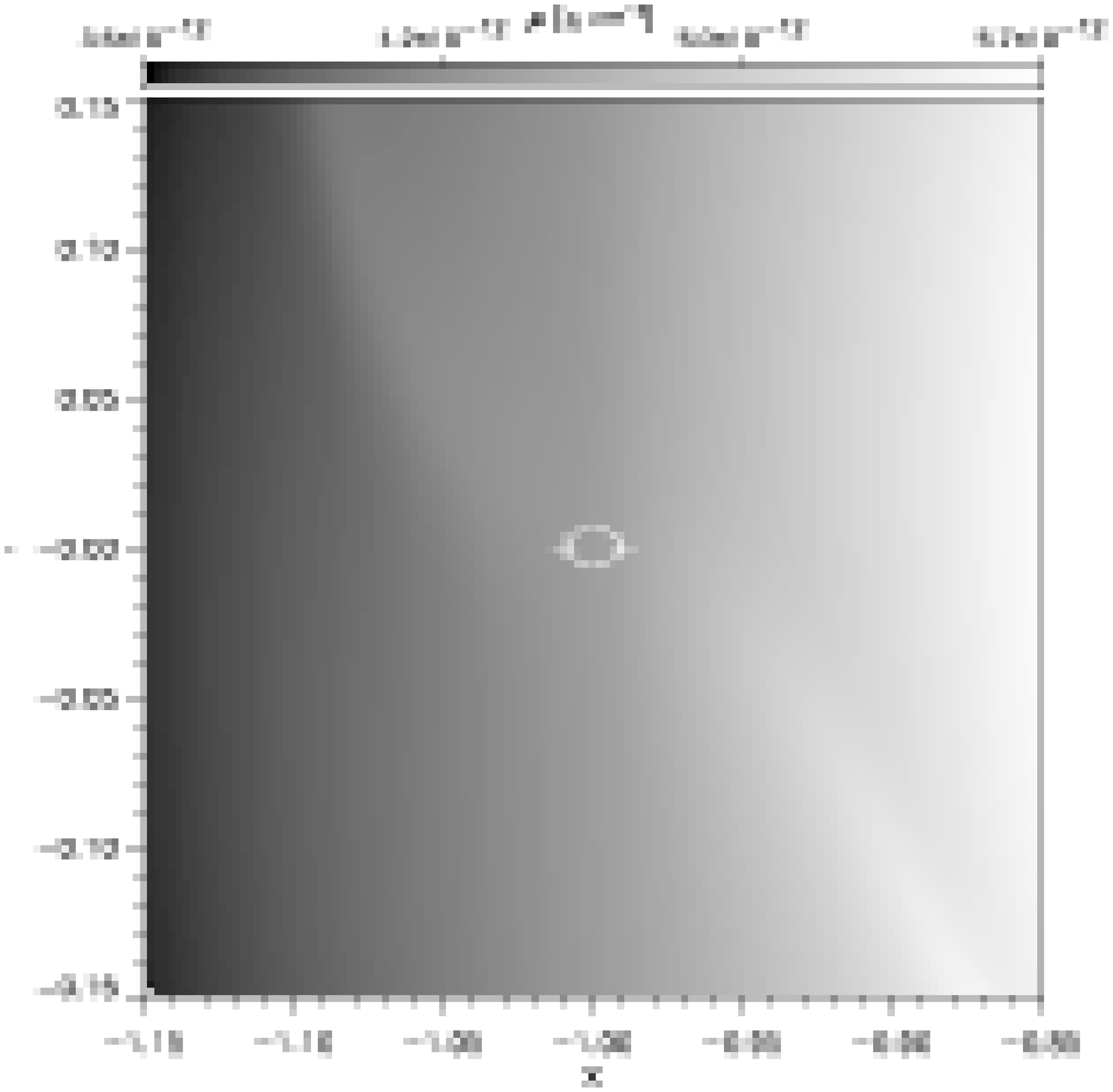,width=7.5truecm}}
\caption{\label{h1} Same as Figure 5, but at distance $H$ from the disc midplane.  The white line indicates the size of the Roche lobe at the midplane. }
\end{figure*}

\begin{figure*}
\centerline{\psfig{figure=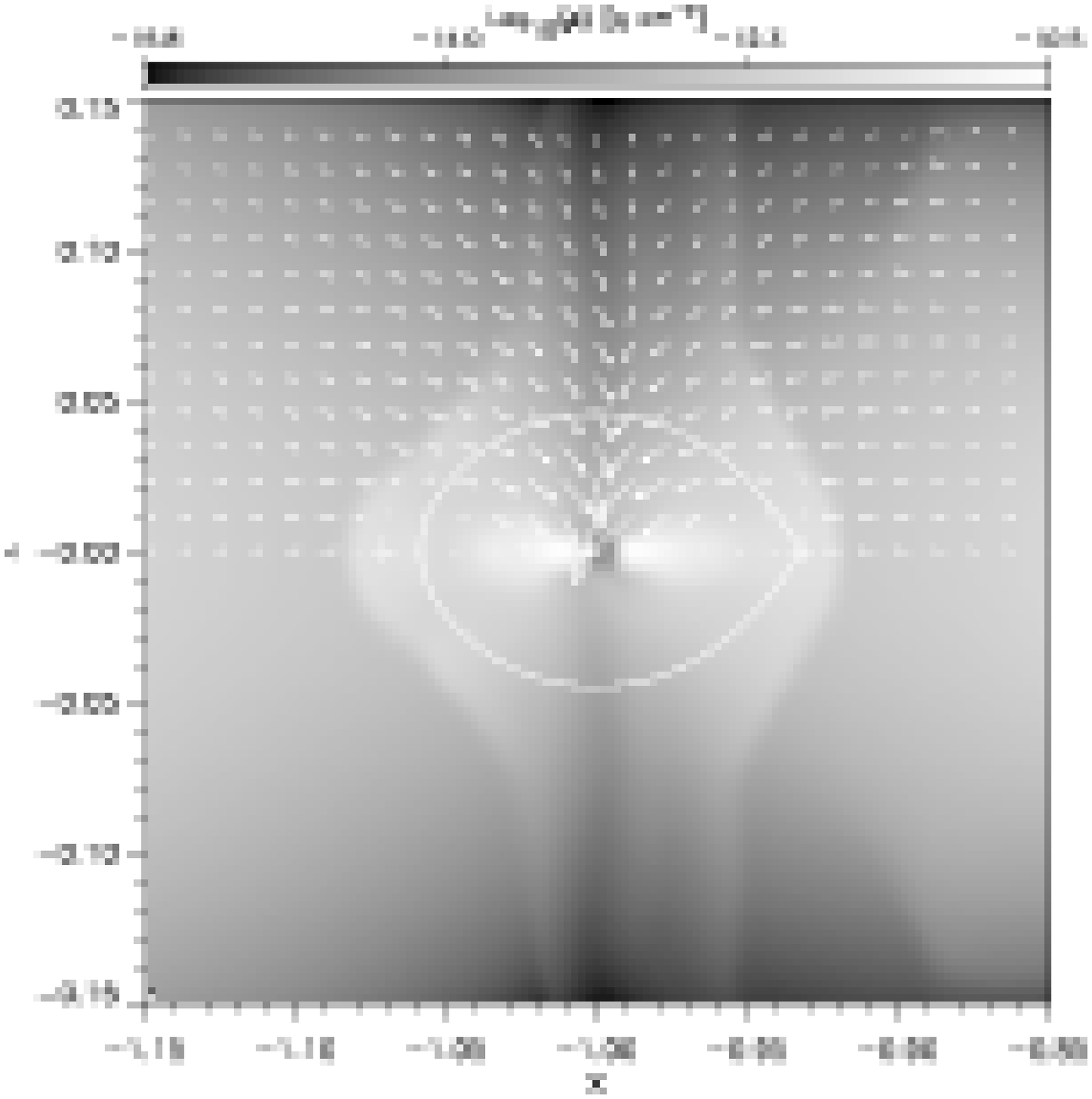,width=7.5truecm}\psfig{figure=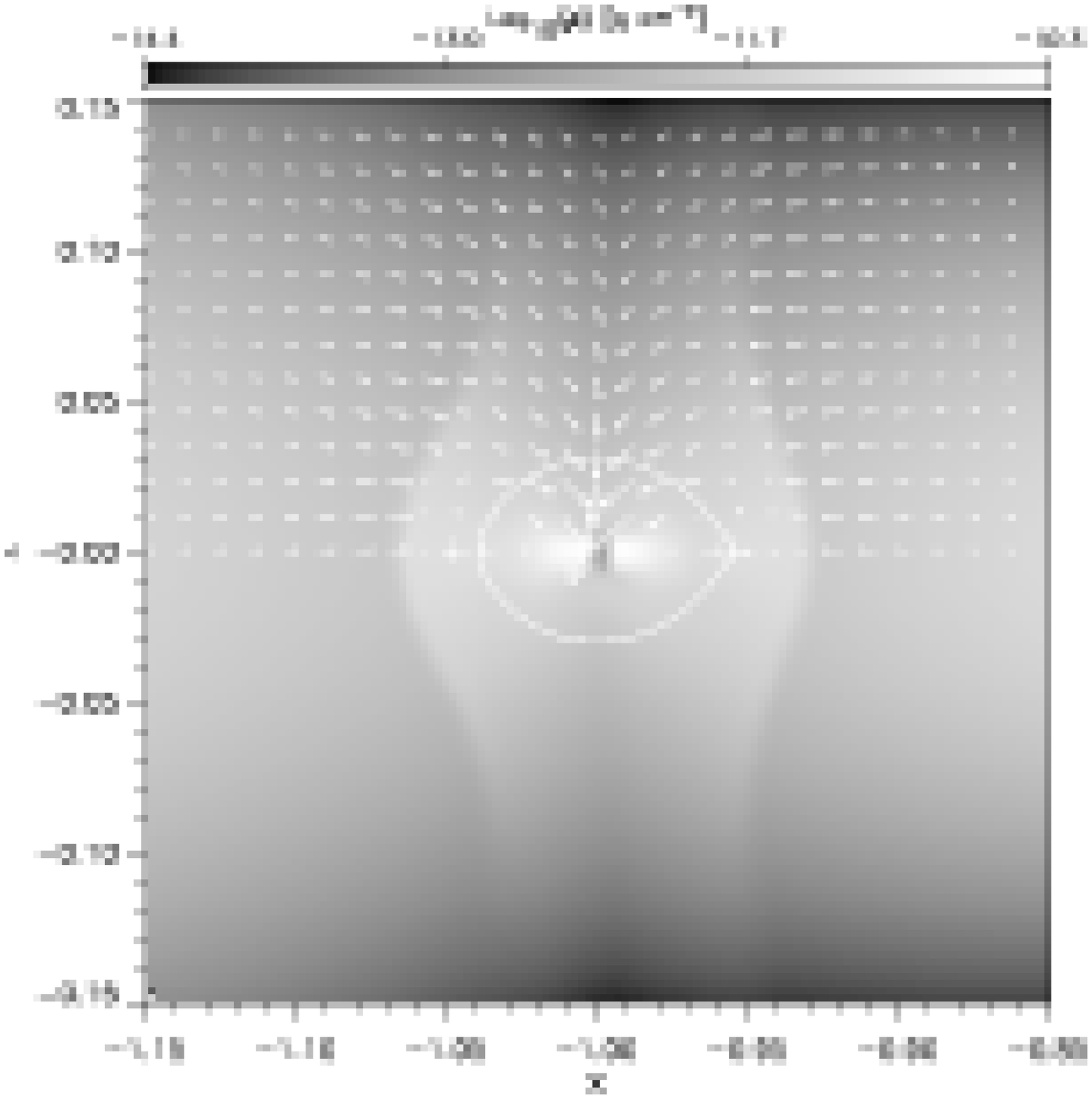,width=7.5truecm}}
\centerline{\psfig{figure=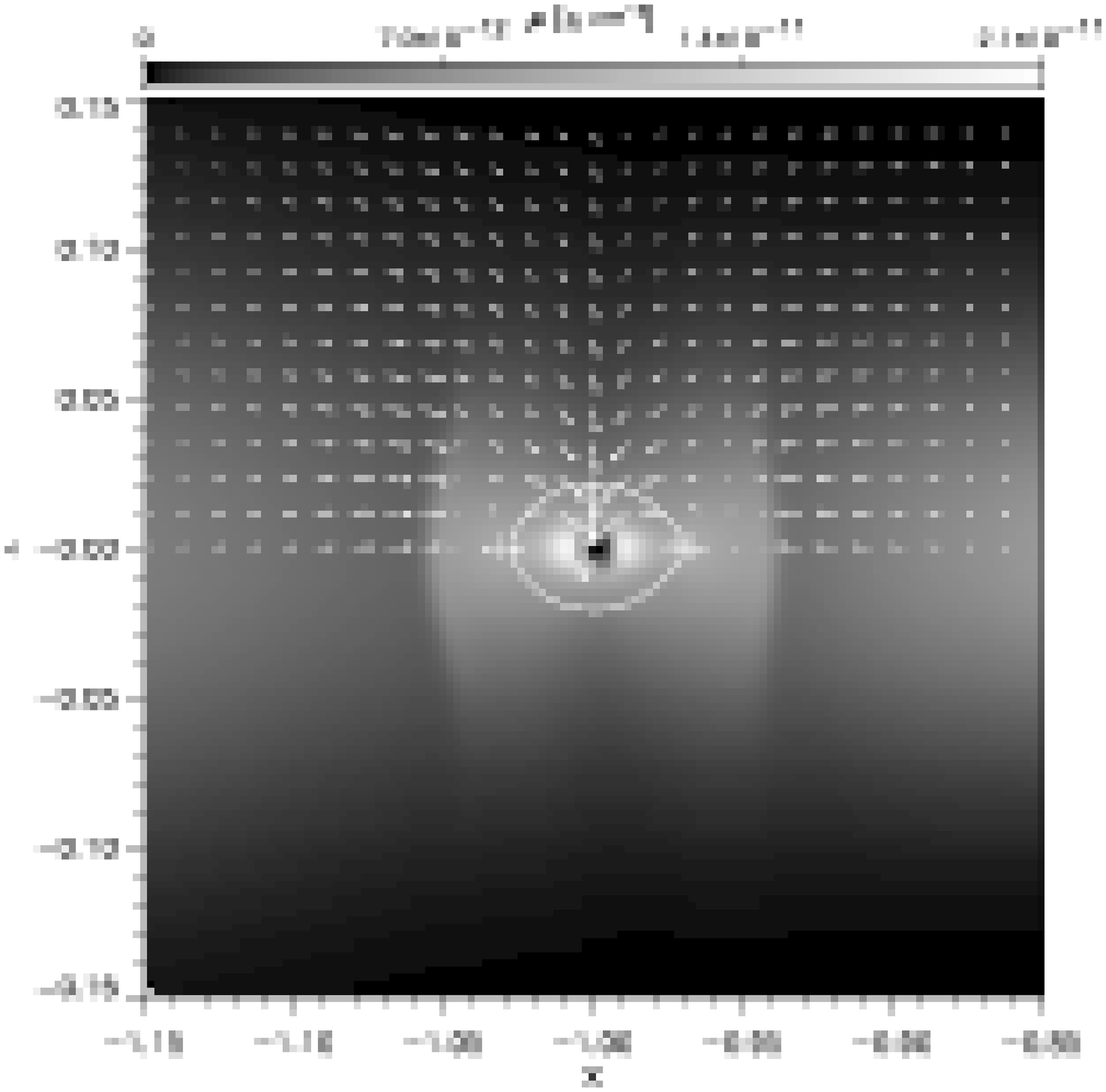,width=7.5truecm}\psfig{figure=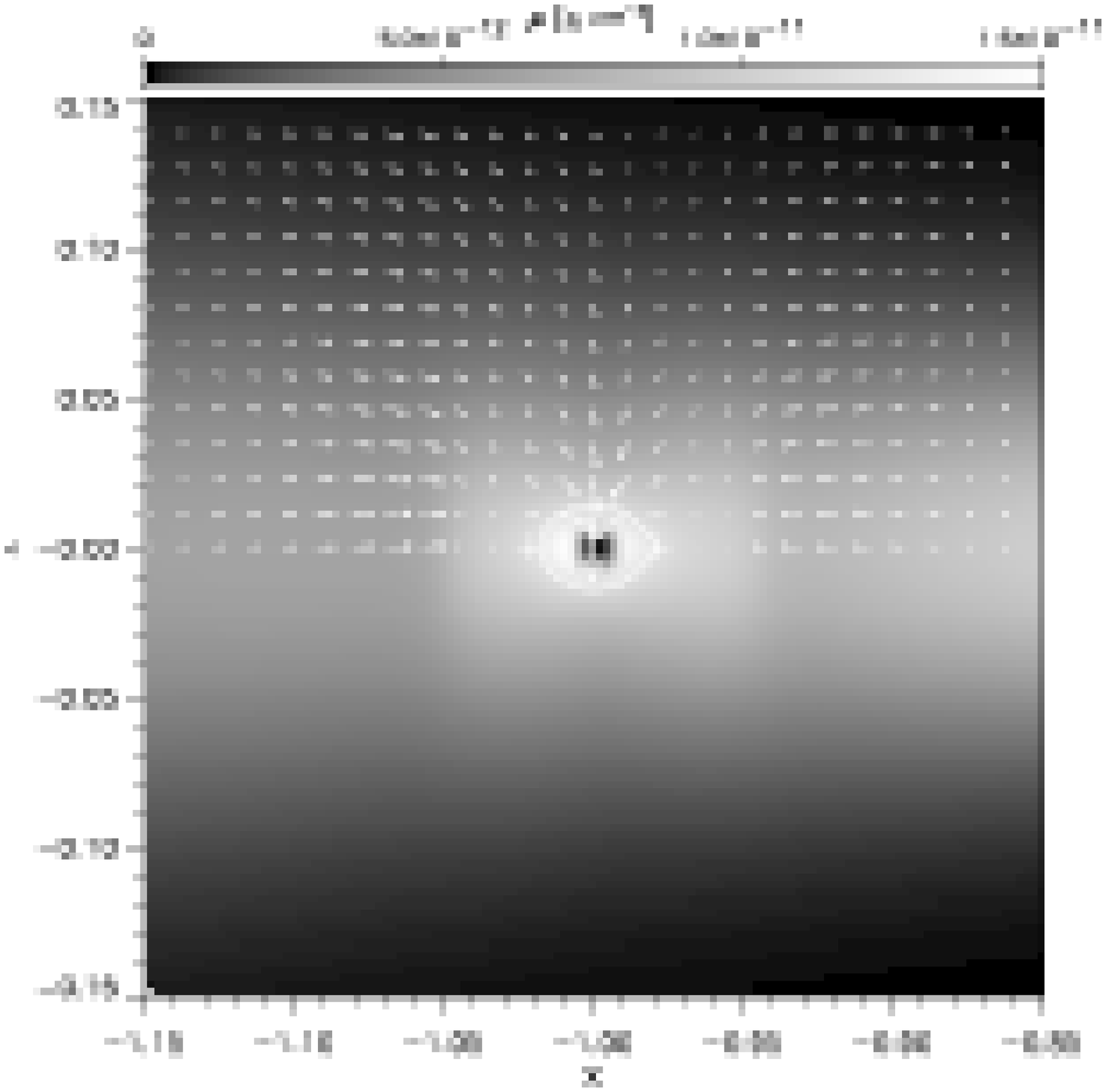,width=7.5truecm}}
\centerline{\psfig{figure=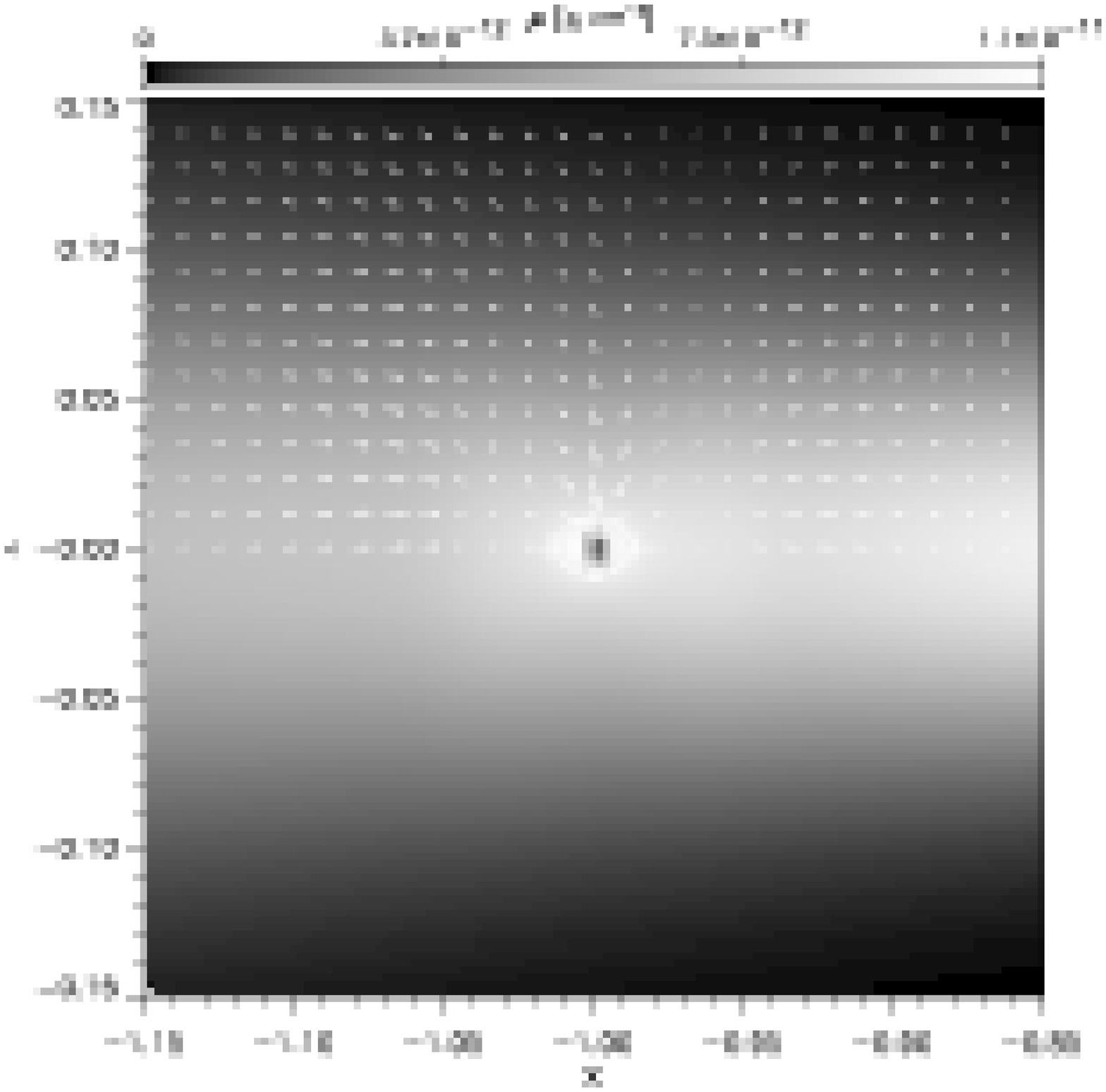,width=7.5truecm}\psfig{figure=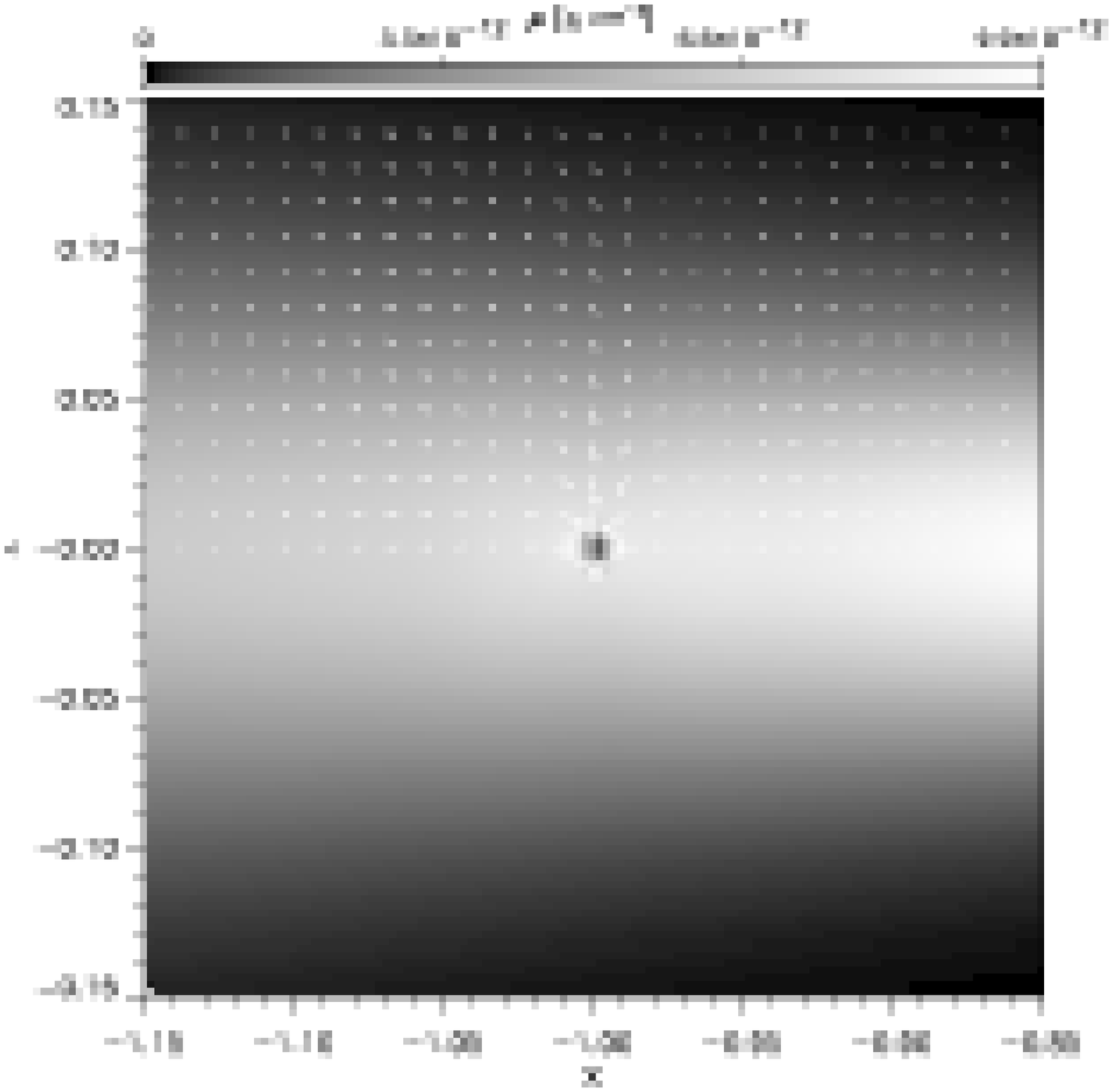,width=7.5truecm}}
\caption{\label{slice} Disc density, $\rho$, and velocity vectors in an $r-z$ slice through the disc at the location of the planet (i.e.\ $\phi=\pi$) for 1, 0.3, 0.1, 0.03, 0.01, and 0.003 \mj\ planets (top-left to bottom-right).  Notice that the top 2 figures are scaled by the logarithm of the density. }
\end{figure*}

\subsubsection{The flow in the vicinity of the planet}
\label{vicinity}

In Figure \ref{zoomsurf}, we plot the surface density in
the vicinity of the planet.  Overlaid on the plots are
the Roche lobes of the planets and their inner (L1) and 
outer (L2) Lagrangian points (crosses).  In Figure 
\ref{hmid}, we plot the density of the disc (greyscale)
and streamlines on the disc midplane 
in the vicinity of the planet.  From these 
figures, we see that the waves are present in the circumstellar
discs, as was found in the previous two-dimensional simulations.
They are initiated near the planet and propagate as shocks.  
Even with the 1 \me\ planet, the density 
jump across the shock is of order 10\%, and the more massive
planets have greater density jumps.  The discontinuities 
in the streamlines across the shocks are obvious for 
planets with masses $\mpl \geq 10$ \me.

The shape of the shocks is independent of the planet's 
mass so long as a deep gap is unable to form 
(i.e.\ for planets with masses $\mpl \lsim 0.1$ \mj).  
For the higher-mass planets, the shocks become more 
curved, the outer shock moves forward, 
and the inner shock moves back.  
Unlike the 1 \mj\ case, the 
shocks generated by the low-mass planets do not extend 
radially all the way to the planet's Roche
lobe.  Instead, their radial extent is determined by the
disc thickness.  The shocks reach no closer to the planet than
a distance of approximately $H$ (i.e.\ they begin at approximately at 
$r=r_{\rm p} \pm H$).  This fact will be important when
we come to consider the torque exerted on the planet by 
the disc.  Although the disturbances are non-linear, 
their form is similar to the surface density perturbations 
expected from linear resonance theory (see Figure 6 of 
Tanaka et al.\ 2002).

Figure \ref{h1} provides the density of the gas in the 
vicinity of the planet, but at height $H$ above 
the midplane.  Figure \ref{slice} gives the density 
and velocity vectors in an $r-z$ slice through the
disc at the location of the planet (i.e.\ $\phi=\pi$).
The results are plotted by
using a Cartesian coordinate system $(x,y,z)$, such that
the the planet is located at $(-1,0,0)$ and
disc midplane is defined by $z=0$.

Notice that we have not plotted streamlines in these 
figures.  It is difficult to plot a set of three-dimensional 
streamlines because they cross each other in projection 
on to a graph.  Streamlines can be plotted at the disc 
midplane where the vertical velocity is zero (by symmetry).
The velocity vectors in Figure \ref{slice}
cannot be connected to form streamlines because information
about the velocity in the deprojected direction ($y$) is required.
Consequently, one cannot conclude that material rapidly drops
to the midplane near $x=1$. Instead, material off the midplane 
flows past a low-mass planet in the $y-$direction.

These figures allow us to determine the structure of 
the shocks outside the Roche lobe of the planet.  
Comparing the locations of the shocks in similar 
panels between Figures \ref{hmid} and \ref{h1}, 
we find that the locations are similar for the 
low-mass planets but, for 1 and 0.3 \mj, the 
shock fronts on the midplane lead the shock fronts 
off the midplane.  Thus, a section perpendicular to
the shock front for the high-mass planets would have a 
`bowshock' shape, whereas for the low-mass planets 
the shocks are essentially vertical planes.
These shock shapes can be seen clearly in Figure 
\ref{slice} with the transition from nearly vertical 
(0.1 \mj) to bowshock-shaped  (1 \mj).  The 
discontinuity in the velocity vectors across 
these shocks is also clear in Figure \ref{slice}.

\begin{figure}
\centerline{\psfig{figure=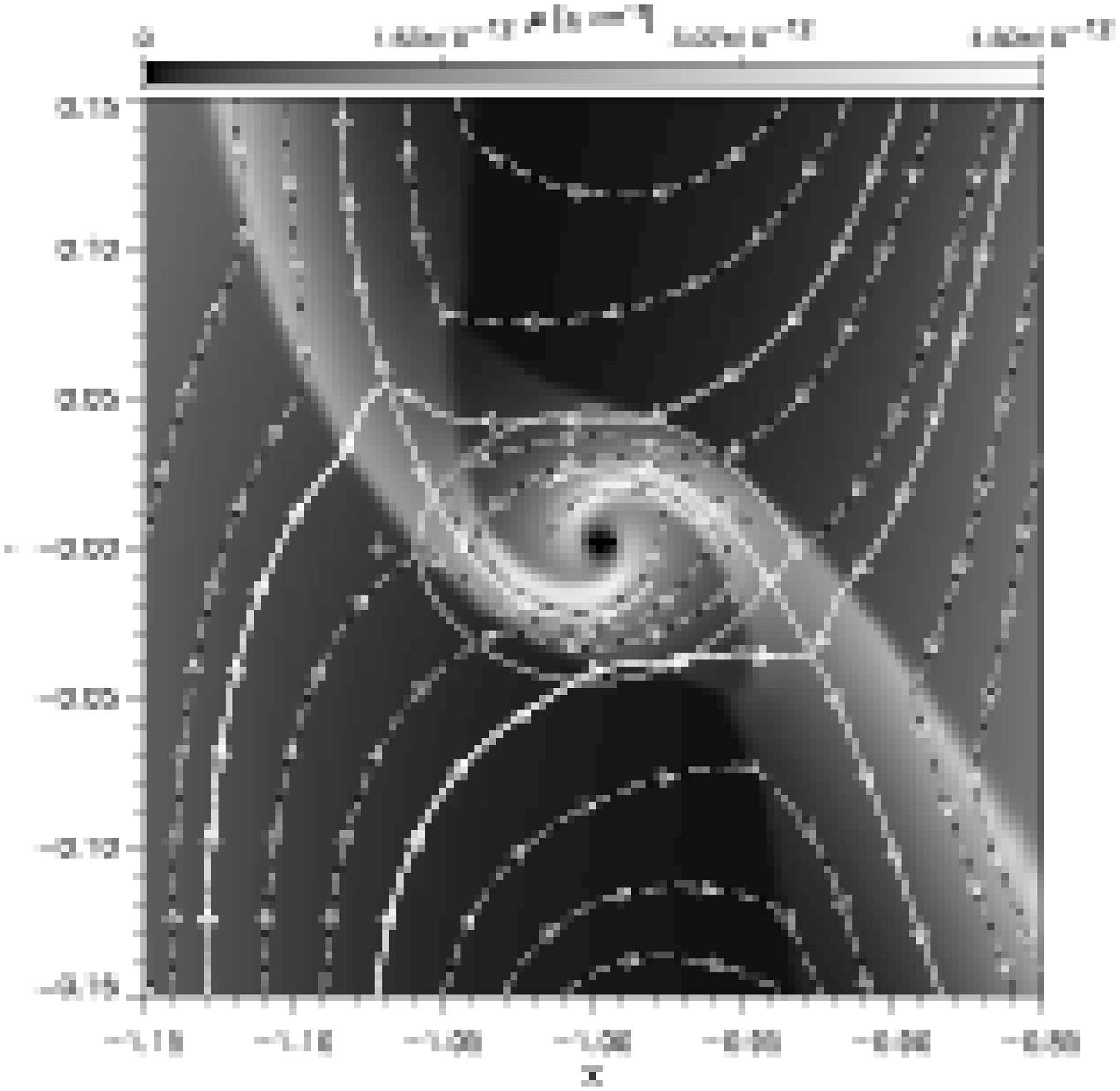,width=7.5truecm}}
\caption{\label{hmid2d} Results from a two-dimensional calculation to be compared with the three-dimensional calculation pictured in the top-left panel of Figure 5.  Disc density, $\rho = \Sigma/(\sqrt{2\pi}H)$, and streamlines on the disc midplane for a 1 \mj\ planet.  The thick white streamlines are the critical streamlines that mark the boundaries between the outer/inner disc, the accretion streams into the Roche lobe, and the horseshoe orbits.  Notice that the streams on to the planet are much broader than in the three-dimensional calculation and there are strong spiral shocks in the circumplanetary disc which are not present in three dimensions.}
\end{figure}

Returning to the density structure and streamlines at 
the midplane (Figure \ref{hmid}), we see that gas at the same 
radius as the planet but outside the planet's
Roche lobe moves on horseshoe orbits, as was reported in
earlier studies (e.g., Bryden et al.\ 1999; Kley 1999;
Lubow et al.\ 1999).  
For the high-mass planets, these horseshoe orbits 
occupy part of the gap in the disc.  The radial extent of 
the horseshoe orbits decreases as the mass of the
planet is decreased.

Between the outer disc and the portion of the horseshoe
orbits ahead of the planet, and between the inner disc 
and the portion of the horseshoe orbits behind the 
planet, there are two streams of material that enter 
the Roche lobe of the planet.  This is the material 
(along with material from above and below the disc 
midplane that is harder to visualise) that is accreted 
by the planet (see also Lubow et al.\ 1999).  For
planets with masses $\mpl \leq 0.1$ \mj, we notice that the 
breadth of these streams generally increases as the 
planet's mass increases, with the breadth of the stream
being of order the Roche radius of the planet.
In Section \ref{accretion}, we use this observation 
to develop a model for the accretion rate of the
low-mass planets.  For the 1 and 0.3 \mj\ planets, 
the streams are significantly narrower than the planet's
Roche lobe.  In fact, for the 0.3 \mj\ planet, 
{\it no streamlines} on the midplane enter the planet's
Roche lobe.  Since this planet has the second highest
accretion rate among the cases we consider, 
material is still being accreted, but 
the flow must be intrinsically three-dimensional with 
the accretion coming from above and below the midplane.
The 0.3 \mj\ planet may be a special case, since the
its Roche radius $r_{\rm R}=0.046$ is almost identical 
to the unperturbed scaleheight of the disc.  However, even with 
the 1 \mj\ planet, the streams entering the Roche 
lobe on the midplane are significantly narrower than 
in the two-dimensional calculations of Lubow et al.\ 
\shortcite{LubSeiArt1999}, again implying that the flow
into the planet's Roche lobe is three-dimensional.
To illustrate further this difference between the 
two-dimensional and three-dimensional accretion streams
on to a Jupiter-mass planet, we have performed a
two-dimensional 1 \mj\ calculation with identical 
resolution in $r$ and $\phi$ to our three-dimensional 
calculation.  The density and streamlines in the 
vicinity of the planet are given in Figure \ref{hmid2d} 
and should be compared to the top-left panel of 
Figure \ref{hmid}.  The accretion streams are much
broader in the two-dimensional case.

\begin{figure*}
\centerline{\psfig{figure=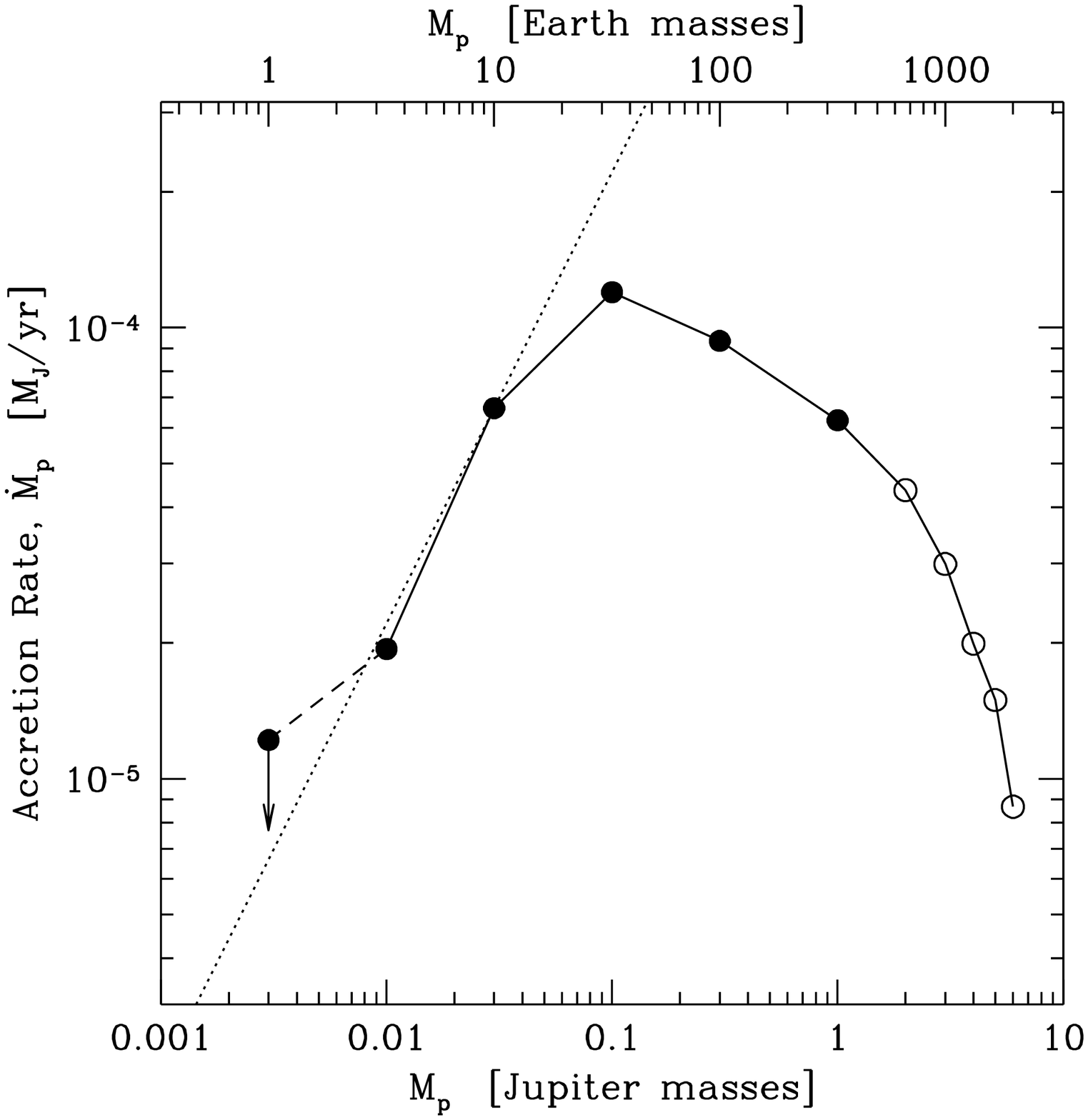,width=8.5truecm}\psfig{figure=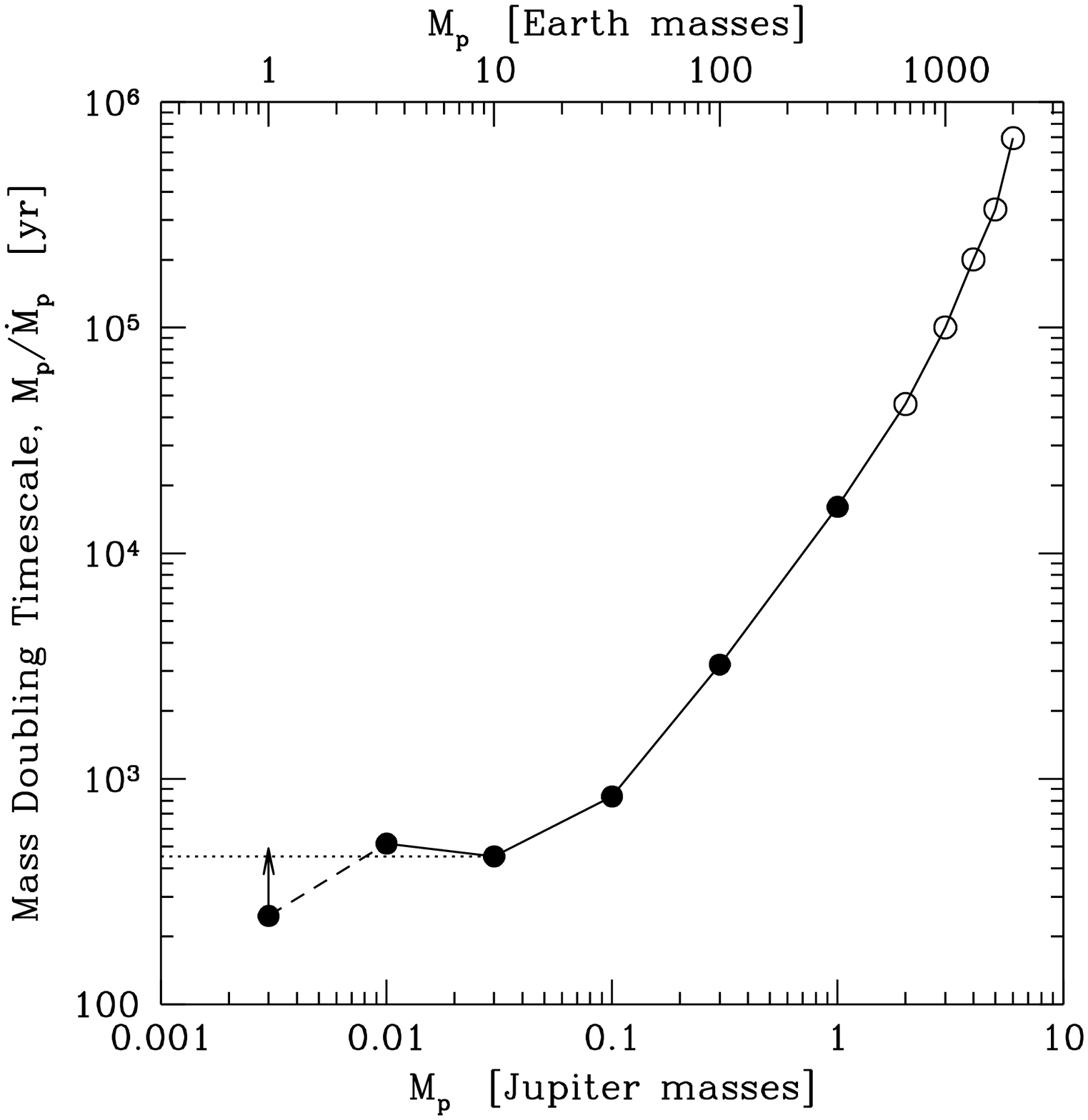,width=8.5truecm}}
\caption{\label{accrate} Left: Accretion rate, $\dot{M}_{\rm p}$, versus planet mass.  Right: Mass doubling timescale, $M_{\rm p}/\dot{M}_{\rm p}$.  Our results are shown by the filled circles.  Open circles are taken from Lubow et al.\ (1999) and scaled to match the accretion rate of our Jupiter-mass planet.  The accretion rate of the lowest-mass planet (1 \me) is slightly over-estimated because gas near the planet's Roche lobe is sucked into the Roche lobe by the evacuated zones (see Figure 5).  Thus, we plot this as an upper limit with an arrow.  Our results show that low-mass planets ($\mpl \lsim 10$ \me) accrete proportional to their mass (equation 8, dotted lines), if the thermal energy of the gas can be radiated away quickly enough.  Thus, the mass doubling timescale is independent of mass for low-mass planets and is $\approx 500$ years for our chosen disc mass.  For higher-mass planets, the accretion rate drops rapidly with increasing mass as the planet opens a gap in the disc.  The rapid fall off in accretion rate for very high masses sets a natural upper limit to the mass of a planet of approximately 10 \mj\ (see also Figure 13).}
\end{figure*}

Finally, we note that flow near the 1 \me\ planet is
only marginally resolved, since the evacuated zones at 
the location of the planet can be seen to affect the 
gas flow at the boundary of the Roche lobe.  Whereas the 
critical streamlines that mark the boundaries of
the accretion streams graze the Roche lobe of the 
3.3 and 10 \me\ planets, gas is sucked into the
Roche lobe of the 1 \me\ planet by the artificial 
pressure gradients.  This effect leads to a slight 
over-estimate of the accretion rate for the 1 \me\ planet
(Section \ref{accretion}).

\subsubsection{The circumplanetary disc}

Once inside the planet's Roche lobe, the gas settles 
into a circumplanetary disc.  In our calculations, 
the discs are resolved only for planets with masses 
$\mpl \geq 0.1$ \mj; the resolution within the Roche lobe 
of lower-mass planets is insufficient to follow 
material around complete orbits.

The scaleheight of the circumplanetary disc is much 
smaller than that of the circumstellar disc, $H$, due to 
the planet's vertical gravity.  This can be 
deduced from the fact that there is no sign of the 
discs in the plots of density at height $H$ above the midplane 
(Figure \ref{h1}).  Furthermore, Figure
\ref{slice} clearly shows the cross sections 
of the discs for the 1 and 0.3 \mj\ cases.
As in two-dimensional calculations, the discs rotate 
in a prograde manner (Lubow et al.\ 1999;
D'Angelo et al.\ 2002).  These circumplanetary 
discs may lead to satellite formation around the 
planets.

Lubow et al.\ \shortcite{LubSeiArt1999} studied 
the flow inside the Roche lobe of an accreting planet.
In two-dimensional calculations of a Jupiter-mass 
planet, they found that the circumplanetary disc 
contained strong shocks that led to rapid accretion 
of the gas through the planet's disc.  They concluded 
that the shocks formed from the collision of the two 
streams passing into the Roche lobe, one from the 
inner disc and one from the outer disc.  
D'Angelo et al.\ \shortcite{DAnHenKle2002} studied 
lower mass planets with a two dimensional nested 
grid code and found these shocks persisted down to 
planet masses of approximately $5$ \me.

We find that such strong shocks in the circumplanetary
disc do not occur in three-dimensional calculations.  
They are an artifact of the two-dimensional 
calculations in which the streams from the inner and 
outer discs are forced to collide without vertical motions.
The 0.3 \mj\ case is a particularly good example of the
difference between two and three-dimensional structures, since
in three dimensions there are no streamlines at the midplane 
that even enter the planet's Roche lobe. 

The disappearance of the spiral shocks in the 
circumplanetary disc when going from two to three-dimensional
simulations is further illustrated for a 1 \mj\ planet
by comparing the density and streamlines in a 
two-dimesional calculation (Figure \ref{hmid2d}) with 
those in the three-dimensional calculation 
(Figure \ref{hmid}, top-left panel).  
Stream collisions do occur at the midplane in both cases.
But the strong spiral shocks present in the two-dimensional 
calculation are greatly diminished in three dimensions.  This 
weakening in three dimensions is evidenced by the fact that 
streamlines within the Roche lobe are more tightly 
wrapped in three-dimensional case.
In two dimensions, the spiral shocks drive the accretion 
through the circumplanetary disc with the gas losing 
angular momentum on each passage through a shock.

We make two suggestions for the effects of the absence of 
strong spiral shocks in the three-dimensional 
circumplanetary discs.  First, accretion through 
the circumplanetary disc may be driven in a similar 
manner to accretion through the circumstellar
disc, rather than by spiral shocks.  
Second, the more quiescent three-dimensional flow
might be more conducive to satellite formation
in circumplanetary discs.

\subsection{Accretion rates}
\label{accretion}

In Figure \ref{accrate}, we plot the final accretion 
rates in Jupiter masses per year.  Lubow et al.\ (1999) 
performed two-dimensional calculations to determine the 
{\it relative} accretion rates of planets with masses 
$\mpl\geq 1$ \mj.  The open circles in Figure 
\ref{accrate} give their 
accretion rates, scaled to match our Jupiter-mass
case.  Two-dimensional calculations are adequate for
modelling accretion on to high-mass planets which open 
well-defined gaps in the disc (Kley et al.\ 2001).
Figure \ref{accrate} then gives accretion rates over
3.3 orders of magnitude in planet mass, from 1 Earth 
mass to 6 Jupiter masses.  As discussed in Section
\ref{vicinity}, the accretion rate on to the 1 \me\ is
slightly over-estimated because of the lack of 
resolution in the vicinity of the planet's Roche lobe.
Thus, the derived accretion rate is plotted as an 
upper limit in Figure \ref{accrate}.

For low-mass planets up to $\mpl \approx 10$ \me\ (0.03 \mj), 
the accretion rate is proportional to the planet's mass.
The accretion rates peak at $\mpl \approx 0.1\ \mj$, 
just as the planet starts to open a gap in the disc
(Figure \ref{surfdens}).  
For higher masses, the accretion rate drops rapidly with 
increasing mass as the gap gets wider 
(Figure \ref{surfdens}).
As discussed by Lubow et al.\ \shortcite{LubSeiArt1999}, 
the rapid decline of accretion rate at high masses 
provides a natural limit of about $10$ \mj\ for 
the mass of a planet on a circular orbit (see Section 4).

The gas mass accretion process is always dominated by 
three-body effects and the relevant capture radius 
is the Roche lobe radius $r_{\rm R}$.  This statement 
can be justified by considering the ratio of the
Roche lobe radius to the Bondi-Hoyle accretion radius
which is approximately $(H/r_{\rm R})^2$.  This ratio is greater 
than unity for planets whose mass is less than 
1 \mj\ for typical disc parameters 
($H/r \simeq 0.05$).  The mass capture rate for a 
low-mass planet can then be estimated by a simple 
argument that
$\dot{M}_{\rm p} \simeq \pi r_{\rm R}^2 \rho v$, 
where $\rho$ is the gas density along the planet's 
orbit and we have used the fact that the breadth of the
accretion streams into the planet's Roche lobe scales
with the Roche radius, $r_{\rm R}$ (Section \ref{vicinity}). 
But since the velocity of the gas relative to the 
planet $v \simeq \Omega_p r_{\rm{R}}$, 
using equation \ref{rocherad},
we can express the accretion rate as
\begin{equation}
\label{macc}
\dot{M}_{\rm p} = b \frac{M_{\rm p}}{M_*} \rho \Omega_{\rm p} r_{\rm p}^3,
\end{equation}
where $b$ is a constant of order unity.
Consequently, we recover the numerical result that 
$\dot{M}_{\rm p} \propto M_{\rm p}$, with $b = 2.30$ 
in the above equation (dotted lines, Figure \ref{accrate}).

The mass accretion efficiency was defined in Lubow et al.\ (1999)
as
\begin{equation}
{\cal E} = \frac{\dot{M}_{\rm p}}{3 \pi \nu \Sigma},
\end{equation}
where $\Sigma$ is the disc surface density just outside
the disc gap. The efficiency measures the ratio of the 
accretion rate on to the planet to the accretion rate 
that would occur in the disc if the planet was absent. 
As in Lubow (1999), we obtain accretion rates of order 
unity. For a 1 \mj\ planet, $\cal{E}$ is about a factor of 
2 larger than the value obtain in two dimensions.
We have also calculated the mass flow rate through $r=r_{\rm p}$
(i.e.\ past the planet).
In all cases, this rate is much less than the accretion rate onto the
planet.  For the 1 \mj\ planet, which opens a well-defined gap in the
disc, the flow through $r=r_{\rm p}$ is less
than 2\% of the planet's accretion rate.  For all the lower-mass planets,
the flow through $r=r_{\rm p}$ is less than 7\% of the planets' 
accretion rates.  Together, these results imply that mass freely flows 
into the gap, but that most is captured by the planet.  Naturally,
the flow past the planet would be expected to increase if the disc's 
viscosity were increased.

%Look at Tajima and Nakagawa 97 to compare rates.

\subsection{Migration rates}
\label{migration}

A planet experiences torques due to its interaction
with the disc.  Resonant torques likely play an important
role (Goldreich \& Tremaine 1980).  In addition, torques
may arise from the gas that flows in the gap, including 
material within the planet's Roche lobe.  For low-mass 
planets that undergo Type I migration, resonant torques 
peak in value at a radial distance of order $H$ from 
$r_{\rm p}$.  Consequently, it is necessary to resolve 
the gas density structure at distances of order $H$ from the 
planet.  Similarly, for planets that open gaps in the disc
and undergo Type II migration (i.e.\ they migrate due
to the disc's viscous evolution), it is necessary to
resolve the flow at distances of order $r_{\rm R} \approx H$
from the planet.

The migration rates due to resonant torques have been
the subject of many linear analyses 
(e.g.\ Goldreich \& Tremaine 1980; Hourigan \& Ward
1984; Ward 1986; Korycansky \& Pollack 1993; Ward 1997;
Tanaka et al.\ 2002).  Ward (1997) considered the motion 
of the planet relative to the disc material, treated the 
disc as being two dimensional, and did not consider 
corotation resonances.  More recently, 
Tanaka et al.\  calculated the Type I
migration rate expected from linear theory taking into
account the three-dimensional nature of the disc and
corotation resonances.  

Tanaka et al.\ \shortcite{TanTakWar2002} give
the Type I radial migration velocity to be
\begin{equation}
\label{tanaka}
v_{\rm I} = - f \frac{M_{\rm p}}{M_*} \frac{r_{\rm p}^2 \Sigma}{M_*} \left(\frac{H}{r_{\rm p}}\right)^{-2} r_{\rm p} \Omega_{\rm p},
\end{equation}
where $\Sigma$ and $H$ are evaluated at $r_{\rm p}$, and 
$f$ is a value of order unity that depends on the 
radial variation of both the disc surface density profile and the
scale height. For our disc parameters, we have $f=3.00$.  
The Type II radial migration velocity is simply the viscous radial 
velocity of the disc
\begin{equation}
\label{type2vel}
v_{\rm II} = - \frac{3\nu}{2 r_{\rm p}} = -\frac{3}{2}\alpha \left(\frac{H}{r_{\rm p}}\right)^2 r_{\rm p} \Omega_{\rm p}.
\end{equation}
The migration timescales are then given by $\tau = r_{\rm p}/|v_{\rm I}|$
and $r_{\rm p}/|v_{\rm II}|$.

\begin{figure}
\centerline{\psfig{figure=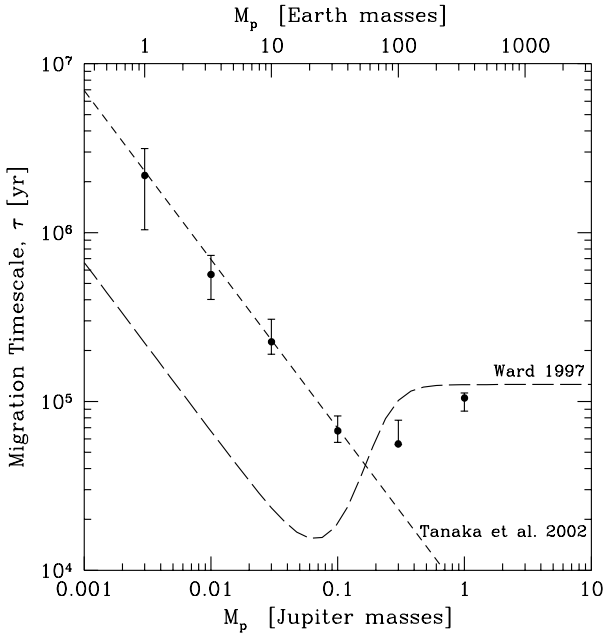,width=8.5truecm}}
\caption{\label{migtime} Migration timescale, $\tau$, versus planet mass.  Our results are shown by the filled circles.  The short-dashed line is the three-dimensional linear prediction of Tanaka et al.\ (2002) which applies for low-mass planets that do not open gaps.  The long-dashed line is the two-dimensional linear prediction of Ward (1997) which applies for both low-mass and high-mass planets.  Our results are calculated by using the net torque from outside the planet's Roche radius $r_{\rm R}$, except for the lowest mass planet (1 \me) where we neglect the torques from inside $2r_{\rm R}$.  The errorbars do not give the uncertainties in our measurements.  Rather, they give the migration timescales that are obtained when neglecting the torques inside $0.5r_{\rm R}$ (lower bar) and $1.5r_{\rm R}$ (upper bar) ($1r_{\rm R}$ and $3r_{\rm R}$ for the 1\me\ planet).  Thus, they indicate the sensitivity of the migration timescale to torques in the vicinity of the Roche lobe.}
\end{figure}

\begin{figure}
\centerline{\psfig{figure=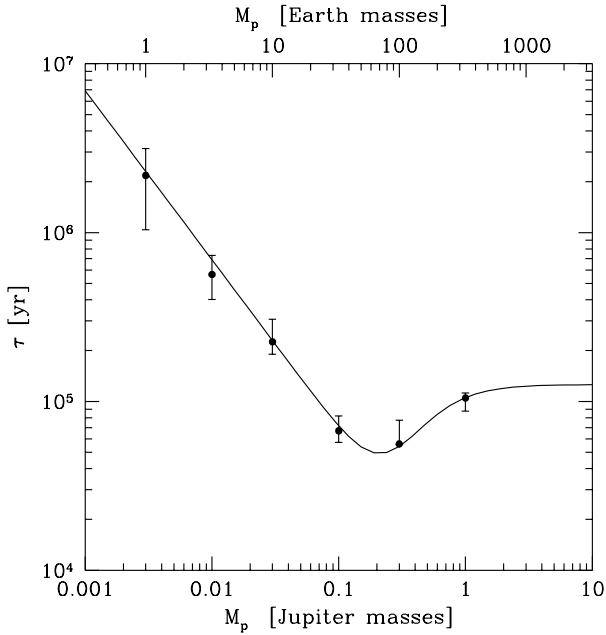,width=8.5truecm}}
\caption{\label{migfit} Migration time, $\tau$, versus planet mass.  Our results are shown by the filled circles (see also Figure 10).  The solid line is an empirical fit to the migration rates (equation 16) that has a transition from the Type I migration rate (Tanaka et al.\ 2002) to the Type II migration rate at $\mpl=0.23~\mj$.}
\end{figure}

\subsubsection{Non-linear results}

Both Ward \shortcite{Ward1997} and 
Tanaka et al.\ \shortcite{TanTakWar2002} assume that
the material near the planet (i.e.\ inside the planet's
Roche lobe) does not exert a net torque on the planet.
In this section, we aim to compare their results with
those of our simulations.
Thus, when evaluating the torque on the planet from our 
numerical calculations, we only include torques from
material at distances greater than some cutoff radius
from the planet, $r_{\rm c}$.  Our default is to set
the radius of this sphere to be equal to the Roche radius of
the planet, $r_{\rm c} = r_{\rm R}$.  We discuss the 
torques inside the Roche lobe in Section \ref{torquedistsec}.

For each of the 6 planet masses, we calculate the 
average torque exerted on the planet over the last
simulated orbital period
(i.e.\ when the calculations have reached steady 
accretion).  The resulting migration timescales $\tau$, are 
plotted with filled circles in Figure \ref{migtime}.
Notice that because the gas flow near the Roche lobe 
of the 1\me\ planet is affected by the evacuated zones 
(Section 3.2.2), we have increased $r_{\rm c}$
to $2r_{\rm R}$ for this case.

As seen in Figure \ref{migtime},
our numerical results are in excellent agreement with
the linear theory of 
Tanaka et al.\ \shortcite{TanTakWar2002} for 
$M_{\rm p}\leq 0.1\mj$.  Although the linear theory of
Tanaka et al.\ is only formally valid for 
$r_{\rm R} \ll H$, the agreement is excellent up to 
$r_{\rm R} \simeq (2/3) H$.  Higher-mass planets migrate
more slowly than predicted by equation \ref{tanaka}, 
with a timescale than converges toward the Type II 
prediction at $M_{\rm p}\gsim 1\mj$.  Our migration
timescales are about a factor of 3 longer than those
reported from the two-dimensional numerical calculations 
of D'Angelo et al.\ \shortcite{DAnHenKle2002}.

We point out, however, that the
corotation resonance is subject to saturation.  For 
our disc parameters saturation may occur for planets of mass
greater than about 8 \me\ (Ward 1992). 
Saturation would result in a reduction of the Type
I migration timescale by less than 30\%. This level of change
does not significantly impact the level of agreement between the
simulations and the theory on the scale of Figure \ref{migtime}.

The errorbars in Figure \ref{migtime} do not 
represent the uncertainties in our 
measurements.  Rather, they give the migration timescales
that are obtained if we change the value of the cut-off 
radius to $r_{\rm c} = 0.5 r_{\rm R}$ or 
$1.5 r_{\rm R}$ ($r_{\rm c} = 1 r_{\rm R}$ or 
$3 r_{\rm R}$ for the 1\me\ case).  
Thus, they indicate the sensitivity of the migration 
timescale to the material just inside or outside of 
the Roche lobe.

Our numerical results show a transition from
Type I to Type II migration that is qualitatively,
but not quantitatively,
similar to that predicted by Ward \shortcite{Ward1997}.
In Figure \ref{migfit}, we provide an empirical fit to our 
results as the migration rate $\tau = r_{\rm p}/|v|$, where
\begin{equation}
v = \frac{v_{\rm I}}{1+(M_{\rm p}/M_{\rm t})^3} + \frac{v_{\rm II}}{1+(M_{\rm t}/M_{\rm p})^3}.
\end{equation}
Parameter $M_{\rm t}$ is the transition mass between 
Type I and Type II migration.  We find that a good fit 
is obtained with $3/5$ of the mass for which $r_{\rm R}=H$, 
i.e., $M_{\rm t} = 1.8 M_* (H/r_{\rm p})^3=0.23~\mj$ (Figure \ref{migfit}).
As can be seen from Figures \ref{migtime} and \ref{migfit}, 
the transition from Type I to Type II migration involves
a much smaller shift in timescales (about a factor of 2), than 
was suggested by Ward (1997) (about a factor of 10).
Note that this shift in timescales is quite sensitive to
$H/r_{\rm p}$ (equations \ref{tanaka} and \ref{type2vel}) and 
increases for thinner discs.

\begin{figure*}
\centerline{\psfig{figure=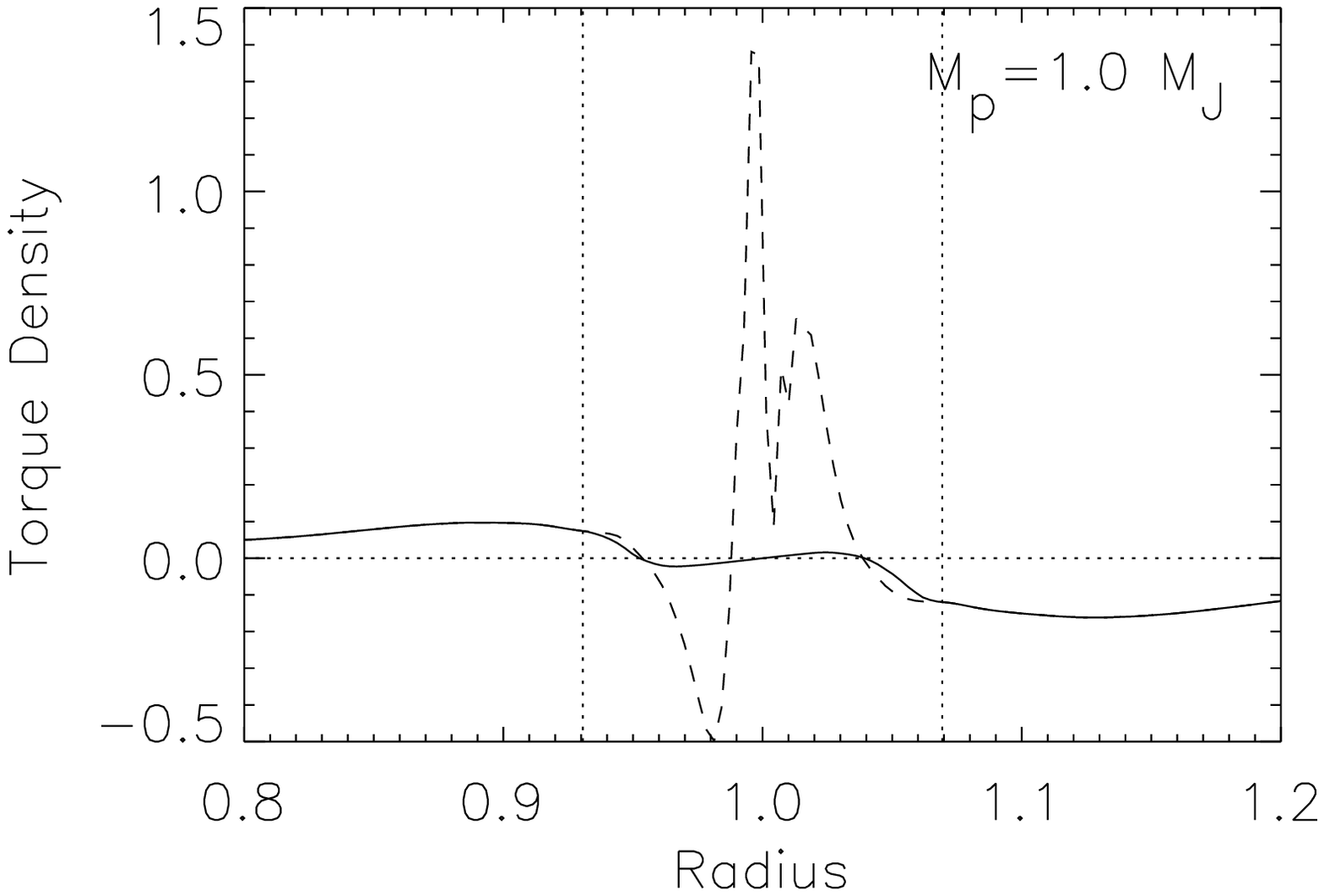,width=4.2truecm}\psfig{figure=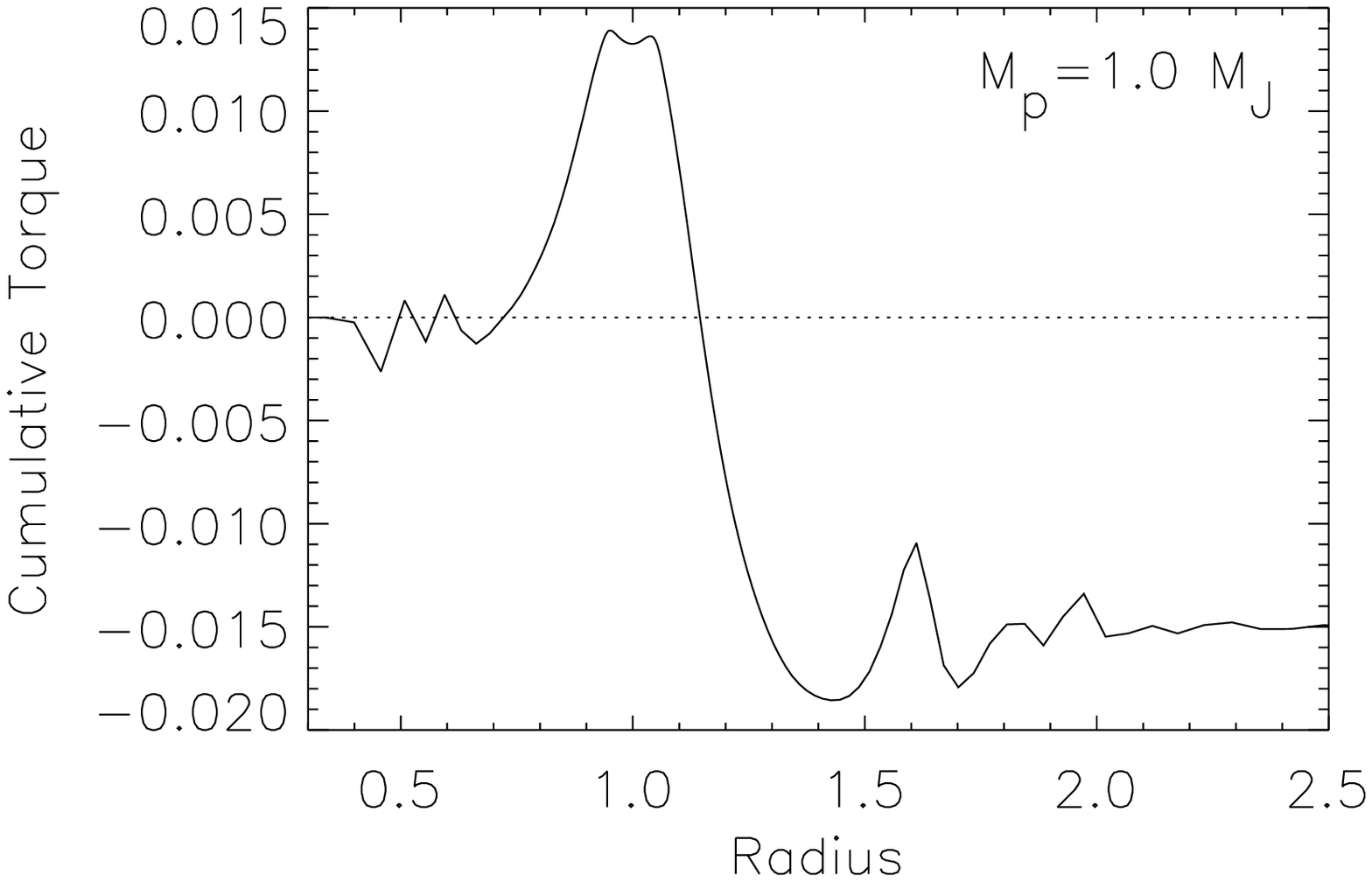,width=4.2truecm}\hspace{1.0cm}\psfig{figure=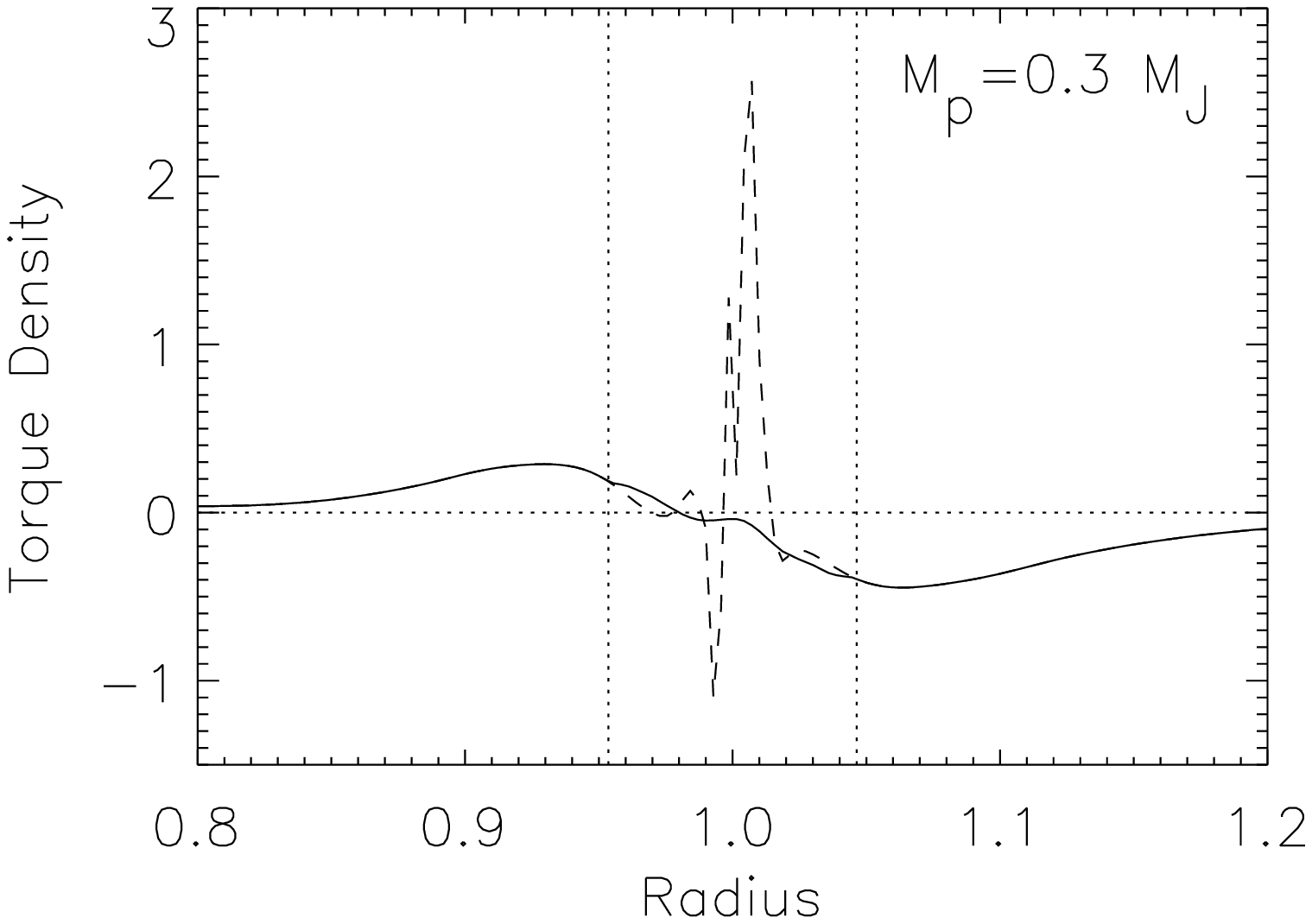,width=4.2truecm}\psfig{figure=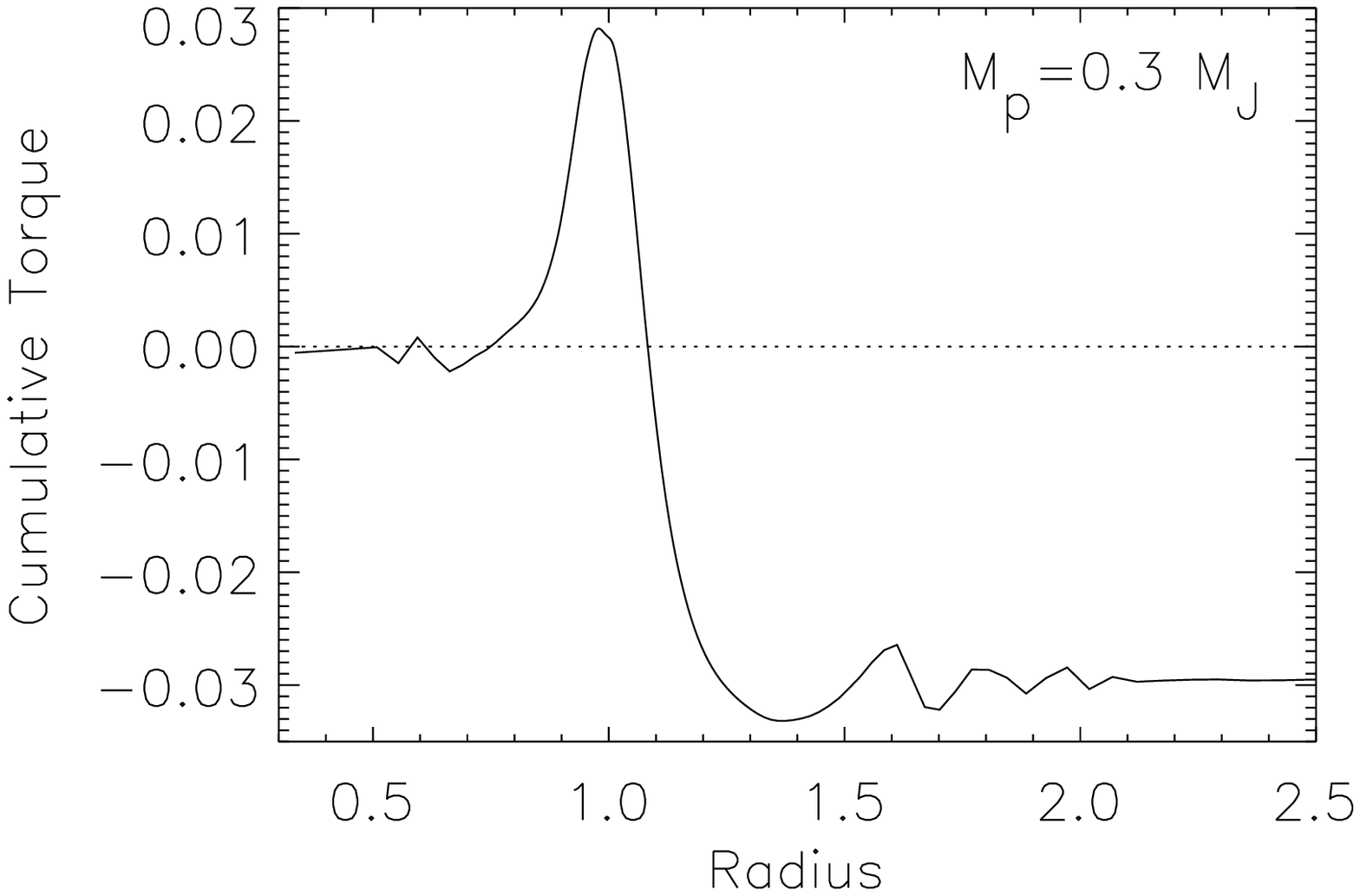,width=4.2truecm}}
\centerline{\psfig{figure=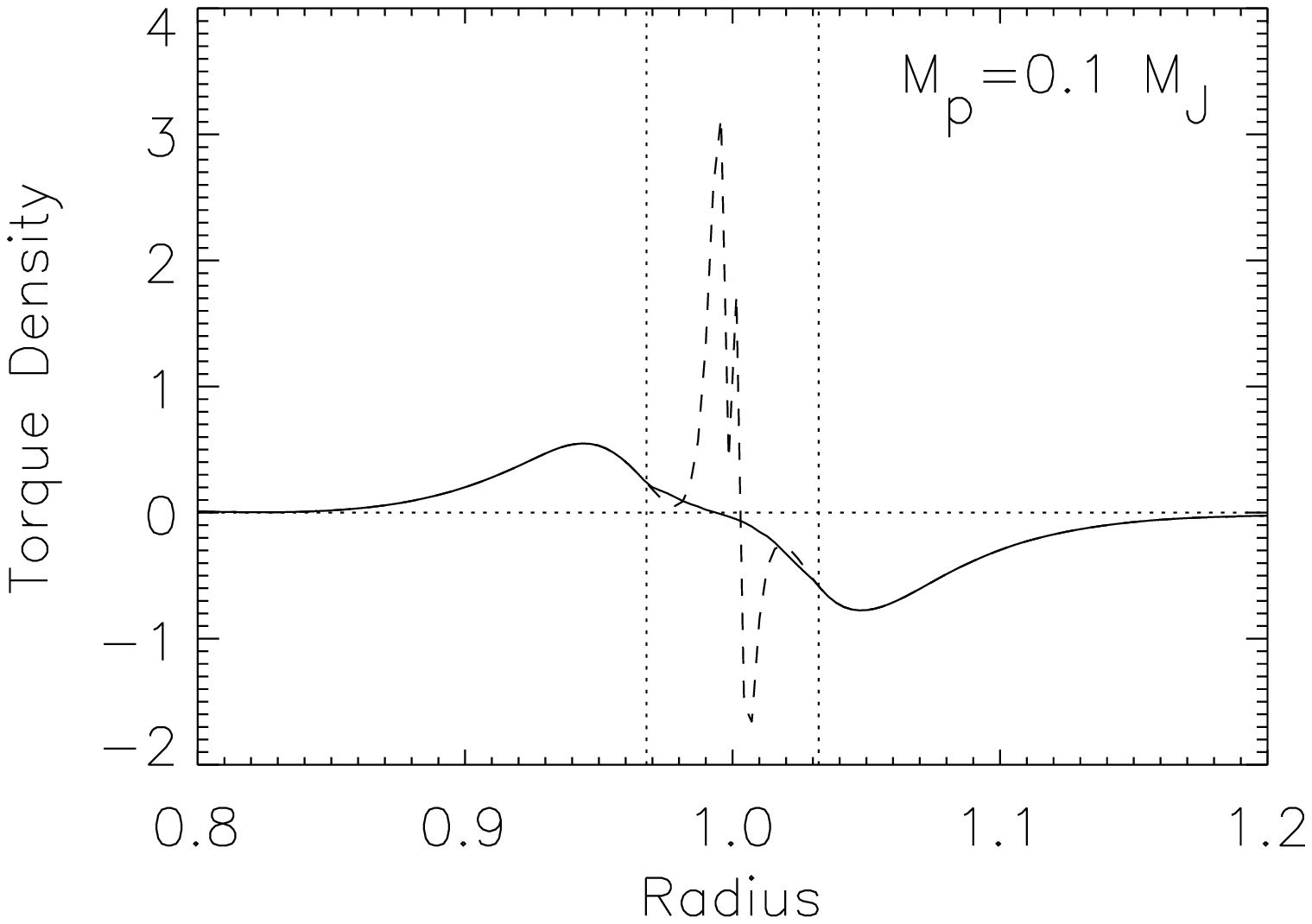,width=4.2truecm}\psfig{figure=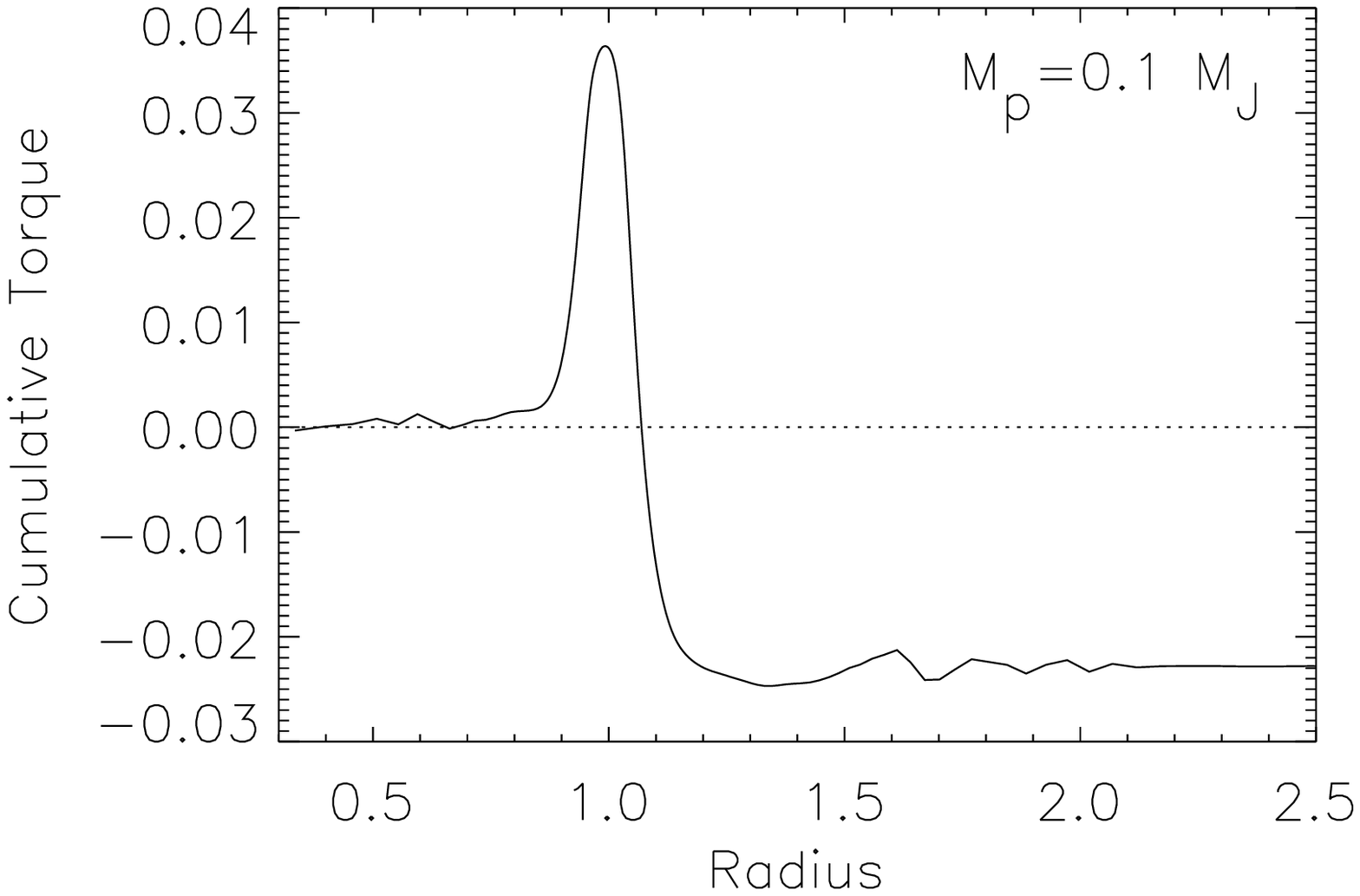,width=4.2truecm}\hspace{1.0cm}\psfig{figure=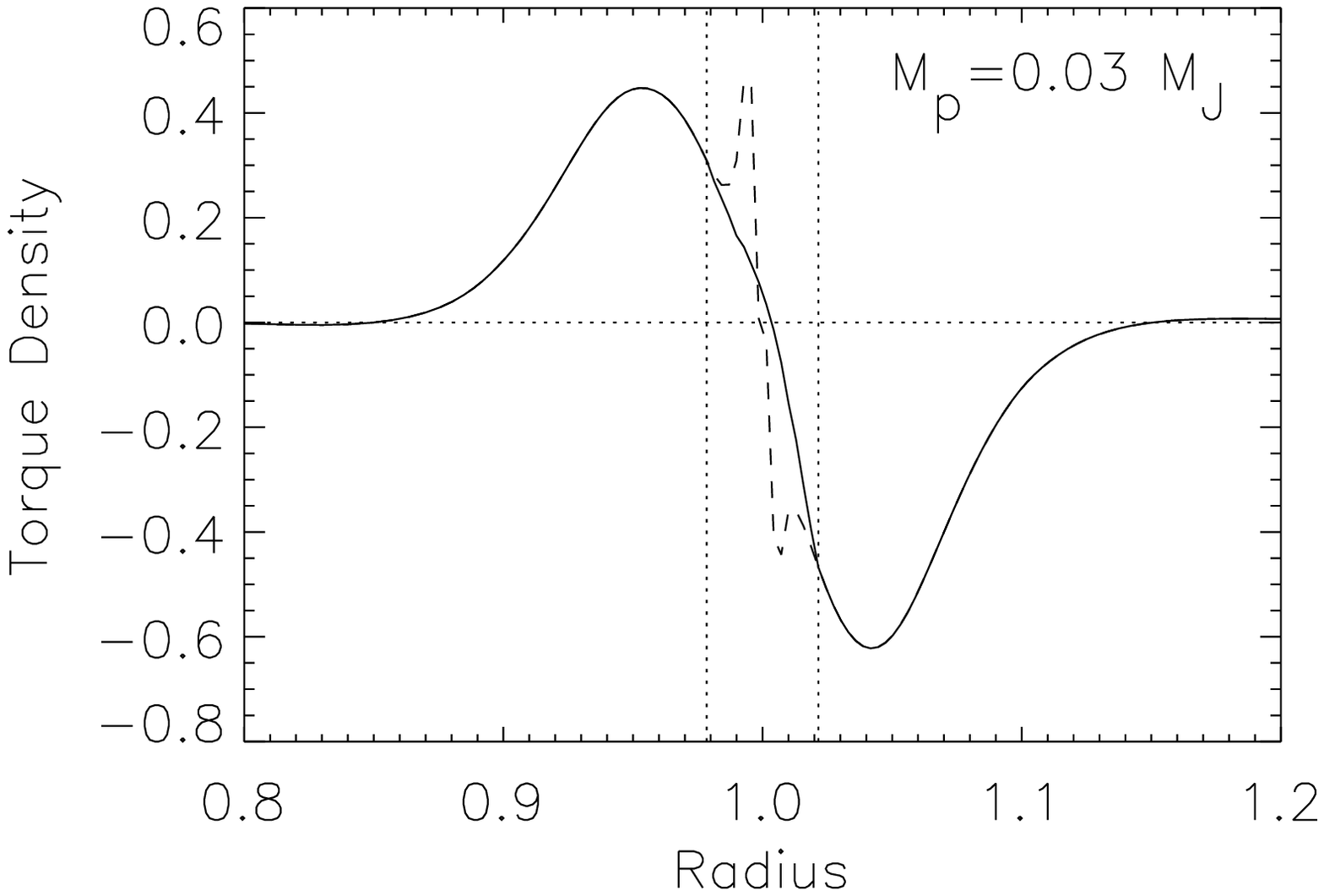,width=4.2truecm}\psfig{figure=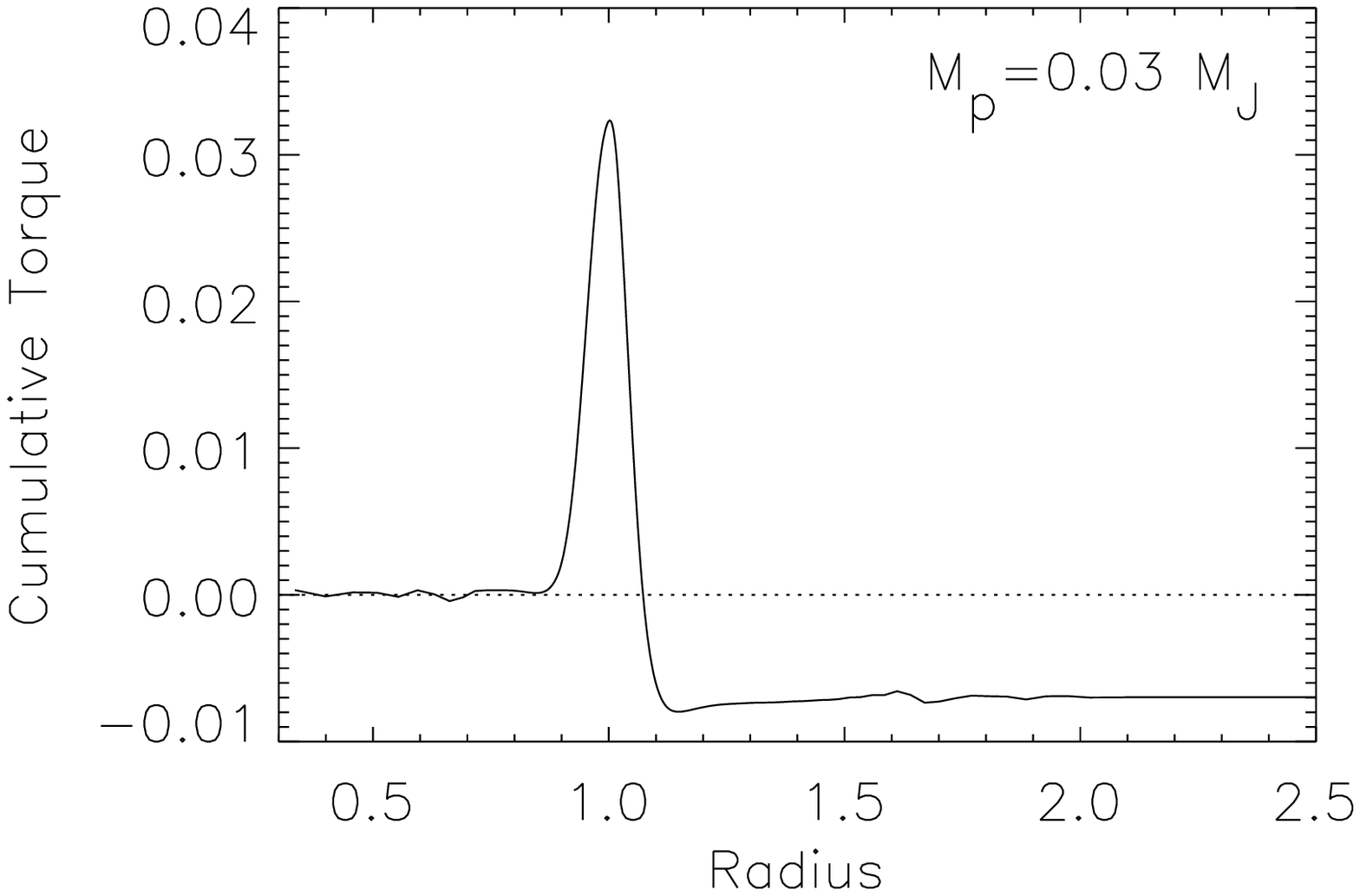,width=4.2truecm}}
\centerline{\psfig{figure=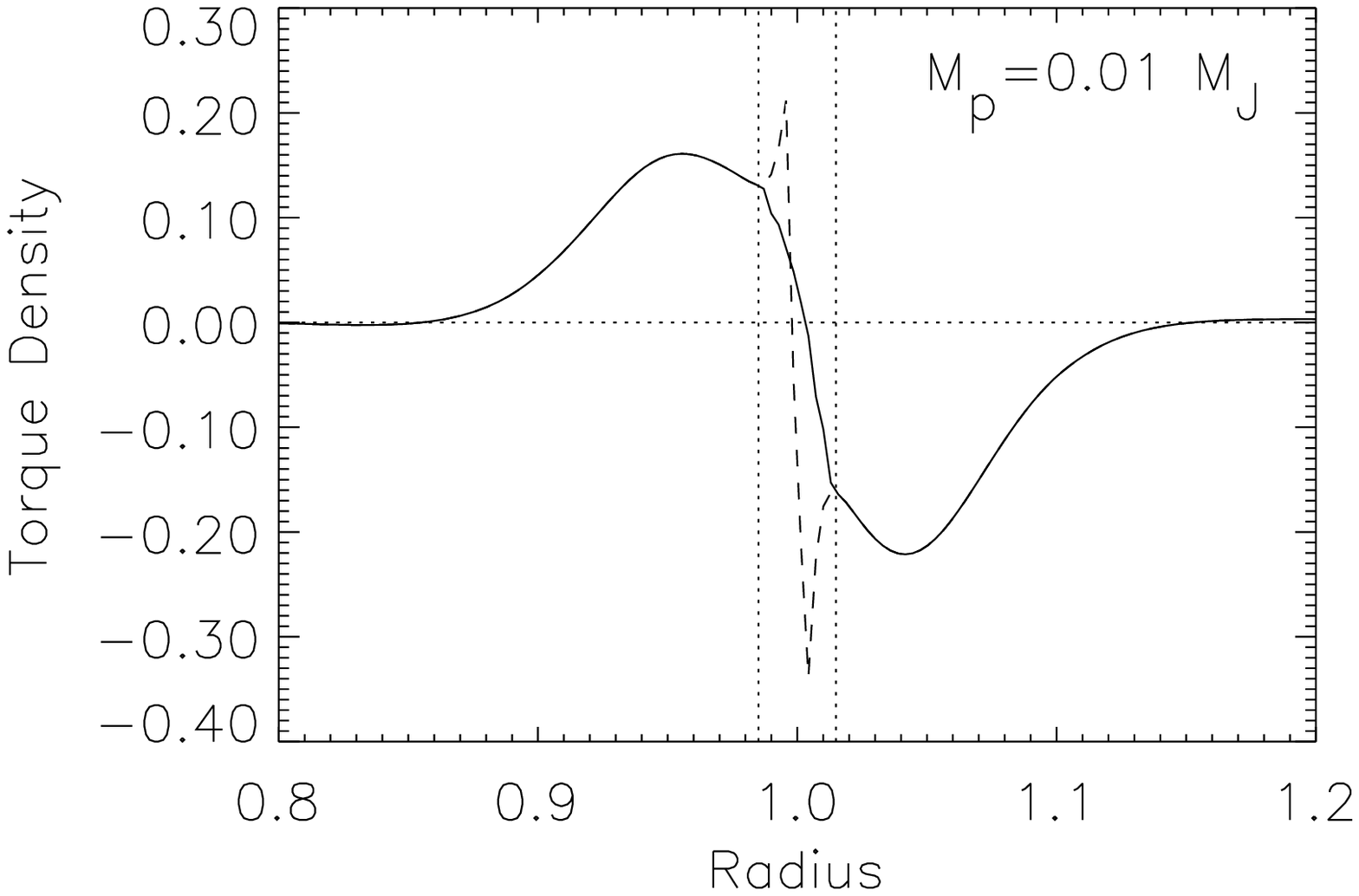,width=4.2truecm}\psfig{figure=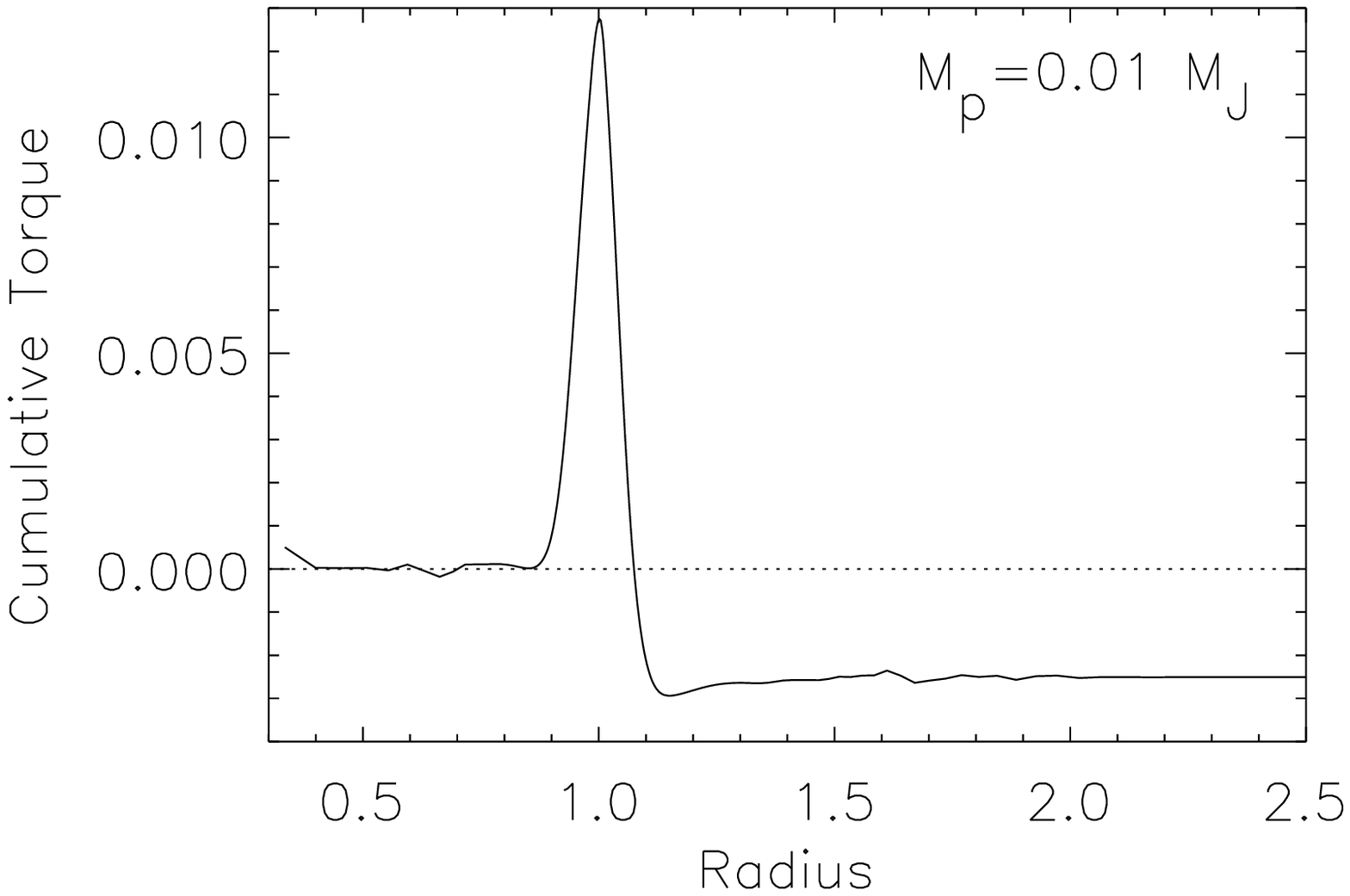,width=4.2truecm}\hspace{1.0cm}\psfig{figure=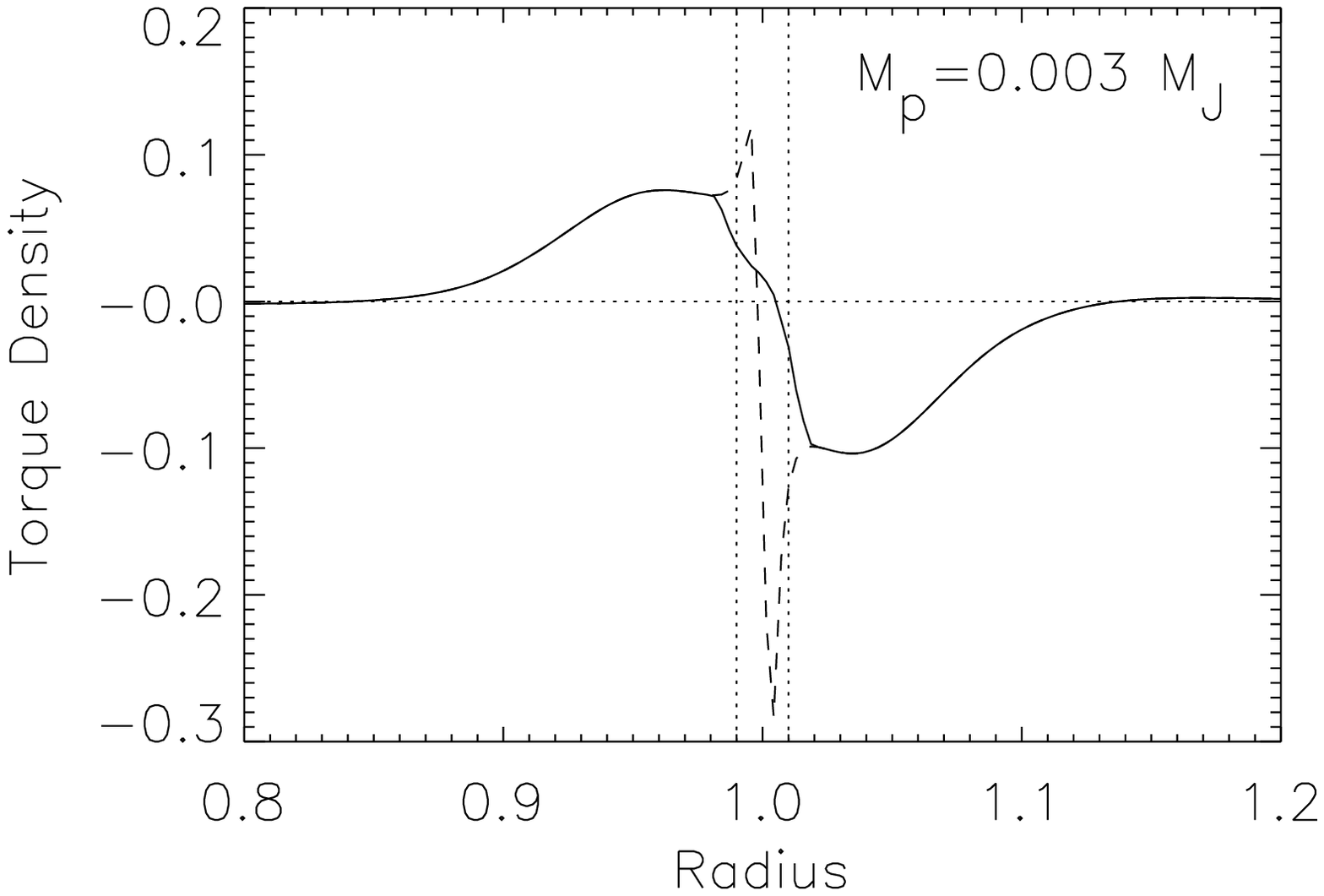,width=4.2truecm}\psfig{figure=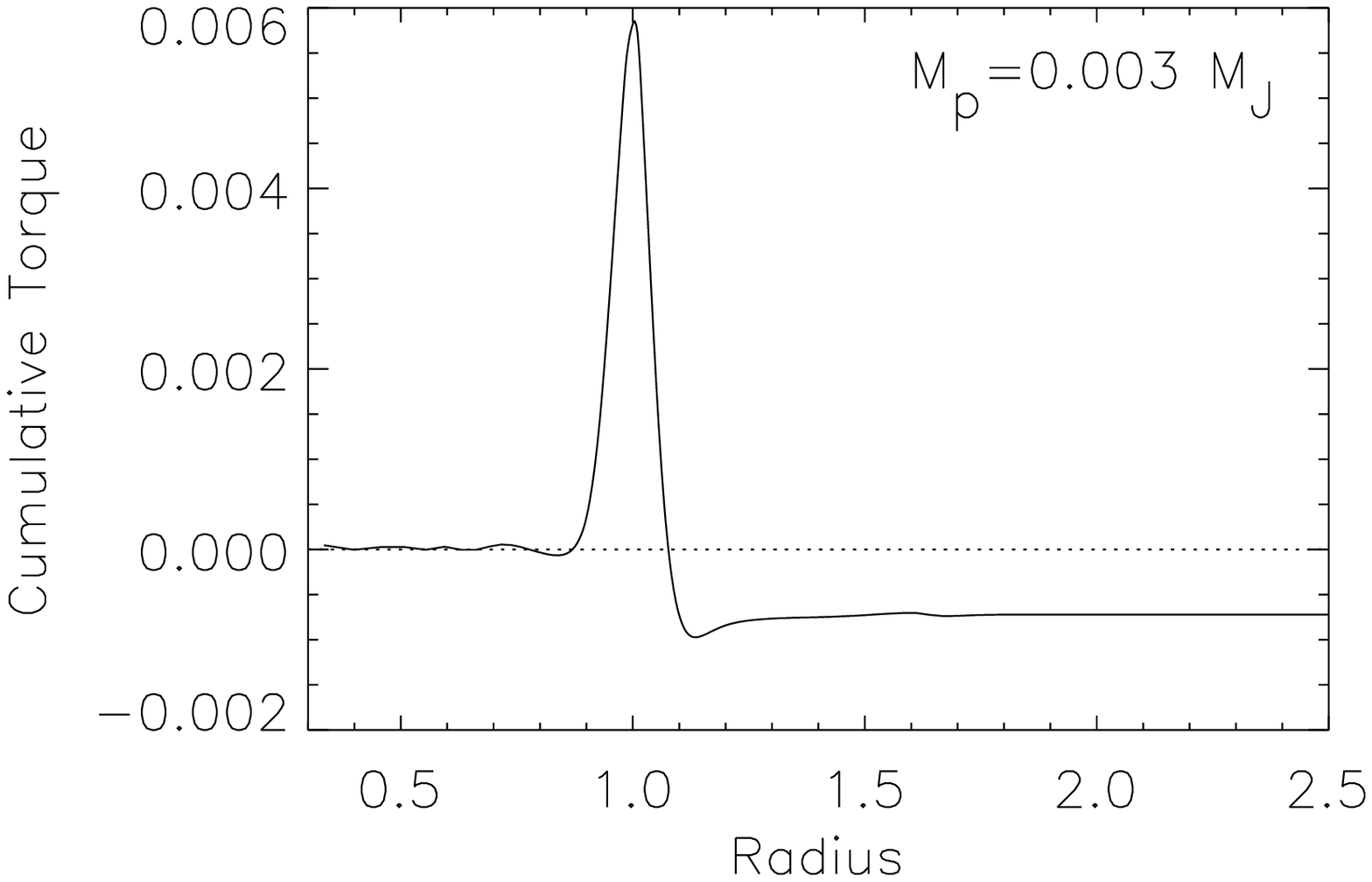,width=4.2truecm}}
\caption{\label{torque} The torque exerted by the disc on planets with masses of 1, 0.3, 0.1, 0.03, 0.01, and 0.003 \mj\ (top-left to bottom-right).  In each case there are two panels.  Left panel: The dashed line gives the torque density (torque per unit radius) as a function of radius calculated using all grid zones.  The solid line gives the torque density calculated only using those grid zones outside of the planet's Roche radius, $r_{\rm R}$.  The torque density is dimensionalised by multiplying by $G M_{\rm p} M_{\rm d}/(4\pi r_{\rm p}^2)$.  The dotted vertical lines show the size of the Roche lobe ($r=r_{\rm p} \pm r_{\rm R}$).  Right panel:  The solid line gives the cumulative torque as a function of radius, neglecting grid zones within $r_{\rm R}$ of the planet, except for the 1\me\ planet.  For the 1\me\ planet we neglect zones within $2r_{\rm R}$.  Note that the radial scale is larger than in the plots of torque density.  The cumulative torque is dimensionalised by multiplying by $G M_{\rm p} M_{\rm d}/(4\pi r_{\rm p})$.  For planets that do not open gaps (i.e.\ $\mpl\leq 0.1\mj$), essentially all of the torque comes from the region within $r=r_{\rm p}\pm 2H$.  In particular, the 2:1 resonance at $r\approx 1.59$ does not contribute significantly to the total torque.  In each case, the overall torque is negative giving inward migration.}
\end{figure*}

\subsubsection{The torque distribution}
\label{torquedistsec}

Linear theory predicts that the strongest resonances
occur at a radial distance of order $H$ from a planet. But
since the resonances have a non-zero width, the strongest torques
are exerted over a region within radial distance of order
$H$ from the planet.  In Figure \ref{torque}, we plot the 
radial distribution of the torque and the cumulative
torque as a function of radius (solid lines).  
For the low-mass
planets that do not begin to open a gap in the disc
($M_{\rm p}\leq 10 \me$), almost the 
entire torque comes from a region $r = r_{\rm p} \pm 2H$.
For higher-mass planets, the radial extent increases.
For 1\mj, the torque comes from 
$r = (1 \pm 0.25) r_{\rm p}$, which is consistent with
a region $r = r_{\rm p} \pm 2(r_{\rm R} + H)$ for the
high-mass planets.  The resulting cumulative torque
distributions are sharply peaked for low-mass planets and
broad for the high-mass planets.  Also visible in most of
the cumulative plots are low-order resonances (e.g. the
2:1 at $r \simeq 1.59 r_{\rm p}$), although they do not
contribute significantly to the total torque.
In all cases, the cumulative torque at large radius 
is negative, indicating inward migration.

As discussed above, in order to compare our results with
linear theory, we have specifically excluded torques
from material inside the Roche lobe of the planet
($r_{\rm c}=r_{\rm R}$).  In any case, as demonstrated 
by the gas flows in Figure \ref{hmid}, we only resolve the
flow inside the Roche lobe for $M_{\rm p}\geq 0.1 \mj$.
The question arises as to whether torques from material
inside the Roche lobe can affect the migration rate of
the planet.

In Figure \ref{torque}, the dashed lines give the 
torque exerted on the planet including all zones on
the numerical grid.  For $M_{\rm p}\geq 0.1\mj$ (for
which the flow inside the Roche lobe is resolved),
the net torque inside the Roche lobe is positive.
In fact, for the 1\mj\ case, the total torque is
slightly positive (i.e., outward migration).  The 
problem with including torques from deep inside the 
Roche lobe is that near the planet the torque per unit mass
becomes very large.  Therefore, small
departures from axisymmetry (either real or due to
finite numerical resolution) can produce a large
net torque that is comparable to the net torque from 
outside the Roche lobe.  As shown by the errorbars 
in Figures \ref{migtime} and \ref{migfit}, including 
torques into $r_{\rm c} = 0.5 r_{\rm R}$ only 
results in changes at the 20\% level 
(for $M_{\rm p}\geq 10 \me$).  Here the flow is still
reasonably well resolved.  Going deeper than this,
however, gives very unreliable results.

\begin{figure*}
\centerline{\psfig{figure=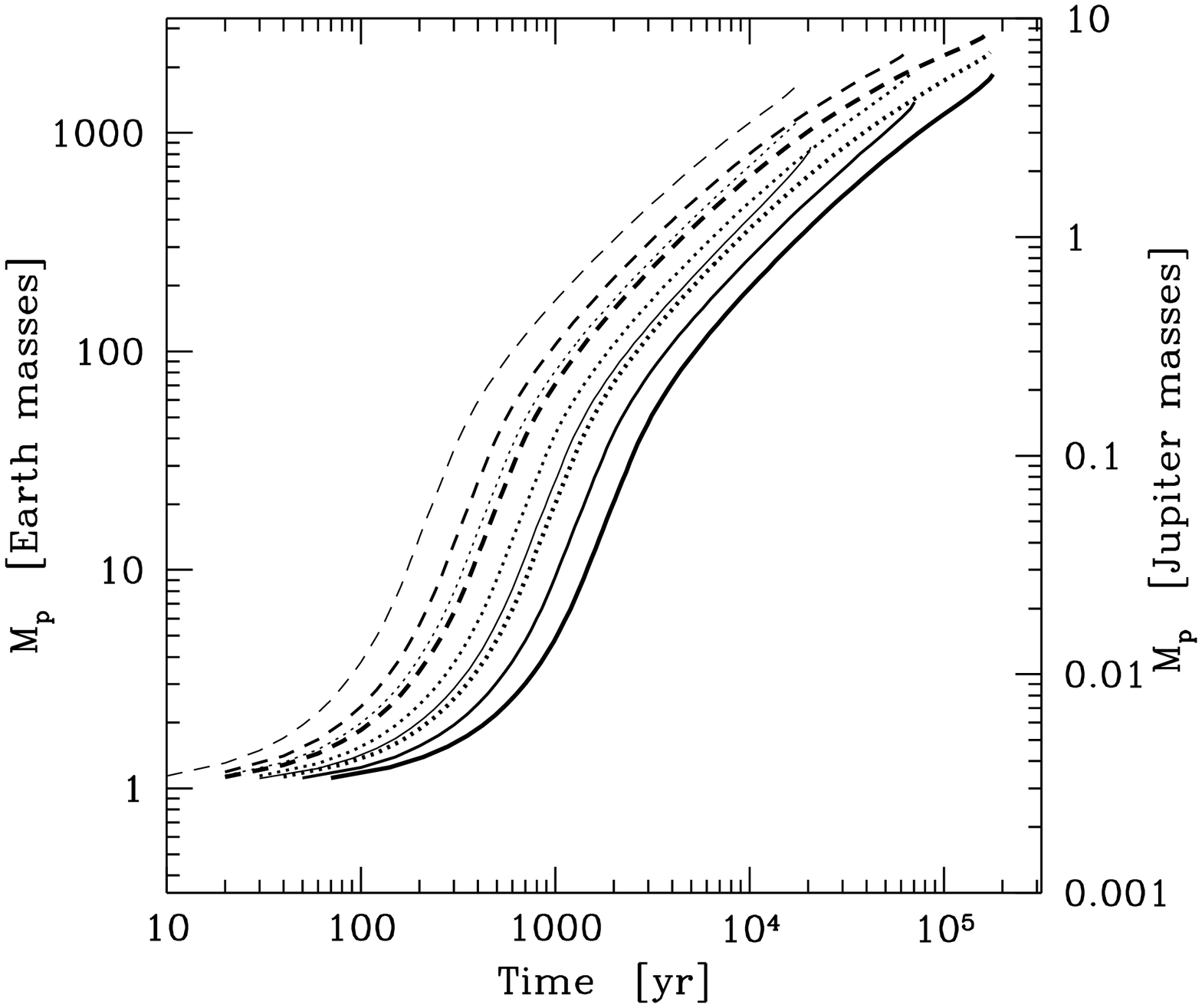,width=8.5truecm}\hspace{0.5cm}\psfig{figure=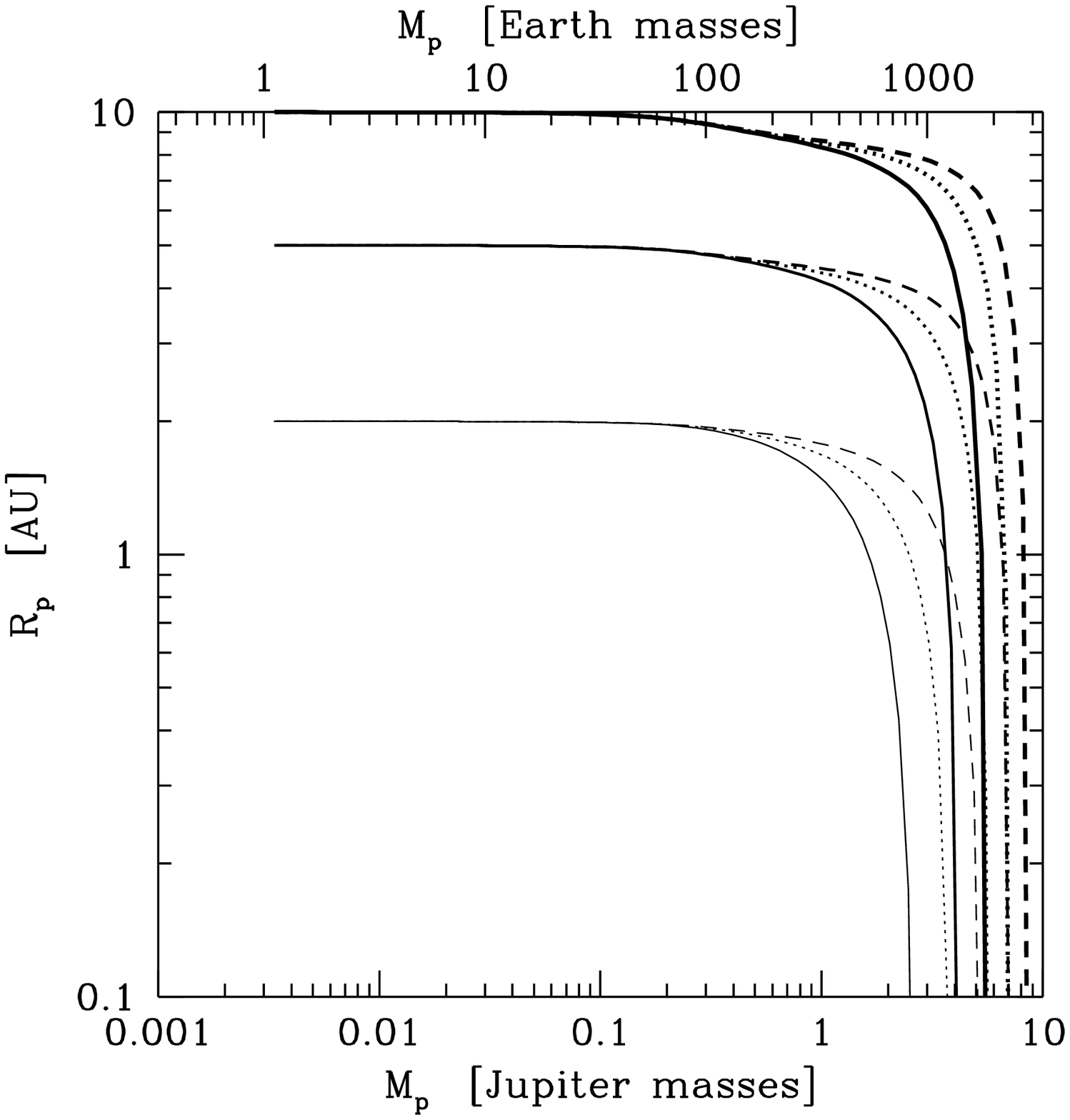,width=8.5truecm}}
\caption{\label{integration} Left: mass of an accreting protoplanet versus time.  Right: the protoplanet's orbital radius versus its mass.  The protoplanet is assumed to begin at an orbital radius of 2 (thin lines), 5 (medium lines), or 10 (thick lines) AU around a 1 \msun\ star.  The line types give the results for different disc surface densities: 75 g~cm$^{-2}$ (solid), 150 g~cm$^{-2}$ (dotted), 300 g~cm$^{-2}$ (dashed) at 5.2 AU.  It is assumed that the protoplanet can accrete gas at the rate at which the disc can provide it.  The evolution of the protoplanet could be stopped at any point by dispersal of the disc.  Notice that there is an upper limit of approximately 10 \mj\ to the mass of a planet.  Also notice that the mass and orbital radius of a giant planet and its gas accretion timescale do not depend significantly on the initial mass of the protoplanet as long as it is less than approximately $ 50 \me$.}
\end{figure*}

The same problem has been encountered in other studies.
Lubow et al.\ \shortcite{LubSeiArt1999} found that the 
contributions to the torque from radii near the planet
were in the opposite sense to the torque from further out
in the disc (i.e.\ radii outside but near the planet gave
a positive net torque and radii inside but near the planet
contributed a negative torque).  
D'Angelo et al.\ \shortcite{DAnHenKle2002} found that
taking into account torques near the planet usually 
resulted
in the migration switching from inward to outward.
They neglected the contribution from zones within
approximately $0.2 r_{\rm R}$ when calculating the net torque.
The problem is worse in these two dimensional calculations
than in three dimensions because of the strong spiral 
shocks within the
circumplanetary disc -- any asymmetry due to numerical
resolution results in a large net contribution.  

A related issue is that the material that is accreted by 
the planet and removed from the calculation carries some
angular momentum.  If this angular momentum were put back
into the orbital angular momentum of the planet it would 
reduce the migration rate of the planet.  The magnitude
of this effect depends on the radius of the region from which the
gas is removed.  In most of our calculations, the amount of angular
momentum that is removed is low.  If it were all added to the 
orbital angular momentum of the planet it would reduce the
migration rates by approximately 2\% in the cases of the 1 and 
0.3 \mj\ planets, by approximately 10\% percent for the 0.1, 0.03 and
0.01 \mj\ planets, and by approximately 50\% percent in the case of 
the 1 Earth-mass planet.  The large effect for the Earth-mass planet
is partially due to its accretion rate being overestimated (Section 3.2.2).

To accurately evaluate the torques from
inside the Roche lobe will require higher resolution,
three-dimensional calculations.  It may also require
a realistic equation of state, since low-mass planets 
may be unable to accrete gas at the rate the disc can
provide it because the thermal energy of the gas cannot
be radiated away quickly enough (see Section 4).  For the 
present, we have presented an accurate determination
of the torque from outside the Roche lobe and find
excellent agreement with the linear theory of Tanaka
et al.\ \shortcite{TanTakWar2002}.

\subsection{Further tests of Type I migration}

The discs in the above calculations had unperturbed scaleheights
$H/r=0.05$ and surface density profiles $\Sigma \propto r^{-1/2}$.
Tanaka et al.\ (2002) predicted how the migration timescales
should vary with the disc's thickness and surface density profile.  
To further test their predictions, we performed
calculations of 10 \me\ planets in discs with different thicknesses
and surface density profiles.

We performed a calculation with $H/r=0.10$.  According to
equation \ref{tanaka}, the planet should migrate 4 times 
slower than with $H/r=0.05$.  We find that the planet 
migration timescale increases by a factor of about $4.9\pm 1.4$,
where the estimated error comes from using values of $r_{\rm c}$ 
ranging between $r_{\rm c}=0.5r_{\rm R}$ and $1.5r_{\rm R}$.
We also note that the torque density 
and cumulative torque distributions are similar to those
in the $\mpl=0.03\mj$ panels of Figure \ref{torque}, except
they are spread over twice the radial range around the planet.
Thus, the torque is generated almost entirely
within the radial range $r = r_{\rm p} \pm 2H$, as discussed in
the previous section.

We performed a calculation with 
$\Sigma \propto r^{-3/2}$.  According to Tanaka et al.\ 
\shortcite{TanTakWar2002}, the planet should migrate 36\%
faster than with $\Sigma \propto r^{-1/2}$.  We find
that the planet migration timescale decreases by a factor
of about $1.8\pm 0.2$.  The distribution of the torque is similar
to the $\mpl=0.03\mj$ panels of Figure \ref{torque}.

Given the difficulties in measuring the torques, these
results are in reasonable agreement with the predictions of
Tanaka et al.\ \shortcite{TanTakWar2002}.

\section{Discussion}

The accretion and migration rates from the previous 
sections can be used to investigate the timescale for 
the gas accretion phase of giant planet formation.  
In the core-accretion model of giant planet formation,
giant planets are thought to form through gas accretion 
on to a solid core of about 10 \me\ (e.g., 
Mizuno 1980; Hayashi, Nakazawa \& Nakagawa 1985).  Runaway gas accretion
is thought to occur at higher masses, greater than about 
50 \me\ (Pollack et al.\ 1996).  
The overall formation timescale may be dominated by the
phase between the onset of gas accretion and the
beginning of runaway gas accretion.  During this stage,
the accretion on to the planet is limited by the energy losses
in the contracting envelope. 
This model is subject 
to several major assumptions, and there are issues
with the apparently small observationally inferred core mass of 
Jupiter (Pollack et al.\ 1996;
Wuchterl et al.\ 2000).  In any case, the accretion rates
during this intermediate stage are typically much smaller
than the rates we plotted in Figure \ref{accretion}. 
Consequently, the accretion rates we obtain
are valid, within the framework of the current calculations of
the core-accretion model, only during the
runaway gas accretion phase.

As a protoplanet accretes, it will migrate through the disc.  
Its rate of migration is given by Figure \ref{migfit}.
Thus, we can investigate the timescale and result 
of giant planet formation by starting with a 
core and integrating its mass and orbital radius 
forward in time.
The results of such integrations are given in Figure 
\ref{integration}.  The evolution of the protoplanet 
could be stopped at any point along the curves 
by dispersal of the disc.  We plot results for cores
beginning at three different orbital radii and for 
three different disc surface densities 
(i.e.\ three different disc masses).  
We have assumed that the disc surface 
density varies as $\Sigma \propto r^{-1/2}$, but the results 
are very insensitive to the index; 
$\Sigma \propto r^{-3/2}$ gives almost identical results.

For convenience, we assume the initial core mass is 
1 \me\ and ignore the possibility that the planet may not
be able to accept mass at the rates we obtain.  Notice, however,
the results we obtain are {\it insensitive to the value of
the threshold mass for rapid gas accretion}, as long as
it is less than approximately 50 \me.  
The time required for the gas accretion to produce a 
giant planet is almost independent of the initial 
core mass, because the mass doubling timescale for a 
low-mass core is much shorter than that for a 
high-mass planet (Figure
\ref{accrate}, right panel).  Thus, only a small
fraction of the total time (Figure \ref{integration}) 
is spent while it is a 
low-mass core.  Similarly, the orbital radius
of the planet is independent of the initial core mass
because the migration timescale of a low mass object is
much longer than its mass doubling timescale.
Thus, the timescale of runaway gas accretion phase and the 
final mass and orbital radii of a giant planet
are essentially independent of whether one assumes that
runaway gas accretion begins at 1 \me\ or 50 \me. 

We find the time required for a core to accrete to
1 \mj\ is only 
$t_{\rm acc}=2.0\times 10^3 - 2.0\times 10^4$ years.
None of the objects migrates significantly during 
this period, and the final orbital radius of the 
1 \mj\ planet simply depends on the initial orbital radius 
of the core.  

Much more of a problem is the formation of very massive
planets.  To form a 4 \mj\ planet requires an order
of magnitude longer 
($t_{\rm acc}=2.0\times 10^4 - 2.0\times 10^5$ years).
Cores that begin rapid gas accretion at 2 AU
migrate into the star before they reach 4 \mj, unless
the disc is massive ($\Sigma \gsim 200$ g~cm$^{-2}$ at 5.2 AU).
Cores that begin at larger radii and survive all migrate 
significantly during this period.
The formation of giant planet with $M_{\rm p}>7~\mj$
is only possible for cores that begin at large
distances in massive discs.  To form a giant planet of
10 \mj\ requires either a very massive circumstellar
disc, a large initial orbital radius, or both.

The above rates represent the fastest possible accretion and the
least possible migration.  Even then, the formation
of planets with masses $M_{\rm p} \simeq 10$\ \mj\ 
is difficult or impossible.  Thus, there is a natural
limit to the mass of a giant planet of $\simeq 10$\ \mj,
as suggested in Lubow et al.\ \shortcite{LubSeiArt1999}.
This is in good agreement with the mass distribution of 
extra-solar planetary systems (e.g.\ Basri \& Marcy 1997; 
Mayor, Queloz, \& Udry 1998; Mazeh, Goldberg, \& Latham 1998;
Halbwachs et al.\ 2000;
Zucker \& Mazeh 2001;
Jorissen, Mayor, \& Udry 2001;
Tabachnik \& Tremaine 2002).  However,
it is worth pointing out that our results apply only
to planets with circular orbits.  Planets with highly eccentric 
orbits may accrete significantly even for masses $\mpl >10$ \mj.
Thus, the eccentric orbits of many observed extrasolar planets 
may assist them in growing to large masses.

\section{Conclusions}

We simulated the three-dimensional interaction of a young planet,
ranging in mass from 1 \me\ to 1 \mj, with a gaseous disc by means of
the ZEUS hydrodynamics code. The disc was vertically isothermal
with an unperturbed disc thickness ratio was
$H/r = 0.05$ and disc turbulent viscosity parameter was $\alpha = 4
\times 10^{-3}$. We have analyzed the flow patterns, the accretion
rates, and migration rates.  Each planet was assumed to remain in a
circular orbit and accrete gas without expansion on the scale of its
Roche lobe. To incorporate the latter assumption into the simulations,
we removed mass from the grid zones that immediately surround the planet
at each timestep.

  Only planets with masses $\mpl\gsim 0.1$ \mj\ produced significant
perturbations in the disc's surface density; $0.03$ \mj\ (10 \me) is
insufficient (see Figure \ref{surfdens}).  The flow near the planet is fully
three-dimensional (see Figure \ref{slice}).  The flow 
at the disc midplane generally involves material that passes by the
planet, two gas streams that penetrate the Roche lobe and supply material
to a circumplanetary disc, and material on
horseshoe orbits (Figure \ref{zoomsurf}).  
These features are similar
to those found previous two-dimensional studies (e.g., Bryden et al.\ 1999;
Kley 1999; Lubow et al.\ 1999), but with some
important differences. 

The shocks in the circumplanetary flow
are much weaker in the three-dimensional case than in the
two-dimensional case (see Figures \ref{hmid} and \ref{hmid2d}).
The circumplanetary disc in three dimensions is likely to
behave more like a standard accretion disc, rather than
being subject to strong shock-driven accretion, as is the
case in two-dimensions.  

The gas streams at midplane are narrower in the three-dimensional
case. However, the overall efficiency of accretion in three dimensions
is still high.  The accretion rate peaks at approximately 0.1 \mj,
but a 1 \mj\ planet still accretes mass at a rate greater than the
usual local viscous rate. 
This result suggests that the lower accretion occurring
at the midplane in the three dimensional calculations 
is compensated by accretion occurring from
above the midplane. For small mass planets, the accretion occurs with a
cross-section whose length scale is of order the size of the Roche lobe
(see Figures \ref{zoomsurf} and \ref{accrate}).  This leads to the
accretion rate increasing in proportion to the planet's mass (equation 8).

We investigated the disc torques on planets and the resulting migration
timescales. The migration timescales obtained
 from torques exerted outside the planet's Roche lobe are in excellent
agreement with recent linear theory (Tanaka et al.\ 2002), as seen
in Figure \ref{migtime}. The transition from Type I (non-gap) migration
to Type II (gap) migration occurs at $\mpl \approx \mj/4$. The transition
is smooth with only about a factor of 2 difference in rates.
The torque outside the Roche lobe is concentrated in a region
of distance of order $H$ from the planet, as expected from linear theory.

It is not yet possible to determine the torques from within the
Roche lobe of a planet. The problem is that the torque per unit
mass increases as the distance from the planet decreases. Consequently,
small numerical  fluctuations in the density close to the planet
give rise to numerical noise in the torque (see Figure \ref{torque}).
On the other hand, it is not clear that strong torques can arise
close to the planet, since the outer portions of
the circumplanetary disc in three dimensions
appear to be quite smooth.

We considered the orbital and mass evolution of a planet, based
on the orbital migration and mass accretion rates we obtained (Figure
\ref{integration}). For low-mass protoplanets, less than about 50 \me,
the accretion rate may be limited by the response of the planet (see
discussion in Section 4). In the absence of this limit, a planet
could gain considerable mass with little migration. Starting
with a higher mass core of say 50 \me, higher mass
planets, up to about 2 \mj, can also grow with little migration.
Masses as high as 10 \mj\ do not appear possible
(as was also suggested in Lubow et al.\ 1999), because the planet
will migrate inward over a large distance before accreting enough
material.

\section*{Acknowledgments}

The computations reported here were performed using the 
U.K.\ Astrophysical Fluids Facility (UKAFF) and the 
GRAND computer.  MRB is grateful for support from the 
Space Telescope Science Institute's Visitor's Programme
and the Institute of Astronomy's Visitor's Programme.  
SHL acknowledges support from NASA grants NAG5-4310 and 
NAG5-10732.

\end{document}